\documentclass[lettersize,journal]{IEEEtran}

\usepackage{soul,xcolor}
\usepackage{bm}
\include{graphics}
\include{amsmath}
\usepackage[normalem]{ulem}
\usepackage{multirow,enumitem}
\usepackage{algorithm}
\usepackage{algpseudocode}
\usepackage{verbatim}
\usepackage[latin1,utf8]{inputenc}
\usepackage{amsmath}
\usepackage{amsfonts}
\usepackage{amssymb}
\usepackage{mathrsfs}
\usepackage{booktabs}
\usepackage{url}
\usepackage{ifthen}
\usepackage{xspace}
\usepackage{dsfont}
\usepackage{bbm}
\usepackage{cite}
\usepackage{epsfig}
\usepackage{epstopdf}
\usepackage{array}
\usepackage{algpseudocode}
\usepackage{algorithm}
\usepackage{breqn}
\usepackage{esint}
\usepackage{amsthm}
\usepackage{setspace}
\usepackage{lipsum}
\theoremstyle{plain}

\theoremstyle{definition}

\theoremstyle{remark}

\usepackage{footnote}
\theoremstyle{plain}

\usepackage{fancybox,framed}

\usepackage{arydshln}

\usepackage{todonotes}

\usepackage{stmaryrd}
\usepackage{algorithm,algcompatible}
\algnewcommand\INPUT{\item[\textbf{Input:}]}%
\algnewcommand\OUTPUT{\item[\textbf{Output:}]}%
\usepackage{lipsum}
\usepackage{mathtools}
\usepackage{cuted}
\usepackage{dblfloatfix}
\usepackage{graphicx}
\usepackage{capt-of}
\usepackage{varwidth}

\usepackage{relsize}

\newcommand{\mathleft}{\@fleqntrue\@mathmargin0pt}
\newcommand{\mathcenter}{\@fleqnfalse}
\newcommand{\add}[1]{#1}

\newcommand{\RN}[1]{%
	\textup{\uppercase\expandafter{\romannumeral#1}}%
}
\ifCLASSINFOpdf

\else

\fi

\begin{document}
%		\relscale{0.94}
	\setulcolor{red}
	\setul{red}{2pt}
	\setstcolor{red}
	\title{Compressed Training for Dual-Wideband Time-Varying Sub-Terahertz Massive MIMO}
%	\title{Dual-Wideband Sub-Terahertz Massive MIMO: A Compressed Training Framework}
%	\title{Dual-Wideband Time-Varying\\ Sub-Terahertz Massive MIMO Systems:\\ A Compressed Training Framework}
	%	\title{Compressed Sensing-Aided Channel Training for Wideband Sub-Terahertz Massive MIMO Systems}
	\author{Tzu-Hsuan~{Chou}, Nicol\`{o} {Michelusi}, David J. {Love}, and James V. {Krogmeier}
		% \nm{coauthors?}
%		\begin{singlespace}			
		\thanks{This work was supported in part by the National Science Foundation under grants CNS-1642982, CCF-1816013, EEC-1941529 and CNS-2129015. A preliminary version of this paper was presented at IEEE Globecom 2021 \cite{chou2021wideband}.}
		\thanks{T.-H. Chou is with Qualcomm Inc., San Diego, CA, USA; email: tzuhchou@qti.qualcomm.com.}
		\thanks{D. J. Love and J. V. Krogmeier are with the School of Electrical and Computer Engineering, Purdue University, West Lafayette, IN, USA; emails: \{djlove, jvk\}@purdue.edu.}% 
		\thanks{N. Michelusi is with the School of Electrical, Computer and Energy Engineering, Arizona State University, AZ, USA; email: nicolo.michelusi@asu.edu.}%
	}

	\maketitle
	\begin{abstract} 
		6G operators may  use millimeter wave (mmWave) and sub-terahertz (sub-THz) bands to meet the ever-increasing demand for wireless access.
		Sub-THz communication comes with many existing challenges of mmWave communication and adds new challenges associated with the wider bandwidths, more antennas, and harsher propagations.
		Notably, the  frequency- and spatial-wideband (dual-wideband) effects are significant at sub-THz.
		This paper presents a compressed training framework to estimate the time-varying sub-THz MIMO-OFDM channels.
		A set of frequency-dependent array response matrices are constructed, enabling  channel recovery from multiple observations across subcarriers via multiple measurement vectors (MMV).
		Using the temporal correlation, MMV least squares (\text{LS}) is designed to estimate the channel based on the previous beam support, and  MMV compressed sensing (\text{CS}) is applied to  the residual signal.  We refer to this as the MMV-LS-CS framework. 
		Two-stage (TS) and  MMV FISTA-based (M-FISTA) algorithms are proposed for the MMV-LS-CS framework.
%		\nm{, attaining a trade-off between estimation accuracy and execution time (or complexity?)}\tc{There is a 200-word abstract constraint, so I am not able to add it in.}.
		Leveraging the spreading loss structure, a channel refinement algorithm is proposed to estimate the path coefficients and time delays of the dominant paths.
		To reduce the computational complexity and enhance the beam resolution, a sequential search method using hierarchical codebooks is developed. % in MMV-LS-CS.      
		Numerical results demonstrate the improved channel estimation accuracy of \text{MMV-LS-CS} over state-of-the-art techniques.
	\end{abstract}

	
	%	\nm{The paper \cite{xing2021millimeter} is on IEEE xplore! Please cite IEEE not arxiv.}

	\section{Introduction}
	Future wireless applications will require networks to provide high rates, reduced power consumption, and  low latency in a wide range of deployment scenarios  \cite{andrews2014will}.  
	These requirements motivate policy and technical research on making spectrum at higher frequencies, outside  the popularly used sub-6 GHz spectrum, commercially available for wireless broadband thanks to large bandwidth availability. 
	In 5G, this technical work culminated in the standardization and deployment of communications in the millimeter wave  (mmWave) spectrum.  
	
%	\subsection{Motivations}	 
	In the sub-terahertz (sub-THz) spectrum, roughly defined as $100-300$ GHz, there is approximately $21.2$ GHz available for wireless broadband  \cite{rappaport2019wireless}.  Making these bands ``usable'' promises to alleviate backhaul and access concerns far into the future \cite{rappaport2019wireless,xing2021millimeter,ju2021millimeter}. 
	Unfortunately, sub-THz communication poses signal processing and wideband communication challenges unique to the spectrum.  
	Notable challenges include the need for i) a dramatic increase in bandwidth and antenna aperture and ii) signal processing techniques that leverage the unique propagation features. 
%	Most notables are the need for a dramatic increase in antenna aperture and bandwidth, and for signal processing techniques to deal with the unique propagation features. 
	As dictated by Friis'  equation, the spreading loss increases with the operating frequency, inducing a more serious issue in sub-THz than in mmWave or sub-6 GHz systems \cite{lin2015adaptive}. 
	Caused mainly by water vapor, molecular absorption loss becomes more pronounced as the frequency increases \cite{jornet2011channel}.
%	e.g., due to propagation in water vapor, becomes more pronounced as the frequency increases \cite{jornet2011channel}.
	Also, the reflection coefficient \cite{piesiewicz2007scattering} should be considered when modeling the pathloss of the sub-THz channel in non-line-of-sight (NLOS).
	By combining these effects together, the overall pathloss is a frequency-selective attenuation.
%	By combining the spreading loss, molecular absorption loss, and reflection coefficient, the overall pathloss is a frequency-selective attenuation.
	To compensate for such severe signal attenuation, sub-THz systems require highly-directional beamforming via massive multiple-input multiple-output (MIMO) \cite{larsson2014massive}.
%	sub-THz systems require highly-directional beamforming to compensate for the severe signal attenuation, which can be achieved using massive multiple-input multiple-output (MIMO) \cite{larsson2014massive}.
%	 \cite{larsson2014massive,bjornson2019massive}.
	Thanks to the short wavelength of sub-THz signals, massive MIMO is possible even with a small form factor.
	However, the hardware limitations (e.g., high power consumption and cost) of the radio frequency (RF) chains in massive MIMO make digital baseband precoding infeasible.
	An analog precoder overcomes the hardware limitation but suffers from performance degradation.
	To achieve larger precoding gains with a limited number of RF chains, hybrid  transceiver structures have been proposed and investigated in \cite{6847111,dovelos2021channel}.
	Our work focuses on the design of a channel training algorithm under a hybrid transceiver structure.
%	Nevertheless, the channel training overhead
%	\nt{hybrid precoder to be added.}

	\textbf{Orders-of-Magnitude Larger Bandwidths and Wideband Effects:}	
	 \emph{Frequency-wideband} and \emph{spatial-wideband} effects of the MIMO channel will be exacerbated in sub-THz systems. 
	The \textit{frequency-wideband effect} is caused by  the delay spread of the multipath channel responses.
%	 and the accompanying  frequency-selective channel responses.
%	In addition, t
	The \emph{spatial-wideband} effect arises from differences in time delays across the antenna aperture.  
	This primarily arises because the angles of arrival of the  propagation paths observed by the receiver vary within the operating frequencies, known as the \emph{beam squint effect} \cite{wang2018spatial,wang2019beam,wang2019block,lin2020tensor,dovelos2021channel,tan2021wideband}.
	Most existing works focusing on sub-6 GHz or mmWave systems \cite{6847111,marzi2016compressive,venugopal2017channel,zhou2017low,araujo2019tensor,gao2015spatially,gao2016channel,choi2005interpolation,pande2006weighted,pande2007reduced} consider only the frequency-wideband effect but ignore the spatial-wideband effect since it is negligible when the available bandwidth is not very wide, which is a reasonable assumption in current and past wireless deployments.
However, sub-THz communication systems may use one order of magnitude more bandwidth than mmWave communication systems \cite{rappaport2019wireless}, hence are more severely affected by the wideband effect.
Motivated by this fact,
 in this paper, we are concerned with both effects, meaning that we consider systems with \emph{dual-wideband} (also known as \emph{spatial-frequency wideband}) effects. 
 To this end, we develop an accurate channel model capturing the dual-wideband effect for sub-THz communication systems. 
% Thus, \hl{we must define}\nm{are you doing this? You need to be more explicit about this. "Tho this end, we develop an ...."} an accurate channel model for sub-THz communication \nm{, "capturing these effects"}.
%	A few recent works \cite{wang2018spatial,wang2019beam,wang2019block,lin2020tensor} \hl{investigate the combined effects on communication and channel estimations in mmWave systems with  bandwidths up to $1$GHz.}
% \nm{how are you different from these works?? Maybe this is not the right place to mention them (this phrase does not seem to be well connected to the rest) but it should be addressed in related work sec. Also, note that I restructured this paragraph in a more logical way..}

%	\hl{Since sub-THz communication can utilize  one order of magnitude more bandwidth than mmWave communication} \cite{rappaport2019wireless} \hl{and is thus  more severely affected by the wideband effect (including dual-wideband, Friis' equation, molecular absorption loss and NLOS reflection), which must be taken into account with an accurate channel model.}\nm{dont you see that this phrase is grammatically wrong?? PLEASE PROOFREAD THE PAPER!}
	
	Because of the promising beamforming gains, massive MIMO communication systems with dual-wideband effects are employed to compensate for the impairments of sub-THz channels.
%	To achieve promising beamforming gains, massive MIMO must be employed to compensate for the impairments of sub-THz channels. 
%	We consider massive MIMO communication systems with dual-wideband effects.	
%	To achieved\pf their promised beamforming gains, massive MIMO systems must compensate for the sub-THz channel impairments while dealing with  dual-wideband effects caused by the wideband communications.
	MIMO orthogonal frequency division multiplexing (MIMO-OFDM) has been envisioned as a vital tool to combat  the frequency-wideband effect and inter-symbol interference of the multipath channel.	
	For  MIMO-OFDM channel estimation, the works \cite{gao2015spatially,gao2016channel} recover the channel from multiple measurements among subcarriers sharing a common support, known as \emph{multiple measurement vectors} (MMV).		
	Yet, the works \cite{gao2015spatially,gao2016channel} neglect the spatial-wideband effect, which breaks the common support assumption across subcarriers and degrades the estimation accuracy.
	The work \cite{dovelos2021channel} addresses the issue by designing a set of frequency-dependent dictionary matrices to preserve the common support across subcarriers, mitigating the estimation losses from  the spatial-wideband effect.
	However, the work \cite{dovelos2021channel} does not exploit the temporal correlation in the time-varying MIMO-OFDM systems to aid the channel training, which is the focus of this paper.
	
%\tc{The MMV approach for the dual-wideband MIMO-OFDM is developed and addressed in \cite{dovelos2021channel}. Our work focuses on the time-varying case. We develop a MMV-LS-CS algorithm aided by the previous channel support to estimate the time-varying dual-wideband MIMO channel.}
%	\nm{Ok, but the reason I asked is not because you need to tell me, its is because you need to address it IN THE MANUSCRIPT. IT is still not addressed. Say here something like": "Yet, the work 13 does not do XA dn Y, which is the focus of this paper."}

	\textbf{Few dominant paths and channel sparsity:}
	Massive MIMO requires accurate channel state information (CSI) acquired via channel estimation to enable coherent alignment in narrow beam communications for high beamforming gains.
%	\hl{deal with the severe signal attenuation.}\nm{I don't think this is the primary reason for the need for channel estimation. In mMIMO, you need accurate ch-est in order to enable coherent alignment and achieve high BF gain.}
%	To compensate for the severe signal attenuation, massive MIMO requires accurate channel state information (CSI) acquired via channel estimation.
	Yet, traditional MIMO channel estimation techniques are
	impractical due to the prohibitive overhead that comes with (non-adaptive) omni-directional training over a 
	large number of antennas.
%	The burden of MIMO channel estimation 
	This burden can be reduced by exploiting the fact that the MIMO channel in sub-THz bands is determined by the geometry (positions and antenna geometry) of the transmitter and receiver, and exhibits a high degree of channel sparsity, with few propagation clusters.  
	For instance, the work \cite{nguyen2018comparing} reported an average of 6 clusters and 4 multipath components (MPCs) per cluster at 140GHz, compared to 8 clusters and 5 MPCs per cluster at 28GHz.
	One approach is to utilize a predetermined set of beams (e.g., a codebook), usually designed to allocate power in specific directions, and to tailor the channel training algorithm to this lower-dimensional beam set instead of the true higher-dimensional MIMO channel.	
%	beam training algorithm to this lower-dimensional beam set \hl{instead of on} the true higher-dimensional MIMO channel.
	For example, the most-discussed approach \cite{nitsche2014ieee, giordani2018tutorial} is to scan over  some set of candidate beams and estimate the strongest (e.g., through received power measurement), but even this approach  incurs a prohibitively large overhead due to the typically large size of the beamforming codebook.
	This issue is exacerbated in time-varying channels, due to the need for periodic beam training.
	For time-varying channels, one approach is to model the dynamic behavior of the channel as a birth-death process of MPCs  \cite{wu2014non,chen2019time}, which models the temporal correlation of the surviving MPCs.
	The channel sparsity, combined with the slow temporal variations, results in slowly-varying beam support, which could be exploited for MIMO channel estimation \cite{han2017compressed}.
	Channel estimation algorithms exploiting the temporal correlation of the channel, via the use of a common (or slow-varying) channel support over time, have been studied for narrowband MIMO systems
%	\hl{the narrowband MIMO}\nm{narrowband MIMO systems} 
	in \cite{han2017compressed} and for frequency-wideband multiuser MIMO-OFDM systems
%	\hl{the frequency-wideband multiuser MIMO-OFDM} 
	in \cite{gao2015spatially}.
	Nevertheless, these techniques cannot be directly applied to time-varying dual-wideband MIMO-OFDM channels due to the beam squint effect and frequency-dependent path gains, which may harm the estimation performance.
	To address the issues, we propose a new channel training framework (MMV-LS-CS) in time-varying dual-wideband MIMO-OFDM systems with a hybrid transceiver structure and a channel refinement algorithm that leverages the spreading loss structure to improve the estimation performance by deriving the path coefficients and time delays of the dominant paths across the pilot subcarriers.
%	To address these issues, we propose a new \text{channel training framework (MMV-LS-CS)} for time-varying MIMO channels with the dual-wideband effect and a \text{channel refinement algorithm} that leverages the spreading loss structure to improve the estimation performance by deriving the path coefficients and time delays of the dominant paths across the pilot subcarriers.

	\noindent\textbf{Prior Works}

	Over the last decade, researchers have focused on MIMO channel estimation in mmWave and \text{(sub-)THz} bands.
	%	To reduce the training overhead, 
	Several works focused on the beam alignment problem, including \emph{feedback-based schemes} \cite{hussain2018energy, hussain2020mobility, hussain2021learning, 6600706,booth2019multi, love2003equal, love2003grassmannian, love2008overview,8851228}, \emph{data-assisted schemes} \cite{chou2021fast,8101513,gonzalez2016radar, 8642397}, and \emph{multipath estimation} \cite{6847111,marzi2016compressive,han2017compressed,venugopal2017channel,zhou2017low,araujo2019tensor,gao2015spatially,gao2016channel,choi2005interpolation,pande2006weighted,pande2007reduced,wang2018spatial,wang2019beam,dovelos2021channel,wang2019block,lin2020tensor}.
	{Feedback-based schemes} adapt the beam training according to the feedback information sent from the receiver in an online fashion \cite{hussain2018energy} or leverage the UEs' mobility as in \cite{hussain2020mobility,hussain2021learning}.
	{Data-assisted schemes} perform the beam training by using side information from other available sources, e.g., GPS positional information \cite{chou2021fast}, lower-frequency communication \cite{8101513}, radar \cite{gonzalez2016radar}, and LIDAR \cite{8642397}.
	{Multipath estimation} schemes can exploit the channel sparsity of the MIMO channel via compressed sensing (CS) to acquire the associated channel parameters, e.g., angles of arrival (AOAs), angles of departure (AODs), time delays, and path gains.	
	The channel training proposed in this work is a form of {multipath estimation}.

	The \emph{narrowband} MIMO channel estimation problem has been investigated in \cite{6847111,marzi2016compressive}.	
	The work \cite{6847111} proposes an adaptive algorithm for mmWave massive MIMO channel estimation	using a hierarchical multi-resolution codebook.	
	With an adaptive structure, the work \cite{marzi2016compressive} proposes channel estimation using a compressive beacon codebook with different pseudorandom phase settings of the antenna arrays.
	In contrast to \cite{6847111,marzi2016compressive}, our work focuses on  \emph{wideband} MIMO channel estimation, taking advantage of the abundant bandwidth available in sub-THz bands.

	Initial works on wideband MIMO channel estimation considering the \emph{frequency-wideband} include \cite{venugopal2017channel,zhou2017low,araujo2019tensor,gao2015spatially,gao2016channel,choi2005interpolation,pande2006weighted,pande2007reduced}.
	Assuming channel sparsity, the work  \cite{venugopal2017channel} formulates  MIMO-OFDM channel estimation as a sparse recovery problem solved via orthogonal matching pursuit (OMP).
	The work \cite{zhou2017low} applies a tensor decomposition to the training signal with multiple dimensions corresponding to the beams and subcarriers and proposes a CANDECOMP/PARAFAC decomposition-based algorithm.
	The work \cite{araujo2019tensor} proposes a CS-aided channel estimation using the Tucker tensor as a compressible representation and reconstruction by tensor-OMP (T-OMP). 
	These works, however, do not leverage the \emph{dual-wideband} structure of wideband MIMO-OFDM channels.

    Dual-wideband estimation of MIMO-OFDM channels has been studied in \cite{wang2018spatial,wang2019beam,wang2019block,lin2020tensor,dovelos2021channel,tan2021wideband}.
	The challenges of dual-wideband MIMO in mmWave are outlined in \cite{wang2018spatial}, and  a channel estimation strategy exploiting the asymptotic characteristics of the channel is proposed.
	The work \cite{wang2019beam} presents a CS-aided channel estimation on dual-wideband MIMO-OFDM exploiting  uplink/downlink channel reciprocity.
	The work in \cite{wang2019block} considers the block sparsity of the beam squint effect to design a CS-based channel estimation.
	A tensor-based channel training using the Vandermonde constraint and spatial smoothing method is proposed in   \cite{lin2020tensor}.
	The dual-wideband effect in THz communications with uniform planar arrays (UPAs) is studied in \cite{dovelos2021channel},  and  an algorithm (GSOMP) to recover the channel by  simultaneous OMP exploiting the common support across the subcarriers preserved by the frequency-dependent dictionary matrices is proposed (evaluated numerically in Section \ref{sec_simulation}).
	These	prior works, however, do not address channel estimation over time-varying channels. In fact,	the channel tends to manifest temporal correlation between consecutive frames, which may be leveraged to improve channel estimation.
	In our work, we devise a channel training framework in dual-wideband MIMO-OFDM that uses the temporal correlation and common support across the frequencies to improve the estimation performance.

	\noindent\textbf{Contributions}
		
	We develop a support tracking-based channel training framework utilizing the estimated previous channel support for time-varying sub-THz dual-wideband MIMO-OFDM systems.
	%	We formulate the channel into a third-order Tucker tensor model.
	%	With the hybrid transmitter/receiver structure, the measurement signal is constructed by the $n$-mode product of the channel tensor with the precoded signal matrix, the combining matrix, and the subcarrier selection matrix along its corresponding mode.
	For the {spatial-wideband effect}, a set of frequency-dependent array response matrices are constructed to preserve the common channel support, leading to the MMV formulation of recovering the sparse beamspace channel from multiple observations across subcarriers. % (MMV problem).
	%	Exploing
	In a slowly-varying channel, the channel supports of consecutive frames tend to share many common elements, enabling the LS-CS residual approach \cite{vaswani2016recursive,vaswani2010ls}, that substitutes the \text{CS on the measurements} with the \text{CS on the LS residual signal}.
	We propose channel estimation that incorporates the MMV and LS-CS residual approach, called two-stage MMV-LS-CS (TS). %, called MMV-LS-CS.
	In addition, we formulate a joint MMV-LS-CS optimization problem,
	and solve it using the framework of the Fast Iterative Shrinkage-Thresholding Algorithm (FISTA) \cite{beck2009fast}, called MMV FISTA (M-FISTA).

	\add{In summary, the contributions of this paper are as follows:
	\begin{itemize}
		\item We propose a support tracking-based channel training framework, \textbf{MMV-LS-CS}, for time-varying dual-wideband MIMO-OFDM systems with hybrid transceivers using the estimated previous channel support.
		\item 
		We propose two channel estimation algorithms in MMV-LS-CS framework: \textbf{TS} and\textbf{ M-FISTA}.
		\begin{itemize}
			\item The TS algorithm does MMV-LS estimation on the estimated previous channel support, followed by MMV-CS estimation on the residual to estimate the time-varying channel components.	
			Note that TS can reduce its average execution time by parallel processing.		
			However, inaccurate estimated previous channel support might induce the channel estimation error in MMV-LS, and an increased number of channel elements is expected to be estimated in MMV-CS.			
%			an inaccurate previous channel support induces the wrongly estimated channel elements in MMV-LS, which degrades the estimation performance due to estimation errors or an increased number of elements to be estimated in MMV-CS.
			\item To avoid the issue, the M-FISTA algorithm uses the framework of FISTA to solve a joint version of MMV-LS-CS optimization problem for the channel estimation. 
			Unlike TS, M-FISTA is more resilient to the inaccuracy in the previous estimated channel. However, M-FISTA can not operate with parallel processing to reduce the average execution time.
			\item Numerical results show that M-FISTA achieves a more accurate channel estimation than TS.
			Nonetheless, TS with parallel processing requires only $32$\% of the average execution time of M-FISTA.
			It shows the trade-off of estimation accuracy and computational complexity between	these two algorithms.
%			We observe that M-FISTA achieves a more accurate
%			TS with parallel processing attains a larger NMSE of estimated channel with a smaller average execution time, which exhibits the trade-off of estimation accuracy and computational complexity between
%			these two algorithms.
%			achieves a more accurate estimation performance due to its resilience to the inaccuracy in the previous estimated channel
%			is more resilient to the inaccuracy in the previous estimated channel, which achieves a more accurate channel estimation performance.
%			However, M-FISTA is not suitable to operate with parallel processing for reducing the execution time.
%			, while the framework of FISTA is not suitable to do the parallel processing
%			\nm{, hence achieves more accurate channel estimation at the cost of...}
		\end{itemize}
		\item We propose two additional operations, \textbf{channel refinement} and \textbf{sequential search}, to enhance the channel estimation performance.
		The channel refinement leverages the spreading loss structure to derive the path gains and time delays of the estimated paths across the subcarriers, so the channel on all subcarriers can be reconstructed from the training measurements on the pilot subcarriers.
		The sequential search uses the hierarchical codebooks to reduce the computational complexity of the greedy beam selection in TS and increase the beam resolution in M-FISTA.						
%		\item We propose a sequential search method using hierarchical codebooks {to reduce the computational complexity of the greedy selection and enhance the beam resolution}.
%		\item We propose a channel refinement algorithm that leverages the spreading loss structure to derive the path gains and time delays of the estimated paths across the subcarriers.
	\end{itemize}
	}

	The rest of the paper is organized as follows.
	Section \ref{sec_sys_model} introduces the channel and signal model.
	Section \ref{sec_ch_training_scheme} proposes two channel estimation algorithms (TS and M-FISTA) in MMV-LS-CS training framework.
%	, followed by the additional operations enhancing the estimation performance and complexity analyses in Section \ref{sec_performance_enhance}.
	Section \ref{sec_performance_enhance} develops the operations (channel refinement and sequential search) to enhance the performance, followed by the complexity analyses.
	Section \ref{sec_simulation} shows the numerical results, and Section \ref{sec_conclusion} concludes the paper.

	{\bf Notation:}	Bold lowercase letters $\mathbf{x}$ and bold uppercase letters $\mathbf{X}$ denote vectors and matrices, respectively;
	$\mathbf{X}^\top$, $\mathbf{X}^H$, $\mathbf{X}^{+}$, $\textrm{vec}(\mathbf{X})$, $\textrm{det}(\mathbf{X})$ represent the transpose, conjugate transpose, Moore-Penrose pseudo-inverse, vectorization, and determinant of $\mathbf{X}$, respectively;
	$[\mathbf{X}]_\Gamma$ (respectively, $[\mathbf{x}]_\Gamma$) is the submatrix with columns of $\mathbf{X}$ (the subvector with elements of $\mathbf{x}$) associated with the indices set $\Gamma$; 
	$(\mathbf{X})_{\mathbf{n}}$ is the $\mathbf{n}$-th element of $\mathbf{X}$;
	$\lvert\Psi\rvert$ is the cardinality of the set $\Psi$;	
%	the Kronecker product is denoted as $\otimes$;
		$\otimes$ denotes the Kronecker product;
	$\mathbf{I}_M$ is an $M \times M$ identity matrix;
	$\mathcal{F}\{\cdot\}$ is the continuous Fourier transform.
	$\textrm{bdiag}(\mathbf{X}_1,\dots,\mathbf{X}_N)$ is a block diagonal matrix having the matrices $\mathbf{X}_1,\dots,\mathbf{X}_N$ on its diagonal.

	\section{System Model} \label{sec_sys_model}
	Clearly defining the system model in sub-THz is critical. 
	Section \ref{subsec_ch_model_dualWB} describes a time-varying sub-THz wideband channel model.
	Section \ref{subsec_freq_sel_channel} discusses its frequency selectivity.
	Section \ref{subsec_hybrid_signal_model} introduces the signal model of the hybrid transceiver, and Section \ref{subsec_extended_virtual_rep} describes the extended virtual representation of the MIMO channel.

	\begin{figure}
		\centering
		\includegraphics[scale=0.45]{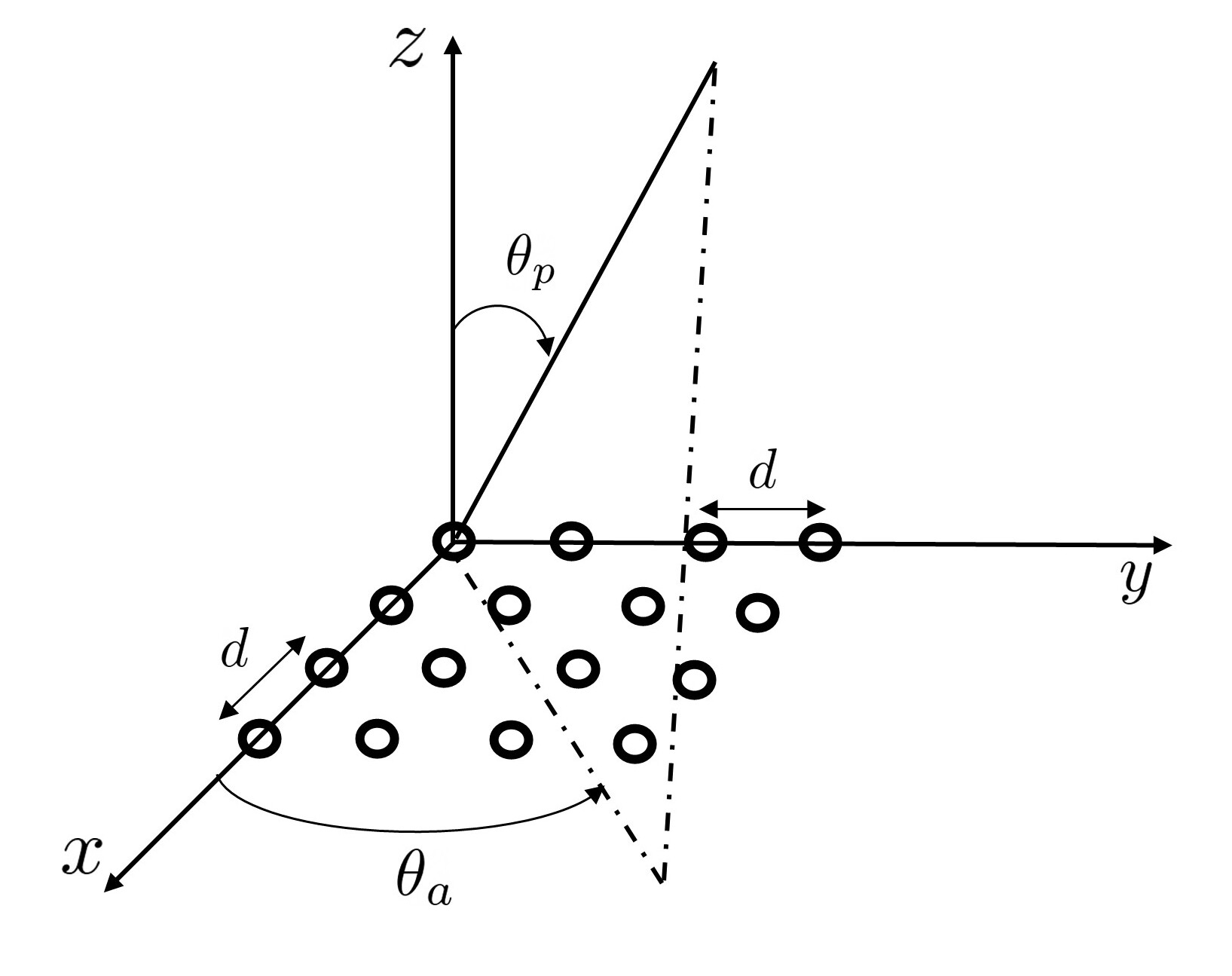}
		\caption{The array geometry of a $4\times 4$ UPA with the physical angle $(\theta_p,\theta_a)$ \cite{balanis2016antenna}.}
		\label{fig:UPA_antenna_array}
	\end{figure}

	\subsection{Channel Model}
	\label{subsec_ch_model_dualWB}
	We consider a MIMO-OFDM system with bandwidth $B$, carrier frequency $f_c$, and $K_o$ subcarriers. 
	We define $\Delta\in[-B/2, B/2]$ as the baseband frequency centered at the carrier frequency $f_c$ so that the actual frequency is $f=f_c+\Delta$.
	Since $f_c$ is fixed, the dependence of the channel propagation quantities (such as attenuation, etc) on the carrier frequency $f_c$ is not shown explicitly.
	The receiver and transmitter employ a UPA with $N_r\!=\!N_{vr}\!\times\! N_{hr}$ antennas and a UPA with $N_t\!=\!N_{vt}\!\times\! N_{ht}$ antennas, respectively, both configured  with antenna spacing $d$, as depicted in Fig.~\ref{fig:UPA_antenna_array}.
%	Both arrays are configured as depicted in Fig. \ref{fig:UPA_antenna_array} with antenna spacing $d$.		
	The receive and transmit antennas are indexed as $\mathbf{n}_r=(n_{vr},n_{hr})$ and $\mathbf{n}_t=(n_{vt},n_{ht})$, respectively.		
	We adopt a wideband geometric massive MIMO channel model with $L$ channel paths between the transmitter and receiver \cite{alkhateeb2016frequency}.	
	The $\ell$-th channel path ($\ell\!=\!1,\dots,L$) is characterized by its frequency-selective attenuation $\mathcal{B}_\ell(\Delta)$,
%	\nm{didnt we say that we only use fc and deltaf?? And we only show the dependence on delta f along keeping fc implicit...}, 
	time delay $\tau_\ell$, physical AOA $\boldsymbol{\theta}_{r}^\ell\!=\!(\theta_{pr}^\ell,\theta_{ar}^\ell)$, and physical AOD $\boldsymbol{\theta}_{t}^\ell\!=\!(\theta_{pt}^\ell,\theta_{at}^\ell)$, specifying the polar and azimuth angles (indexed by ``$\mathit{p}$" and ``$\mathit{a}$", respectively).
	The time-domain response of the $\ell$-th channel path is denoted as $\beta_{\ell}(t;u)$\footnote{We denote the time as $u$ and the delay component of the channel response as $t$.}.
	With the slowly-varying channel, the channel parameters are unchanged within the channel coherence time, so we omit the time variable $u$ of the channel parameters for ease of exposition.
	\add{
%		For the\nm{"For the??" you mean "Due to?"} 
	Due to the frequency selectivity of the interactions with the environment \cite{molisch2005ultrawideband}, $\beta_{\ell}(t)$ models the distortion of the $\ell$-th channel path with $\mathcal{F}\{\beta_{\ell}(t)\}=\mathcal{B}_\ell(\Delta)$,
%	$\int_{-\infty}^{\infty}\beta_\ell(t)e^{-j2\pi f t}\,dt=\mathcal{B}_\ell(f)$, 
	where $\mathcal{B}_\ell(\Delta)$ is the frequency-selective attenuation introduced in Section \ref{subsec_freq_sel_channel}.	}
	The baseband signal at the $\mathbf{n}_r$-th  receive antenna is 
	\begin{equation}\label{eq_original_signal}
	r_{\mathbf{n}_r}(u) = \sum_{\mathbf{n}_t}\sum_{\ell=1}^{L}{\Tilde x}_{{\ell},\mathbf{n}_t}(u-\tau_{\ell,\mathbf{n}_r,\mathbf{n}_t}) e^{-j 2\pi f_c \tau_{\ell,\mathbf{n}_r,\mathbf{n}_t}} + v_{\mathbf{n}_r}(u),
	\end{equation}
	where $\tau_{\ell,\mathbf{n}_r,\mathbf{n}_t}$ is the time delay of the $\ell$-th channel path between the $\mathbf{n}_t$-th transmit antenna and $\mathbf{n}_r\textrm{-th}$ receive antenna,
	$v_{\mathbf{n}_r}(u)$ is the additive noise at the $\mathbf{n}_r$-th receive antenna, and 
	$\Tilde{x}_{{\ell},\mathbf{n}_t}(u) \!=\! \beta_{\ell} \ast  {x}_{\mathbf{n}_t}(u)$
	is the convolution of the baseband signal ${x}_{\mathbf{n}_t}(u)$ transmitted at the $\mathbf{n}_t$-th antenna with the frequency-selective attenuation of the $\ell$-th channel path $\beta_{\ell}(t)$. 
%	\add{For the $\ell$-th channel path, $\beta_\ell(t)$ is assumed the same between all possible pairs of the transmit and receive antennas due to the fact that path gains are large-scale fading channel components.}
	\add{Note that $\beta_\ell(t)$ is assumed the same for the $\ell$-th channel path between all possible pairs of the transmit and receive antennas, due to the fact that path gains are large-scale fading channel components.}
%	\nm{weird spacing}

	Due to the increasing scale of the antenna arrays, the time delay of waves traveling across the array aperture is non-negligible, so we denote the time delay as \cite{tse2005fundamentals} 
	\begin{equation}\label{eq_path_delay}
	\tau_{\ell,\mathbf{n}_r,\mathbf{n}_t} = \tau_\ell + \tau_{\mathbf{n}_r}(\boldsymbol{\theta}_{r}^\ell) - \tau_{\mathbf{n}_t}(\boldsymbol{\theta}_{t}^\ell),
	\end{equation}
	where $\tau_\ell$ is the reference path delay of the $\ell$-th channel path on the first transmit and receive antenna pair ($\mathbf{n}_t\!=\!\mathbf{n}_r\!=\!(1,1)$).
	The propagation delay of the $\mathbf{n}_r$-th receive ($\mathbf{n}_t$-th transmit) antenna across the UPA aperture with respect to the $(1,1)$-th receive (transmit) antenna is denoted as 
	$\tau_{\mathbf{n}_\zeta}(\boldsymbol{\theta}_{\zeta}^\ell)\!=\!\frac{d}{c}\Big((n_{h\zeta}-1)\cos\theta_{a\zeta}^\ell\sin\theta_{p\zeta}^\ell+(n_{v\zeta}-1)\sin\theta_{a\zeta}^\ell\sin\theta_{p\zeta}^\ell\Big),$ $\zeta\!\in\!\{r,t\}$ \cite{balanis2016antenna}.			
	By applying the continuous Fourier transform on \eqref{eq_original_signal}, we obtain the signal at the baseband frequency $\Delta$ as	
	\begin{align*}
	%	\label{scalar_rx_signal}
	&(\mathbf{R}(\Delta))_{\mathbf{n}_r}
	=\mathcal{F}\{r_{\mathbf{n}_r}(u)\}=\\
	&\sum_{\mathbf{n}_t} \sum_{\ell=1}^{L}\alpha_\ell (\Delta)
	e^{-j 2\pi (f_c + \Delta) \left(\tau_{\mathbf{n}_r}(\boldsymbol{\theta}_{r}^\ell) - \tau_{\mathbf{n}_t}(\boldsymbol{\theta}_{t}^\ell)\right)} 
	e^{-j 2\pi \Delta \tau_\ell}(\mathbf{X}(\Delta))_{\mathbf{n}_t}\\
	&+(\mathbf{V}(\Delta))_{\mathbf{n}_r},
	\end{align*}
	where	$(\mathbf{X}(\Delta))_{\mathbf{n}_t}\!\!=\!\mathcal{F}\{{x}_{\mathbf{n}_t}(u)\}$, $(\mathbf{V}(\Delta))_{\mathbf{n}_r}\!\!=\!\mathcal{F}\{v_{\mathbf{n}_r}(u)\}$, and we define the baseband path coefficient 
	\begin{equation}\label{equiv_path_gain}
	\alpha_\ell(\Delta)\triangleq\mathcal{B}_\ell(\Delta) e^{-j 2\pi f_c \tau_\ell}.
	\end{equation} 
	By stacking up the MIMO signal on UPA antennas, we obtain the frequency-dependent input-output relationship for the MIMO channel as
	\begin{equation}
	\mathbf{r}(\Delta) = \mathbf{H}(\Delta)\mathbf{x}(\Delta) + \mathbf{v}(\Delta),
	\end{equation}
	where $\mathbf{r}(\Delta)= \textrm{vec}(\mathbf{R}(\Delta))\in\mathbb{C}^{N_r\times 1}$ is the received signal vector, $\mathbf{x}(\Delta)=\textrm{vec}(\mathbf{X}(\Delta))\in\mathbb{C}^{N_t\times 1}$ is the transmit signal vector, and
	${\mathbf{v}(\Delta)=\textrm{vec}(\mathbf{V}(\Delta))\in\mathbb{C}^{N_r\times 1}}$ is the additive noise vector.
	The frequency response of the baseband MIMO channel $\mathbf{H}(\Delta)\in\mathbb{C}^{N_r\times N_t}$ is
	\begin{align}
		\nonumber
	\label{freq_MIMO_ch}
	&\mathbf{H}(\Delta)= \sqrt{N_r N_t}\times
	\\
	& \sum_{\ell=1}^{L}\alpha_\ell(\Delta) \mathbf{b}_{N_r}(\psi_{hr}^{\ell},\psi_{vr}^{\ell};\Delta)
	\mathbf{b}_{N_t}^{H}(\psi_{ht}^{\ell},\psi_{vt}^{\ell};\Delta)
	e^{-j 2\pi \Delta \tau_{\ell}},
	\end{align}
	where $\psi_{hr}^{\ell}$ (respectively, $\psi_{ht}^{\ell}$) is the horizontal spatial AOA (AOD) of the $\ell$-th path,  $\psi_{vr}^{\ell}$ $(\psi_{vt}^{\ell})$ is the vertical spatial AOA (AOD) of the $\ell$-th path,  
	defined as ${\psi_{h\zeta}^{\ell} = \frac{d}{\lambda_c}\cos\theta_{a\zeta}^{\ell} \sin \theta_{p\zeta}^{\ell}}$ and $\psi_{v\zeta}^{\ell} = \frac{d}{\lambda_c}\sin\theta_{a\zeta}^{\ell} \sin \theta_{p\zeta}^{\ell}$, for $\zeta\in\{r,t\}$;
	$\mathbf{b}_{N_r}(\psi_{hr},\psi_{vr};\Delta)$, $\mathbf{b}_{N_t}(\psi_{ht},\psi_{vt};\Delta)$ denote the receive and transmit spatial-frequency UPA vectors, respectively, given by
	\begin{equation}
	\label{eq_spatial_freq_vec}
	\mathbf{b}_{N_\zeta}(\psi_{h\zeta},\psi_{v\zeta};\Delta) = \mathbf{a}_{N_{h\zeta}}(\psi_{h\zeta};\Delta) \otimes \mathbf{a}_{N_{v\zeta}}(\psi_{v\zeta};\Delta),
	\end{equation}
	for $\zeta\in\{r,t\}$, and the array response vector having $N$ antennas along each dimension is defined as
	\begin{align}
	\nonumber
	&\mathbf{a}_{N}(\psi;\Delta)\\
	&=\frac{1}{\sqrt{N}}\left[1,e^{-j2\pi \left(1+\frac{\Delta}{f_c}\right)\psi},\dots,e^{-j2\pi (N-1)\left(1+\frac{\Delta}{f_c}\right)\psi}\right]^\top.
	\label{steering_vector}
	\end{align}

	\subsection{Frequency-Selectivity of the MIMO Channel in Sub-THz Bands}\label{subsec_freq_sel_channel}
	Eq. \eqref{freq_MIMO_ch} models the frequency-domain channel $\mathbf{H}(\Delta)$, which is frequency-selective due to the \textbf{frequency-wideband effect}, \textbf{spatial-wideband effect}, and \textbf{path gain}.
	The frequency-wideband effect originates from the frequency-selective channel response caused by the time delays of the multipath fading channel and   has been widely investigated in the existing works \cite{venugopal2017channel,zhou2017low,araujo2019tensor,gao2015spatially,gao2016channel,choi2005interpolation,pande2006weighted,pande2007reduced}.

	The spatial-wideband effect arises from the distinct time delays across the antenna array for a given channel path (as in \eqref{eq_path_delay}). 
	Given a channel path associated with the array response vector $\mathbf{a}_{N}(\psi;\Delta)$ having a spatial angle $\psi$ (as in \eqref{steering_vector}), its effective spatial angle is $(1+\frac{\Delta}{f_c})\psi$, inducing a frequency-dependent phase shift $\frac{\Delta}{f_c}\psi$ that is  called  the beam squint effect \cite{wang2019beam}. 
	A system that encounters both the frequency- and spatial-wideband effects is a dual-wideband system.
	Sub-THz systems are more susceptible to the dual-wideband effect owing to the several orders of magnitude increase in bandwidths in the higher frequency spectrum.

	For the path gain of the channel, signals propagating in sub-THz bands suffer from a spreading loss $L_{spread}$,  absorption loss $L_{abs}$, and  reflection	coefficient $\chi_\ell$.
%	 (with $\chi_\ell=1$ for a LOS path).	
	\add{The equivalent path gain of the $\ell$-th channel path (as in \eqref{equiv_path_gain}) in sub-THz bands \cite{lin2015adaptive} is defined as  
	\begin{equation}\label{freq_dependent_gain}
	\lvert\alpha_\ell(\Delta)\rvert^2 \!=\! \lvert\mathcal{B}_{\ell}(\Delta)\rvert^2 = \lvert\chi_\ell(\Delta)\rvert^2 L_{spread}(\Delta,D_\ell)L_{abs}(\Delta,D_\ell),
	\end{equation}
	where $\Delta$ is the baseband frequency 
	and $D_\ell$ is the distance covered by the $\ell$-th path.}	
	The spreading loss (viewed as the path loss in conventional wireless communication) models  the attenuation incurred during wave propagation  \cite{lin2015adaptive} and follows Friis' transmission formula
	\begin{equation}
	L_{spread}(\Delta,D_\ell)=\left(\frac{c}{4\pi (f_c+\Delta) D_\ell}\right)^2,
	\end{equation}
	where $c$ is the speed of light.
	The absorption loss is the signal attenuation suffering from the molecular absorption in sub-THz bands, mainly due to  water vapor molecules \cite{jornet2011channel,lin2015adaptive}, defined as
	\begin{equation}
	L_{abs}(\Delta,D_\ell)=e^{-D_\ell\kappa_{a}(f_c+\Delta) },
	\end{equation}
	where $\kappa_a(\cdot)$ is the frequency-dependent molecular absorption coefficient and depends on the propagation medium at a molecular level \cite{jornet2011channel}.
	The reflection coefficient of the LOS path  $(\ell=1)$ is assumed to be ${\chi_1(\Delta)=1}$.
	For the NLOS paths $(\ell=2,\dots,L)$, we consider the single-bounce reflected rays model with the reflection coefficient defined according to \cite{piesiewicz2007scattering,lin2015adaptive} as 
	\begin{align}
	\nonumber
	\chi_\ell(\Delta)=&\frac{Z(\Delta)\cos \varphi_{i,\ell}-Z_o\cos\varphi_{r,\ell}}
	{Z(\Delta)\cos\varphi_{i,\ell}+Z_o\cos\varphi_{r,\ell}}\\
	&\times \exp\left({-\frac{1}{2}\left(\frac{4\pi (f_c+\Delta) \sigma_{r} \cos\varphi_{i,\ell}}{c}\right)^2}\right),
	\end{align}
	where $Z(\cdot)$ is the wave impedance of the reflecting material and a function of the frequency, $Z_o=377\ \Omega$ is the  wave impedance in free space, $\varphi_{r,\ell}=\arcsin(\frac{Z}{Z_o}\sin\varphi_{i,\ell})$ is the angle of refraction, $\varphi_{i,\ell}$ is the angle of incidence (or reflection),   and $\sigma_{r}$ is the standard deviation of the reflecting surface characterizing the material roughness.

\begin{figure}
		\centering			
		\includegraphics[scale=0.42]{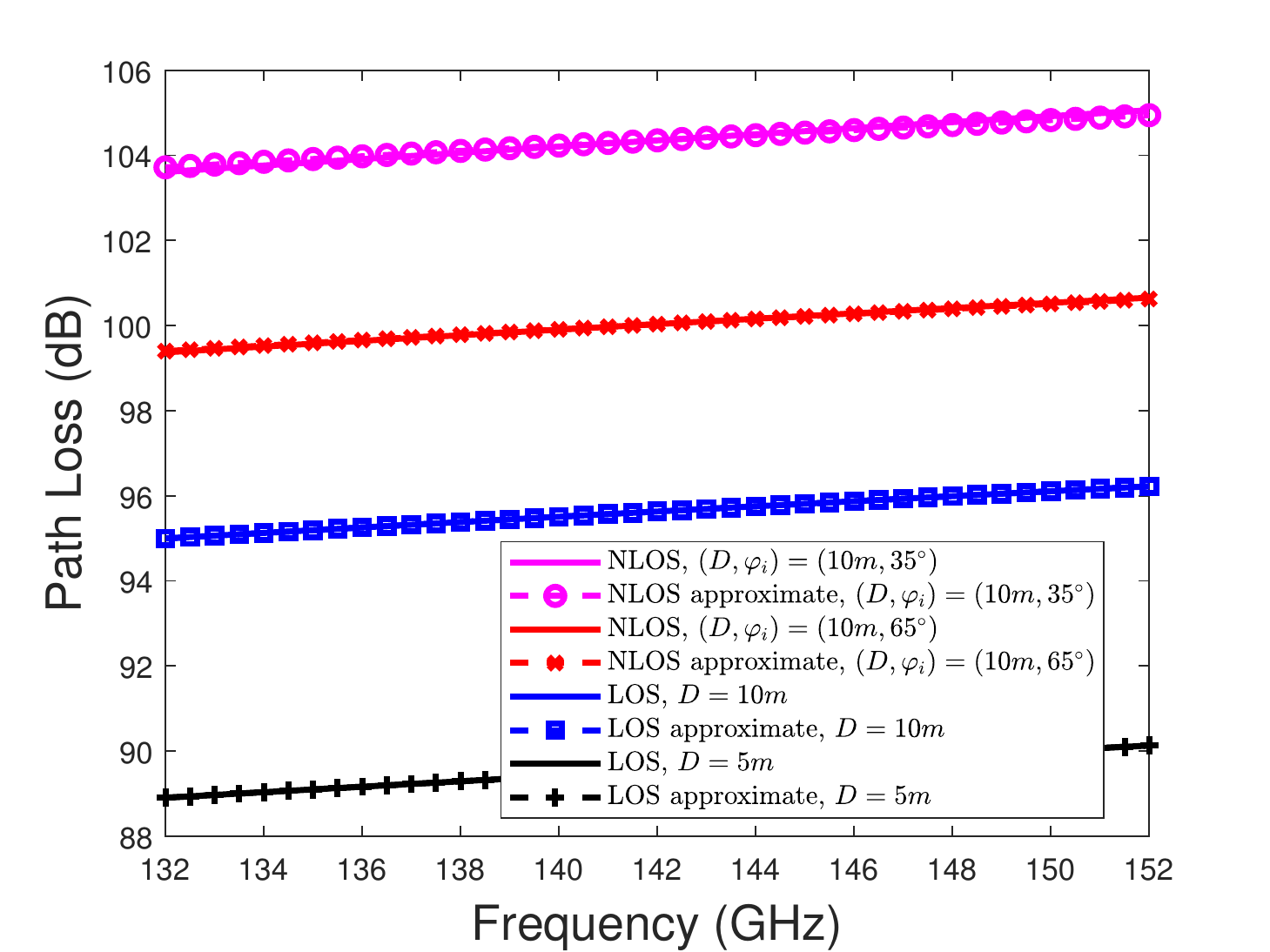}
		\caption{
			The overall pathloss $1/|\alpha_\ell(\Delta)|^2$ (dB) versus the operating frequency $(f_c+\Delta)$ in sub-THz bands, with the distance $D$ and $f_c=142$ GHz. 
			For the NLOS path, the angle of reflection $\varphi_{i}$, the refractive index = $2.24-j0.025$, and $\sigma_r = 0.088\times 10^{-3} $m are considered \cite{lin2015adaptive,jornet2011channel,han2014multi}.} 
		\label{fig:pathloss_freq}
	\end{figure}

	By combining these effects together,
	the overall pathloss $1/|\alpha_\ell(\Delta)|^2$ exhibits the frequency-dependent behavior depicted in Fig \ref{fig:pathloss_freq}. 
	Also, we plot a curve that uses the approximation $\varrho(1+\Delta/f_c)^2$ (up to a scaling factor $\varrho$) which shows a good fit, so we approximate the baseband path coefficient as $\alpha_{\ell}(\Delta) \approx \frac{\alpha_\ell^\prime }{1+\Delta/f_c}$, where $\alpha_\ell^\prime$ is the reference path coefficient having the path gain at the carrier frequency ($\lvert\alpha_{\ell}^\prime\rvert \!=\! \lvert\alpha_{\ell}(0)\rvert$), with frequency-independent phase shift \cite{lin2015adaptive}.
%	 at the carrier frequency ($\alpha_{\ell}(0)= \alpha_{\ell}^\prime$), with frequency-independent phase shift \cite{lin2015adaptive}.
%	\nm{but, with this approximation, arent you basically saying that the frequency dependence of the refelction loss and absorbption are irrelevant? I.e. the dominant effect is the spreading loss. So, why are you saying that capturing absorbiont and reflection loss is one of your  contributions, but at the same time you treat them as irrelevant?}	\tc{It can be understood in this way. In the previous works, the three effects (spreading loss, absorption loss, and reflection coefficient) are considered to depict the channel model in THz bands ($0.1$ to $10$ THz). However, in our work, we consider the sub-THz communication, roughly defined as $100$ to $300$ GHz. Through the simulation results in Fig. \ref{fig:pathloss_freq}, we observe that the dominant effect is the spreading loss, which justify the proposed approximation of the baseband path coefficient $\alpha_{\ell}(f) \approx \frac{\alpha_\ell^\prime }{1+f/f_c}$.}

	\subsection{Hybrid Transceiver and Signal Model}\label{subsec_hybrid_signal_model}
\begin{figure*}[t!]		\centering 
		\includegraphics[width=1.0\textwidth]{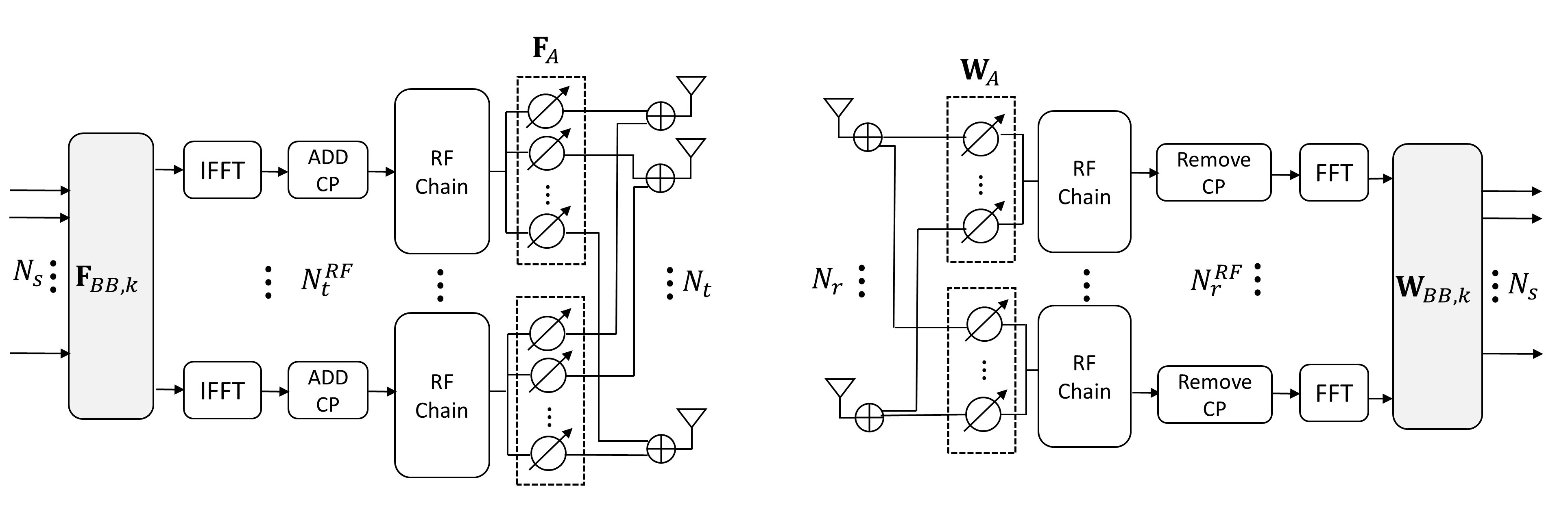}
		\caption{The MIMO-OFDM system with hybrid transceiver architecture \cite{lin2020tensor,dovelos2021channel}.}
		\label{fig:hybrid_MIMO_OFDM}
	\end{figure*}

	To facilitate the trade-off between performance and hardware cost, we consider a single-user hybrid transceiver design for MIMO-OFDM systems with $N_t^{RF}$ and $N_r^{RF}$ RF chains at the transmitter and receiver, respectively, as shown in Fig.~\ref{fig:hybrid_MIMO_OFDM}.
	We can extend our work to the downlink multi-user scenario by viewing the link between the base station and each user as a single-user scenario.
%	Our work can be extended to the downlink multi-user scenario by viewing the link between the base station and each user as a single-user scenario.
	The fully connected network of phase shifters is implemented in the RF precoder and combiner.
	Our proposed training framework in Section \ref{subsection_beam_training_scheme} is general and applicable to the new hybrid structures \cite{dovelos2021channel,tan2019delay} that consider the beam squint effect.
	\add{On a MIMO-OFDM system with $K_o$ subcarriers, we denote the MIMO channel on the $k$-th subcarrier as $\mathbf{H}_k\!=\!\mathbf{H}(\Delta_k)$, where ${\Delta_k\! =\! (k\!-\!\frac{K_o+1}{2})\frac{B}{K_o}}$ is the baseband frequency of the $k$-th subcarrier.}		

	The transmitter sends the signal on the $k$-th subcarrier at the $\nu$-th subframe, denoted as 
	\begin{equation}
	\mathbf{x}_{k,\nu} = \mathbf{F}_{A}\mathbf{F}_{BB,k}\mathbf{s}_{k,\nu}\in\mathbb{C}^{N_t\times 1},
	\end{equation}
	where $\mathbf{s}_{k,\nu}\in\mathbb{C}^{N_s\times 1}$ is the baseband signal on the $k$-th subcarrier at the $\nu$-th subframe,
	%	and $\mathbf{F}_k = \mathbf{F}_{A}\mathbf{F}_{BB,k}$ is the hybrid precoder, where 
	$\mathbf F_{A}\in\mathbb{C}^{N_t\times N^{RF}_t}$ is a frequency-flat analog precoder, and $\mathbf{F}_{BB,k}\in\mathbb{C}^{N^{RF}_t\times N_s}$ is the baseband precoders for subcarrier $k$.
	At the receiver, the signal is received by the frequency-flat analog combiner $\mathbf{W}_A\in\mathbb{C}^{N_r\times N^{RF}_r}$ and then processed through the cyclic prefix removal and the discrete Fourier transform.
	The baseband combiner $\mathbf{W}_{BB,k}\in\mathbb{C}^{N^{RF}_r\times N_s}$ is processed for subcarrier $k$,
%	implemented on the signal for each subcarrier separately,
	yielding the received signal on the $k$-th subcarrier at the $\nu$-th subframe as 
	\begin{equation}\label{comb_rx_signal}
	\mathbf{y}_{k,\nu}=\mathbf{W}_{BB,k}^H \mathbf{W}_A^H\mathbf{H}_k\mathbf{F}_{A}\mathbf{F}_{BB,k}\mathbf{s}_{k,\nu} + \mathbf{W}_{BB,k}^H \mathbf{W}_A^H\mathbf{v}_{k,\nu},
	\end{equation}
	where 
	%	$\mathbf{W}_k = \mathbf{W}_A\mathbf{W}_{BB,k}$ is the overall hybrid combiner and	
	$\mathbf{v}_{k,\nu}\in\mathbb{C}^{N_r\times 1}$ is the additive noise vector,
	with independent and identically distributed (i.i.d.) zero-mean complex Gaussian components with variance $\sigma_n^2$.
	{Note that the hybrid precoder ($\mathbf{F}_A$/$\mathbf{F}_{BB,k}$) and combiner ($\mathbf{W}_A$/$\mathbf{W}_{BB,k}$) can be chosen differently on distinct subframes.}
%	\nt

	\subsection{Extended Virtual Representation of the MIMO Channel}
	\label{subsec_extended_virtual_rep}
	%	\hl{With the array geometry}\nm{?}, 
	Using the uniform antenna array assumption, the MIMO channel can be formulated as an extended virtual representation \cite{heath2016overview}.
	For the UPA at the receiver, we consider the physical AOA of interest as $(\theta_{pr},\theta_{ar})\in [-\pi/2,\pi/2)\times [-\pi,\pi)$.
	Assuming half wavelength antenna spacing $(\frac{d}{\lambda_c}=\frac{d}{c/f_c}=0.5)$,
	%	($d/\lambda_c=1/2$, where $\lambda_c=c/f_c$), 
	the beam direction region of interest is  $(\psi_{hr},\psi_{vr})\in[-0.5,0.5)\times [-0.5,0.5)$.
	We assume that the horizontal (vertical) spatial AOAs take values from the uniform grid $\mathcal{G}_{hR}$ $(\mathcal{G}_{vR})$ of size $G_{hr}\geq L$ $(G_{vr}\geq L)$, given by
	\begin{equation}\label{receive_grid}
	\mathcal{G}_{\rho R} = \bigg\{\psi_{i_{\rho}} = \frac{i_{\rho}-\frac{G_{\rho r}+1}{2}}{G_{\rho r}},\ i_{\rho} = 1,\dots,G_{\rho r}\bigg\},
	\end{equation}
	for $\rho\in\{h,v\}$.
	%	We assume that the horizontal and vertical spatial AOA take values from the uniform grid $\mathcal{G}_{hR}$ of size $G_{hr}\geq L$ and $\mathcal{G}_{vR}$ of size $G_{vr}\geq L$, defined as
	%	\begin{align}\label{receive_grid_h}
	%	\mathcal{G}_{hR} = \left\{\psi_{i_{h}} = \frac{i_{h}-\frac{G_{hr}+1}{2}}{G_{hr}},\ i_{h} = 1,\dots,G_{hr}\right\},\\
	%	\label{receive_grid_v}
	%	\mathcal{G}_{vR} = \left\{\psi_{i_{v}} = \frac{i_{v}-\frac{G_{vr}+1}{2}}{G_{vr}},\ i_{v}= 1,\dots,G_{vr}\right\}.
	%	\end{align}
	%	\begin{align}\label{receive_grid_h}
	%		\mathcal{G}_{hR} = \left\{\psi_{i_1} = -\frac{1}{2}+\frac{i_1-1}{G_{hr}},\ i_1 = 1,\dots,G_{hr}\right\},\\
	%		\label{receive_grid_v}
	%		\mathcal{G}_{vR} = \left\{\psi_{i_2} = -\frac{1}{2}+\frac{i_2-1}{G_{vr}},\ i_2= 1,\dots,G_{vr}\right\}.
	%	\end{align}
	We define the grid of the receive UPA vectors as $\mathcal{G}_R=\mathcal{G}_{hR}\times \mathcal{G}_{vR}$, whose size $\lvert\mathcal{G}_R\rvert$ is $G_r=G_{hr}G_{vr}$.	
	%	Assuming the spatial AOAs take values from an uniform grid $\mathcal{G}_R$
	%	\begin{equation}
	%		\mathcal{G}_R = \left\{(\psi_{g_{hr}},\psi_{g_{vr}}):\right\}
	%	\end{equation}	
	We construct the receive array response matrix on the $k$-th subcarrier by collecting the spatial-frequency UPA vectors with the AOAs taking values on $\mathcal{G}_R$ as 
	\begin{equation}\label{rx_array_matrix}
	\mathbf{A}_{R,k}(:,i_r) = \mathbf{a}_{N_{hr}}(\psi_{i_{h}};\Delta_{k}) \otimes \mathbf{a}_{N_{vr}}(\psi_{i_{v}};\Delta_{k})\in\mathbb{C}^{N_r\times G_r},
	\end{equation}
	where $i_r\!=\!(i_{h}-1)G_{vr} + i_{v}$	with $(\psi_{i_{h}},\psi_{i_{v}})\!\in\!\mathcal{G}_R$.	
	Similarly, for the UPA at the transmitter, we assume the horizontal (vertical) spatial AODs take values from the uniform grid $\mathcal{G}_{hT}$ ($\mathcal{G}_{vT}$) of size $G_{ht}\geq L$ ($G_{vt}\geq L$).
	The grids $\mathcal{G}_{hT}$ and $\mathcal{G}_{vT}$ are constructed as in \eqref{receive_grid} by substituting $(G_{ht}, G_{vt})$ for $(G_{hr},G_{vr})$, and the grid of the transmit UPA vectors is defined as $\mathcal{G}_T=\mathcal{G}_{hT}\times \mathcal{G}_{vT}$, whose size $\lvert\mathcal{G}_T\rvert$ is $G_t=G_{ht}G_{vt}$.
	By collecting the spatial-frequency UPA vectors with the AODs taking values on $\mathcal{G}_T$, we write the transmit array response matrix on the $k$-th subcarrier as
	\begin{equation}\label{tx_array_matrix}
	\mathbf{A}_{T,k}(:,i_t) = \mathbf{a}_{N_{ht}}(\psi_{i_{h}};\Delta_{k}) \otimes \mathbf{a}_{N_{vt}}(\psi_{i_{v}};\Delta_{k})\in\mathbb{C}^{N_t\times G_t},
	\end{equation}	
	where $i_t=(i_{h}-1)G_{vt} + i_{v}$ with $(\psi_{i_{h}},\psi_{i_{v}})\in\mathcal{G}_T$.

	In this work, we consider a frame-based system, where each frame consists of $T_c$ subframes (channel uses).
	Assuming the frame duration is smaller than the channel coherence time, the channel remains constant in each frame, i.e., the common block-fading assumption.
	At the {$F\textrm{-th}$} frame, the MIMO channel on the $k$-th subcarrier has an extended virtual representation \cite{heath2016overview} as
	\begin{equation}\label{virtual_channel_expression}
	\mathbf{H}_k^{(F)} = \mathbf{A}_{R,k}\mathbf{D}_k^{(F)}\mathbf{A}_{T,k}^{H},
	\end{equation}
	where $\mathbf{D}_k^{(F)}\in\mathbb{C}^{G_r\times G_t}$ is the beamspace channel matrix whose non-zero elements are located in  positions corresponding to the spatial AOAs/AODs of the channel paths.
	Due to the mismatch between the spatial AOAs/AODs and the corresponding quantized
	values, a grid-mismatch error may exist but can be diminished if the grid sizes $(G_{hr},G_{vr}, G_{ht}, G_{vt})$ are chosen sufficiently large, which comes with a prohibitive computational complexity.
	In Section \ref{sec_seq_search}, we will design a sequential search method using hierarchical codebooks that allows mitigating the grid-mismatch problem with reduced computational complexity.
	The impact of grid-mismatch is numerically evaluated in Fig. \ref{fig:Exp1_NMSE_SNR}, which shows the degradation of the channel estimation using small grid sizes.
	With the compact antenna deployments and the limited scattering of the sub-THz channel, the MIMO channels are spatially correlated and focus on certain spatial directions.
	{The work \cite{nguyen2018comparing} shows that the channel of a sub-THz band (140GHz) has an average of 6 clusters and 4 MPCs/cluster, 
	whereas  the mmWave band (28GHz) has an average of 8 clusters and 5 MPCs/cluster.}
	Hence, the beamspace channel matrix $\mathbf{D}_k^{(F)}$ tends to be  sparse.
	We assume that $\mathbf{D}_k^{(F)}$ has at most $L \!<\! G_rG_t$ non-zero elements, and the remaining elements are negligible.

	We define the \text{channel support} of the $F$-th frame as
	%\begin{equation}
	$\Omega^{(F)} = \{(i_r,i_t):\mathbf{{D}}_k^{(F)}(i_r,i_t)\!\neq\! 0\},$
	%	\nm{doesnt $\Omega $ depend on time as well?}
	%\end{equation}	
	which is the indices set of the dominant elements in $\mathbf{{D}}_k^{(F)}$.
	The channel support is independent of the subcarrier $k$ since we construct the array response matrix $\mathbf{A}_{R,k}$, ($\mathbf{{A}}_{T,k}$) based on the same uniform grid $\mathcal{G}_R$ ($\mathcal{G}_T$) containing the quantized spatial AOAs (AODs).
	%	\nt{To be revised. State that the sub-THz channel has fewer clusters due to the high pathloss.
	%	Besides, the massive MIMO is required to compensate the pathloss, leading to a narrower beamwidth. 
	%	Thus, the beamspace channel has more nonzeros elements which are slowly moving.}
	For the massive MIMO system, the channel support is mainly determined by the deployment geometry, hardware, and  the time-varying effects of the propagation environment on the electromagnetic wave \cite{wu2014non,chen2019time}.
	%	\nm{before continuiing, explain in a few lines why the support varies slowly and provide reference}

	\begin{figure}[t]
		\centering
		\includegraphics[scale=0.38]{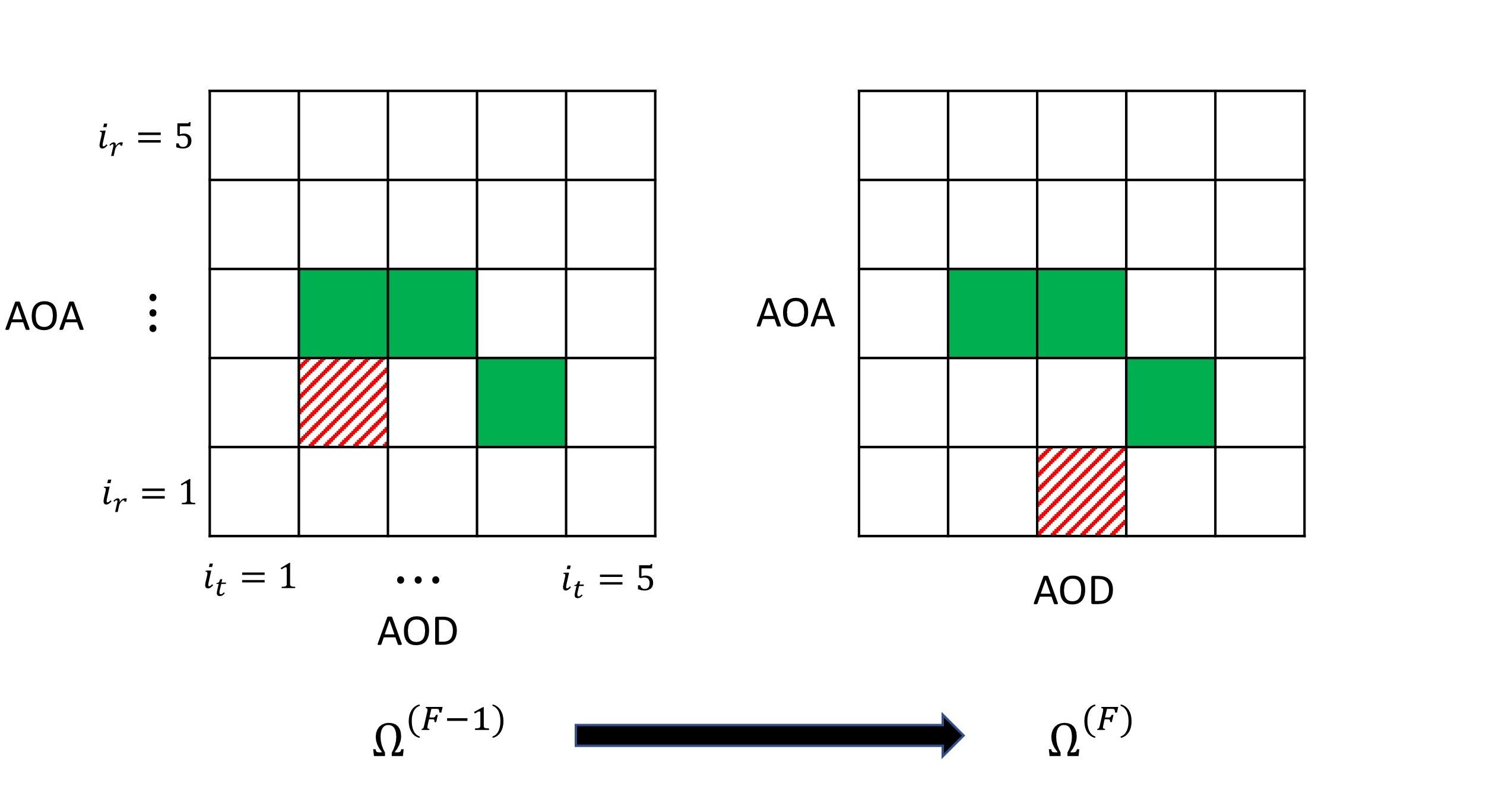}
		\caption{The evolution of beamspace channel with $(L, L_{cm}) = (4, 3)$ is illustrated. 
			Between two consecutive frames, the green and red elements represent the common and changing elements, respectively.
			%				\nm{can you label the x and y axes? AOA/AOD?}
		}
		%			\captionof{figure}{The proposed channel training protocol.}
		\label{fig:dense_sparse_illustration}
%		\vspace{-1em}
	\end{figure}

%	\begin{figure}[t]
%		\centering
%		\begin{minipage}{.45\textwidth}
%			\centering
%			\includegraphics[scale=0.38]{Dense_sparse_illustration_V15.jpg}
%			\caption{The evolution of beamspace channel with $(L, L_{cm}) = (4, 3)$ is illustrated. 
%				Between two consecutive frames, the green and red elements represent the common and changing elements, respectively.
%				%				\nm{can you label the x and y axes? AOA/AOD?}
%			}
%			%			\captionof{figure}{The proposed channel training protocol.}
%			\label{fig:dense_sparse_illustration}
%		\end{minipage}%
%		\hfill
%		\begin{minipage}{.45\textwidth}
%			\centering
%			\includegraphics[scale=0.55]{execution_flow_V12.jpg}
%			\caption{\add{The proposed channel training protocol.}}
%			\label{fig:channel_training_flow_chart}
%		\end{minipage}
%	\end{figure}

	In a frame-based system, the time-varying channel is approximated as fixed over the duration of a frame but may change across subsequent frames. 
	With the temporal correlation, the channel support varies slowly over time, meaning that $\Omega^{(F-1)}$ and $\Omega^{(F)}$ share many common elements \cite{han2017compressed}.
	We assume $L_{cm}$ is the minimum number of channel elements shared between $\Omega^{(F-1)}$ and $\Omega^{(F)}$, denoted as $\lvert\Omega^{(F)}\cap \Omega^{(F-1)}\rvert \!\geq\! L_{cm}$, i.e., the AOA-AOD pairs that remain fixed between the consecutive frames.
	%	In this work, we consider the scenario that the channel  
	Given a fixed number of channel paths $L\geq L_{cm}$, there are at most $L-L_{cm}$ paths changing from one frame to the next.
	%\nm{instead of newly added elements, isnt it better to say that they change from one fading block to the next?}
	%newly added elements which are not contained in the previous channel support set.
	%Note that the common paths means that
	%	, such as $|\Omega^{(i)}\backslash \Omega^{(i-1)}|\leq L-L_{cm}.$
	In Fig. \ref{fig:dense_sparse_illustration}, an example of the channel support evolution is illustrated, where the colored and white elements denote the dominant (non-zero) and negligible (zero) channel elements, respectively.
	%	Because of the slow channel variation, 
	For a slowly-varying channel with $L\!=\!4$ paths, the channel supports ${\Omega}^{(F-1)}$ and ${\Omega}^{(F)}$ share $L_{cm}\!=\!3$ common elements (shown in green), so only one path may change from one frame to the next (shown in red). This structure enables
	%\nm{can you show the shared elements more clearly in the figure? There is no need to show the two grids on the right, you can just show different colors for the shared elements and those that change, and add a pointer for the text descriptions}
	the LS-CS approach \cite{vaswani2016recursive,vaswani2010ls} to reduce the training overhead, discussed in Section \ref{sec_ch_training_scheme}.

	\section{Proposed LS-CS Channel Training} \label{sec_ch_training_scheme}
	In this section, we propose support tracking-based channel training for dual-wideband MIMO-OFDM systems.
	Section \ref{subsection_beam_training_scheme} depicts the beam training model, and Section \ref{subsection_ch_training_scheme} introduces the channel training protocol.
	Section \ref{subsec_tensor_LS_CS} proposes a two-stage MMV-LS-CS (TS) channel estimation, and Section \ref{subsec_joint_MMV_LS_CS} proposes an MMV FISTA-based (M-FISTA) channel estimation.

	\subsection{Beam Training Scheme} \label{subsection_beam_training_scheme}	
	With each frame divided into $T_c$ equal-sized subframes,  we assume that $T_p$ subframes are used for pilot-based channel training and the remaining $T_c\!-\!T_p$ subframes are used for data transmission. %, illustrated in Fig. \ref{fig:channel_block}.	
	For the pilot transmission, we exploit $K_p\!<\!K_o$ subcarriers with a comb-type arrangement,
%	\nm{are you used the remining subcarriers  for data tx? It is not clear, since you say that Tp subframes are dedicated to data Tx.}
	i.e., $\mathcal{P} \!=\! \left\{1+(i-1)\delta_p:i=1,\dots,K_p,\ \delta_p = \lceil{K_o}/{K_p}\rceil\right\}$, and the remaining subcarriers are used for data transmission.
%	\sst{During the $T_p$ channel training subframes, with only $K_p$ subcarriers for the pilot transmission, the channel resource of the remaining $K_o-K_p$ subcarriers can be used for the data transmission, while the hybrid transceiver with the CSI on the current frame is not available.}
%\nm{In your simulations, are you indeed using these remaining subcarriers for data transmission? I think you should, it further improves the spetral efficiency.}\tc{simulation results are updated.}
	On the $k$-th subcarrier, the transmitter sends the precoded signal at the $\nu$-th subframe, and the receiver combines the measurement signal at the $q$-th stream by the combining vector $\mathbf{w}_{k,q}$.	
%	At the $\nu$-th subframe, the transmitter sends the precoded signal $\mathbf{x}_{k,\nu}$ on the $k$-th subcarrier, and the receiver combines the measurement signal at the $q$-th stream on the $k$-th subcarrier by the combining vector $\mathbf{w}_{k,q}$.
	Assuming that the transmitter employs distinct precoded pilots of $T_p$ subframes in length and the receiver combines the signal into $Q_p$ streams, the $Q_p\times T_p$ combined signal on the $k$-th subcarrier, $k\in\mathcal{P}$, is denoted as 
	\begin{equation}
	\label{IO_matrix_beamtraining_virtual}
	\mathbf{Y}_{k} 
	= \mathbf{W}_k^{H}\mathbf{H}_k \mathbf{X}_k + \mathbf{\Tilde V}_k=(\mathbf{A}_{R,k}^{H}\mathbf{W}_k)^{H}\mathbf{D}_k(\mathbf{A}_{T,k}^{H}\mathbf{X}_k)+\mathbf{\Tilde V}_k, 
	\end{equation}
	where $\mathbf{W}_k=[\mathbf{w}_{k,1},\dots,\mathbf{w}_{k,Q_p}]$ and $\mathbf{X}_k=[\mathbf{x}_{k,1},\dots,\mathbf{x}_{k,T_p}]$ are the measurement and pilot matrices, respectively.
	We denote the combined noise matrix as $\mathbf{\Tilde V}_k=\mathbf{W}_k^H\mathbf{V}_k$, where $\mathbf{V}_k=[\mathbf{v}_{k,1},\dots,\mathbf{v}_{k,T_p}]$.
	With the extended virtual representation of the MIMO channel \eqref{virtual_channel_expression}, we have a combined signal form \eqref{IO_matrix_beamtraining_virtual}. 
	For the channel training, we adopt the random beamforming method \cite{zhou2017low,lin2020tensor} 
	to design the pilot and measurement matrices, i.e.,
 $\mathbf{X}_k{(m,n)}\!=\!\frac{e^{j\zeta_{m,n}}}{\sqrt{N_t}},\ \zeta_{m,n}\!\sim\!\mathcal{U}(0,2\pi)$
 and
 $\mathbf{W}_k{(m,n)}\!=\!\frac{e^{j\eta_{m,n}}}{\sqrt{N_r}},\ \eta_{m,n}\!\sim\!\mathcal{U}(0,2\pi)$.
	With sufficiently large $N_r$ and $N_t$, the elements of $\mathbf{A}_{R,k}^{H}\mathbf{W}_k$ and $\mathbf{A}_{T,k}^{H}\mathbf{X}_k$ are approximately i.i.d. $\mathcal{CN}(0,1/N_r)$ (respectively, $\mathcal{CN}(0,1/N_t)$) according to the central limit theorem \cite[Appendix B]{zhou2017low}, leading to the successful sparse recovery condition \cite{candes2006stable}.

	The hybrid transceiver is widely employed in mmWave and sub-THz communications, so we design the measurement and pilot matrices adopting the random beamforming method (discussed in the previous paragraph) based on the hybrid transceiver structure.
	For simplicity, we assume the number of subframes $T_p$ and the number of streams $Q_p$ satisfy $T_p\!=\!N_{ST} N_t^{RF}$ and $Q_p\!=\!N_{SR} N_r^{RF}$, for some $N_{ST},N_{SR}\!\in\!\mathbb{N}$, but the analysis can be generalized by considering $\mathbf{X}_k$ and $\mathbf{W}_k$ in \eqref{IO_matrix_beamtraining_virtual}.
	Then, we design a sequence of measurement matrices
	$\mathbf{W}_k \!=\!\left[\mathbf{W}_{k,1},\cdots,\mathbf{W}_{k,N_{SR}}\right]\!\in\!\mathbb{C}^{N_r\times Q_p}$
	and of pilot matrices
	$\mathbf{X}_k \!=\!\left[\mathbf{X}_{k,1},\cdots,\mathbf{X}_{k,N_{ST}}\right]\!\in\!\mathbb{C}^{N_t\times T_p}$ so as to match the beamforming values generated randomly as explained in the previous paragraph.
	With the hybrid receiver structure, each submatrix $\mathbf{W}_{k,i}\!=\!
	\mathbf{W}_{A,i}\mathbf{W}_{BB,k,i}
	\!\in\!\mathbb{C}^{N_r\times N_r^{RF}},i\!=\!1,\dots,N_{SR}$ is obtained by setting $\mathbf{W}_{BB,k,i}\!=\!\mathbf{I}_{N_r^{RF}}$ and $\mathbf{W}_{A,i}{(m,n)}\!=\!\frac{e^{j\eta_{m,n}}}{\sqrt{N_r}},\ \eta_{m,n}\!\sim\!\mathcal{U}(0,2\pi)$.
	Similarly, each submatrix $\mathbf{X}_{k,i}\!=\!
	\mathbf{F}_{A,i}\mathbf{F}_{BB,k,i}\mathbf{S}_{k,i}
	\!\in\!\mathbb{C}^{N_t\times N_t^{RF}},i=1,\dots,N_{ST}$ is obtained by setting $\mathbf{F}_{BB,k,i}\!=\!\mathbf{S}_{k,i}\!=\!\mathbf{I}_{N_t^{RF}}$ and $\mathbf{F}_{A,i}{(m,n)}\!=\!\frac{e^{j\zeta_{m,n}}}{\sqrt{N_t}},\ \zeta_{m,n}\!\sim\!\mathcal{U}(0,2\pi)$.
	Our work focuses on the design of the channel training, and this configuration justifies our proposed channel training in the hybrid transceiver structure.

	\subsection{Channel Training Protocol} \label{subsection_ch_training_scheme}
\begin{figure}[t]
		\centering
		\includegraphics[scale=0.55]{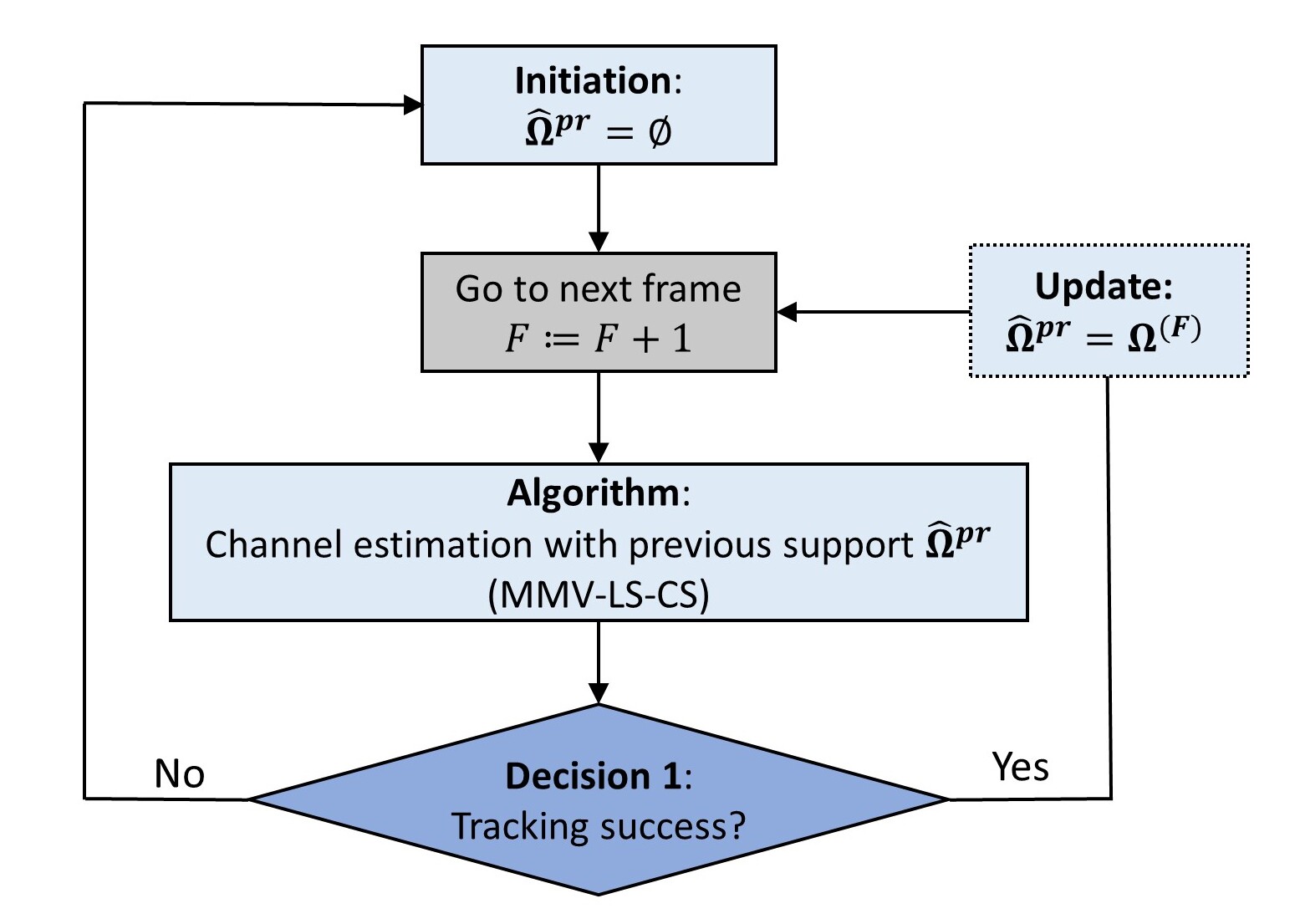}
		\caption{\add{The proposed channel training protocol.}}
		\label{fig:channel_training_flow_chart}
	\end{figure}

	We introduce the protocol of the support tracking-based channel training by exploiting the temporal correlation, shown in Fig.~\ref{fig:channel_training_flow_chart}.
	\add{In each frame, support tracking-based channel training is used to estimate the channel aided by the previous channel support.
	Note that the quality of the estimated previous channel support is crucial to the channel estimation performance. 
	An inaccurate previous channel support might deteriorate the performance of the support tracking-based approach because more channel elements are required to be estimated in the CS stage. 
	In the first frame, we initialize the previous channel support as an empty set since no prior CSI is available.
	In the following frames, the previous channel support is derived from the estimated channel in the preceding frame.
	To ensure satisfactory quality, we check the residual signal of the support tracking-based channel training.
	If its magnitude is smaller than a predefined threshold, we continue the support tracking-based channel training on the next frame.
	Otherwise, the estimated previous channel support is considered unreliable, and the channel training procedure is reset by setting the previous channel support as an empty set.}

	\subsection{Two-Stage MMV-LS-CS Channel Estimation (TS)} \label{subsec_tensor_LS_CS}
	Our goal is to develop an approach to reduce the overhead of channel training using the \textit{estimated previous channel support} $\hat{\Omega}^{pr}$, where the real previous channel support ${\Omega}^{pr}$ is unknown.
	We estimate the spatial AOAs/AODs of the dominant channel paths and then refine the time delays and path coefficients of the estimated paths.
	Given the training signal \eqref{IO_matrix_beamtraining_virtual} with the measurement and pilot matrices $(\mathbf{W}_k,\mathbf{X}_k)$ and the relationship $\textrm{vec}(\mathbf{ABC})=(\mathbf{C}^\top\otimes\mathbf{A})\textrm{vec}(\mathbf{B})$,	we have the vectorization of $\mathbf{Y}_k$ as
	\begin{equation}\label{LS_eq}
	\mathbf{y}_k=\textrm{vec}(\mathbf{Y}_k)= \mathbf{\Theta}_k\mathbf{z}_k + \mathbf{\Tilde v}_k,
	\end{equation}
	where $\mathbf{\Theta}_k = (\mathbf{A}_{T,k}^{H}\mathbf{X}_k)^{\top}\otimes(\mathbf{A}_{R,k}^{H}\mathbf{W}_k)^{H}$ is the dictionary matrix on the $k$-th subcarrier, $\mathbf{z}_k=\textrm{vec}(\mathbf{D}_k)$ and $\mathbf{\Tilde v}_k=\textrm{vec}(\mathbf{\Tilde V}_k)$. 
%	\nm{yuou are already using $\mathbf{s}_k$ for transmitted symbols. Please use smt different to denote tx symbols, such as z?}
	%	\nm{should this analysis in (22) and (23) go right after (20)??}
	We introduce the \textbf{two-stage MMV-LS-CS (TS)} algorithm for the time-varying MIMO-OFDM channel estimation, which separates the channel training into two stages: \textbf{MMV-LS} and \textbf{MMV-CS}.
	MMV-LS exploits the slowly-varying channel by performing the	LS estimate on the estimated previous channel support.
	Next, \text{MMV-CS} applies a CS-based approach to the LS residual, expected to be sparse since most dominant channel elements have been estimated in MMV-LS.
	The \text{TS} algorithm is shown in \textbf{Algorithm \ref{alg_tensor_LS_CS}} and operates as follows.

	First, \text{MMV-LS} estimates the beamspace channel elements corresponding to the estimated previous channel support $\hat{\Omega}^{pr}$.
	We denote the set of column indices (of $\mathbf{\Theta}_k$) corresponding to $\hat{\Omega}^{pr}$ as $\mathcal{Q}=\{(i_t-1)G_r + i_r: (i_r,i_t)\in\hat{\Omega}^{pr}\}$.
	The beamspace channel vectors share common support across the subcarriers, enabling the MMV to estimate the dominant paths.
%	Given the frequency-dependent $\mathbf{\Theta}_k$, the beamspace channel vectors share a common support, enabling the MMV to estimate the dominant paths.
	The MMV-LS algorithm is initialized as $\Gamma=\emptyset$ \text{(line 3)}.	
	Then, $\Gamma$ is reconstructed recursively by collecting the indices from $\mathcal{Q}$, which leads to the minimum residual error after orthogonalization, until $\lvert\Gamma\rvert=L_{cm}$ (lines 4-7).
	The estimated beamspace channel in MMV-LS is derived as
	$[\mathbf{\hat{z}}_k^{LS}]_{\Gamma}=[\mathbf{\Theta}_k]_{\Gamma}^{+}\textbf{y}_k$ and $[\mathbf{\hat{z}}_k^{LS}]_{\Gamma^c} = 0$ (line 8), where $\Gamma^c = \{1,\dots,G_rG_t\}\backslash \Gamma$.

	Secondly, \text{MMV-CS} executes a CS-based approach on the LS residual signal obtained by subtracting the effect of the MMV-LS estimated beamspace channel $\mathbf{\hat{z}}_k^{LS}$ (line 10), given as
	\begin{equation*}
	\mathbf{y}_k^{CS}=	\textbf{y}_k-\mathbf{\Theta}_k\mathbf{\hat{z}}_k^{LS}=\mathbf{\Theta}_k \mathbf{z}_k^{CS} + \mathbf{\Tilde v}_k,
	\end{equation*}
	where $\mathbf{z}_k^{CS} = \mathbf{z}_k - \mathbf{\hat{z}}_k^{LS}$ is expected to be sparser than $\mathbf{z}_k$
%	\nm{than what?} 
	since most dominant path components ($L_{cm}$ out of $L$) are expected to be detected in $\mathbf{\hat{z}}_k^{LS}$, so that $\mathbf{z}_k^{CS} $ is expected to have only $(L-L_{cm})$ non-zero components.
	The sparse recovery problem on the pilot subcarriers is formulated as
	\begin{equation}\label{eq_MMV_CS}
	\arg\min\limits_{\mathbf{z}_k^{CS}}\ \sum\nolimits_{k\in\mathcal{P}}\lVert\mathbf{z}_k^{CS}\rVert_1,\  
	\textrm{s.t. }\lVert \mathbf{y}_k^{CS} - \mathbf{\Theta}_k\mathbf{z}_k^{CS}\rVert_2\leq \epsilon,
	\end{equation}
	where $\epsilon>0$ is a constant threshold.
	Among many available sparse recovery algorithms using the common channel support across subcarriers, we adopt the simultaneous OMP algorithm (SOMP) \cite{tropp2006algorithms}.
	The algorithm is initialized as $\Upsilon\!=\!\emptyset$ and the residual $\mathbf{r}_{k} \!=\! \mathbf{y}_k^{CS},\ k\in\mathcal{P}$ (line 10).
	Then, the set $\Upsilon$ is constructed by collecting the index of the column (of $\mathbf{\Theta}_k$) having the largest correlation with $\mathbf{r}_k$ (line 13).
%	, until attaining the maximum iteration number or the mean squared error between the current and previous residuals is less than a constant threshold (line 12).
	In each iteration, the residual $\mathbf{r}_{k}$ is updated by removing the channel effect of the indices in $\Upsilon$ (line 14-15).
	The TS algorithm stops when it reaches $t_{max}$ iterations or the relative error of the residual signal $\varepsilon_t <\epsilon$ (line 12).
%	With $\Upsilon^c = \{1,\dots,G_rG_t\}\backslash \Upsilon$, 
	The estimated beamspace channel in MMV-CS is derived as $[\mathbf{\hat{z}}_k^{CS}]_{\Upsilon}=[\mathbf{\Theta}_k]_{\Upsilon}^{+}\textbf{y}_k^{CS}\textrm{ and }[\mathbf{\hat{z}}_k^{CS}]_{\Upsilon^c} = 0$ (line 18), where $\Upsilon^c = \{1,\dots,G_rG_t\}\backslash \Upsilon$.

	\begin{algorithm}[t] \scriptsize{\Large}
		\caption{Two-stage MMV-LS-CS Channel Estimation (TS).}
		\label{alg_tensor_LS_CS}
		\begin{algorithmic}[1]
			% \REQUIRE 
			\INPUT  measurement $\mathbf{Y}_k$, dictionary matrix $\mathbf{\Theta}_k$,
			estimated previous channel support $\hat{\Omega}^{pr}$
			%			, the set of pilot subcarriers $\mathcal{P}$.
			
			\OUTPUT estimated channel $\mathbf{\hat H}_k$ %, estimated channel support $\hat{\Omega}$.
			
			%		\STATE \textbf{Initialization:} $\Gamma_{0}=\emptyset$, $\Upsilon_n=[\emptyset],\ n=1,2,3.$
			%			\STATE Calculate $\mathbf{\Theta}_k= (\mathbf{A}_{T,k}^{H}\mathbf{ X})^{\top}\otimes(\mathbf{A}_{R,k}^{H}\mathbf{W})^{H},\ k\in\mathcal{P}$;			
			\STATE \emph{{MMV-LS} aided by the estimated previous channel support}
			%			\nm{the use of \%\%\%\% looks weird...remove them. Use \emph{} instead for the text. Also, there should not be a line number for comments}
			\STATE $\mathbf{y}_k=\text{vec}(\mathbf{Y}_k),\ k\in\mathcal{P}$; 
			\STATE $\Gamma=\emptyset$; $\mathcal{Q}=\{(i_t-1)G_r + i_r: (i_r,i_t)\in\hat{\Omega}^{pr}\}$;
			%		$\Upsilon_{LS}=\emptyset$;		 
			\WHILE{$\lvert\Gamma\rvert<L_{cm}$ \add{and $\hat{\Omega}^{pr}\neq \emptyset$}}
			%			\STATE $\Gamma \coloneqq  \Gamma \cup \arg\min_{j\in\mathcal{Q}\backslash\Gamma}			\sum_{k\in\mathcal{P}}\lVert \mathbf{y}_{k}-[\mathbf{\Theta}_k]_{\Gamma\cup{j}} [\mathbf{\Theta}_k]_{\Gamma\cup{j}}^{+} \mathbf{y}_{k} \rVert_F^2;$
			\STATE $j^* = \arg\min\limits_{j\in\mathcal{Q}\backslash\Gamma} \sum_{k\in\mathcal{P}}\lVert \mathbf{y}_{k}-[\mathbf{\Theta}_k]_{\Gamma\cup{j}} [\mathbf{\Theta}_k]_{\Gamma\cup{j}}^{+} \mathbf{y}_{k} \rVert_F^2;$
			\STATE $\Gamma \coloneqq  \Gamma \cup j^*;$
			\ENDWHILE
			\STATE  $[\mathbf{\hat{z}}_k^{LS}]_{\Gamma}=[\mathbf{\Theta}_k]_{\Gamma}^{+}\textbf{y}_k\textrm{ and }[\mathbf{\hat{z}}_k^{LS}]_{\Gamma^c} = 0,\ k\in\mathcal{P}$;						
			\item[]			
			\STATE \emph{MMV-CS on the LS residual}
			\STATE $\mathbf{y}_k^{CS}=	\textbf{y}_k-\mathbf{\Theta}_k\mathbf{\hat{z}}_k^{LS},\ k\in\mathcal{P}$;
			$\Upsilon=\emptyset$;  $\mathbf{r}_{k,t} = \mathbf{y}_k^{CS},\ k\in\mathcal{P}$;
			\STATE $\mathcal{J}=\{1,\dots,G_rG_t\}$; 	
			$\varepsilon_{t}=\infty$;			
			\WHILE{$\varepsilon_t >\epsilon$ \textbf{and} $t<t_{max}$}
			\STATE $\Upsilon \coloneqq \Upsilon \cup \arg\max_{j\in\mathcal{J}\backslash \Upsilon}\sum_{k\in\mathcal{P}}
			\lvert\mathbf{\Theta}_k(:,j)^{H}\mathbf{r}_{k,t}\rvert^2;$
			%			\STATE $j^*=\arg\max_{j\in\mathcal{J}\backslash \Upsilon}\sum_{k\in\mathcal{P}}
			%			\lvert\mathbf{\Theta}_k(:,j)^{H}\mathbf{r}_{k,t}\rvert^2;$
			%			\STATE $\Upsilon \coloneqq \Upsilon \cup j^*;$
			\STATE $\mathbf{\hat g}_k = \arg\min_{\mathbf{g}} \lVert \mathbf{y}_k^{CS}-[\mathbf{\Theta}_k]_{\Upsilon}\mathbf{g}  \rVert_F^2,\ k\in\mathcal{P};$
			\STATE $\mathbf{r}_{k,t+1} = \mathbf{y}_k^{CS}-[\mathbf{\Theta}_k]_{\Upsilon}\mathbf{\hat g}_k,\ k\in\mathcal{P};$
			\STATE $\varepsilon_{t+1}=\frac{1}{\lvert\mathcal{P}\rvert}\sum_{k\in\mathcal{P}}\lVert \mathbf{r}_{k,t+1}- \mathbf{r}_{k,t}\rVert_F^2$; $t\coloneqq t+1$;
			%			\STATE $t\coloneqq t+1$;
			\ENDWHILE
			\STATE $[\mathbf{\hat{z}}_k^{CS}]_{\Upsilon}=[\mathbf{\Theta}_k]_{\Upsilon}^{+}\textbf{y}_k^{CS}\textrm{ and }[\mathbf{\hat{z}}_k^{CS}]_{\Upsilon^c} = 0,\ k\in\mathcal{P};$
			\item[]
			%			\STATE \emph{Add the new detections from {MMV-CS}}
			\STATE \emph{Combine the detections from {MMV-LS} and {MMV-CS}}
			\STATE $\Xi\!=\!\{i\!:\!i\in\{1,\dots,G_rG_t\},\ \frac{1}{\lvert\mathcal{P}\rvert}\sum\limits_{k\in\mathcal{P}}\lvert(\mathbf{\hat{z}}_k^{LS} + \mathbf{\hat{z}}_k^{CS})_i\rvert \neq 0\}$;
			%			, where  $\mathbf{\hat{z}}_k^{A} = \mathbf{\hat{z}}_k^{LS} + \mathbf{\hat{z}}_k^{CS},\ k\in\mathcal{P}$;						
			%			\item[]
			\STATE \emph{Derive the estimated current channel support}
			\STATE $[\mathbf{\hat{z}}_k^{det}]_{\Xi}=[\mathbf{\Theta}_k]_{\Xi}^{+}\textbf{y}_k,\  [\mathbf{\hat{z}}_k^{det}]_{\Xi^c} = 0,\ k\in\mathcal{P};$
			\STATE $\mathbf{z}_{eq} = \frac{1}{\lvert\mathcal{P}\rvert}\sum_{k\in\mathcal{P}}\lvert\mathbf{\hat{z}}_k^{det}\rvert$;
			\STATE $\Tilde{\Xi} = \arg\max_{\Tilde{\Xi}\subset\Xi,\  \lvert\Tilde{\Xi}\rvert=L^\prime}\sum_{i\in\Tilde{\Xi}}(\mathbf{z}_{eq})_i$;
			\item[]
			%			\STATE $[\mathbf{\hat{s}}_k^{det}]_{\Xi}=[\mathbf{\Theta}_k]_{\Xi}^{+}\textbf{y}_k\textrm{ and }[\mathbf{\hat{s}}_k^{det}]_{\Xi^c} = 0,\ k\in\mathcal{P};$	
			%			\STATE $\Tilde{\Xi} = \Xi\ \backslash\ \{i:i\in\Xi,\ \frac{1}{\lvert\mathcal{P}\rvert}\sum_{k\in\mathcal{P}}\lvert(\mathbf{\hat{z}}_k^{det})_i\rvert \leq \zeta_{D}\}$, where $[\mathbf{\hat{z}}_k^{det}]_{\Xi}=[\mathbf{\Theta}_k]_{\Xi}^{+}\textbf{y}_k,\  [\mathbf{\hat{z}}_k^{det}]_{\Xi^c} = 0,\ k\in\mathcal{P};$	
			%			\\
			%			\STATE Derive the estimated channel support $\hat{\Omega}$ from $\Tilde{\Xi}$;
			\STATE Reconstruct $\mathbf{\hat{H}}_k$ by \textbf{Algorithm \ref{alg_gain_delay_refine}} with $(\mathbf{Y}_k, \mathbf{\Theta}_k,\Tilde{\Xi})$;			
		\end{algorithmic}
	\end{algorithm}

	As discussed in \cite{vaswani2010ls}, with the $\ell_1$-norm minimization subject to the noise constraint (e.g., MMV-CS in \eqref{eq_MMV_CS}), the estimate $\mathbf{\hat{z}}_k^{CS}$ tends to be biased towards zero.
	Also, MMV-CS might lead to false detections of the components, whose beamspace channel value is zero but is detected as non-zero due to  noise or detection error.
	To reduce the bias and improve performance in practical settings, 
	we derive the channel support by combining the new detection $\mathbf{\hat{z}}_k^{CS}$ in MMV-CS with the estimated $\mathbf{\hat{z}}_k^{LS}$ from MMV-LS (line 20) as 
	\begin{equation*}
	\Xi=\Big\{i:i\in\{1,\dots,G_rG_t\},\ \frac{1}{\lvert\mathcal{P}\rvert}\sum_{k\in\mathcal{P}}\lvert(\mathbf{\hat{z}}_k^{LS} + \mathbf{\hat{z}}_k^{CS})_i\rvert \neq 0\Big\}.	
	\end{equation*}
	The LS estimate on $\Xi$ is computed as
	$[\mathbf{\hat{z}}_k^{det}]_{\Xi}=[\mathbf{\Theta}_k]_{\Xi}^{+}\textbf{y}_k$ and $[\mathbf{\hat{z}}_k^{det}]_{\Xi^c} = 0,\ k\in\mathcal{P},$
	where $\Xi^c = \{1,\dots,G_rG_t\}\backslash \Xi$.
	Next, we derive the estimated channel support by collecting the indices of the largest $L^\prime$ values of $\mathbf{z}_{eq} = \frac{1}{\lvert\mathcal{P}\rvert}\sum_{k\in\mathcal{P}}\lvert\mathbf{\hat{z}}_k^{det}\rvert$, given by
	\begin{equation*}
		\Tilde{\Xi} = \arg\max_{\Tilde{\Xi}\subset\Xi,\  \lvert\Tilde{\Xi}\rvert=L^\prime}\sum\nolimits_{i\in\Tilde{\Xi}}\lvert(\mathbf{z}_{eq})_i\rvert.
	\end{equation*}
	where $L^\prime\!=\! \lfloor mL\rfloor$ with $m\geq 1$. 
	We choose $m\!=\!4$ to account for the potentially missed and false detections.
	The set $\Tilde{\Xi}$ becomes the estimated previous channel support in the next frame.
	
	\add{Lastly, we apply the channel refinement algorithm (\textbf{Algorithm \ref{alg_gain_delay_refine}}, which will be introduced in Section \ref{subsec_gain_delay_refinement}) to reconstruct the estimated channel $\mathbf{\hat H}_k$.
	The channel refinement algorithm improves the channel estimation performance by deriving the path coefficients and time delays of the estimated paths corresponding to the estimated channel support $\Tilde{\Xi}$ from the received signal jointly on the pilot subcarriers.}

	\add{Note that the TS algorithm collects $L_{cm}$ elements from the estimated previous channel support $\hat{\Omega}^{pr}$ in MMV-LS.
	Due to the potential estimation errors in the preceding frame, $\hat{\Omega}^{pr}$ is possibly inaccurate and contains less than $L_{cm}$ correct channel elements.
	With an inaccurate $\hat{\Omega}^{pr}$, MMV-LS would collect wrong channel elements inducing the estimation error, and MMV-CS requires estimating an increased number of channel elements.
	To address this issue, we develop a joint MMV-LS-CS approach (M-FISTA) in the next subsection to do the channel estimation, which does not require the exact number of the correct channel elements in $\hat{\Omega}^{pr}$.}

	\subsection{MMV FISTA-based Channel Estimation (M-FISTA)} \label{subsec_joint_MMV_LS_CS}			
	We propose a joint MMV-LS-CS using the framework of FISTA \cite{beck2009fast} instead of doing the MMV-LS-CS in two stages.
	Given the estimated previous channel support $\Gamma \!=\!\{(i_t-1)G_r \!+\! i_r: (i_r,i_t)\in\hat{\Omega}^{pr}\}$, by staking up the signal \eqref{LS_eq} over the subcarriers, we have the signal model as
	\begin{equation}\label{eq_original_signal_model_FISTA}
	\mathbf{\tilde{y}} = \mathbf{\Phi}_{A}\mathbf{\tilde z}^{LS}+\mathbf{\Phi}_{B}\mathbf{\tilde z}^{CS}+\mathbf{\tilde v},
	\end{equation}
	where $\mathbf{\tilde{z}}^{LS} \!=\! \left[[\mathbf{z}_1]_{\Gamma}^\top, \dots,[\mathbf{z}_{K_p}]_{\Gamma}^\top\right]^\top$,  $\mathbf{\tilde{z}}^{CS} \!=\! \left[[\mathbf{z}_1]_{\Gamma^c}^\top, \dots,[\mathbf{z}_{K_p}]_{\Gamma^c}^\top\right]^\top$,  $\mathbf{\Phi}_A\!=\!\textrm{bdiag}\left(\left[\mathbf{\Theta}_1\right]_{\Gamma},\dots,\left[\mathbf{\Theta}_{K_p}\right]_{\Gamma}\right)$, $\mathbf{\Phi}_B=\textrm{bdiag}\left(\left[\mathbf{\Theta}_1\right]_{\Gamma^c},\dots,\left[\mathbf{\Theta}_{K_p}\right]_{\Gamma^c}\right)$, $\mathbf{\tilde{y}}\!=\![\mathbf{y}_1^\top, \dots,\mathbf{y}_{K_p}^{\top}]^\top$ and $\mathbf{\tilde{v}} \!=\! [\mathbf{\tilde v}_1^\top, \dots,\mathbf{\tilde v}_{K_p}^{\top}]^\top$.
	For ease of exposition, we index the pilot subcarriers as $k=1,\dots,K_p$.
	\add{For a vector $\mathbf{x}\in\mathbb{C}^{MN\times 1}$ with the subvectors $\mathbf{x}(1+i_2 M:M+i_2M)\in\mathbb{C}^{M\times 1},\ i_2=0,\dots,N-1$ sharing the joint sparsity pattern, the mixed $\ell_2/\ell_1$ norm of $\mathbf{x}$ is defined as
	\begin{equation}
	\lVert\mathbf{x}\rVert_{2,1} = \sum_{i_1=1}^{M}\sqrt{\sum_{i_2=0}^{N-1}\left\lvert\mathbf{x}(i_1+i_2M)\right\rvert^2}.
	\end{equation}} 
	We consider $(M,N)=(\lvert\Gamma\rvert,K_p)$ for $\lVert\mathbf{\tilde z}^{LS}\rVert_{2,1}$ and $(M,N)=(\lvert\Gamma^c\rvert,K_p)$ for $\lVert\mathbf{\tilde z}^{CS}\rVert_{2,1}$, respectively.	
	The sparse recovery problem can be formulated as 
	\begin{align}
	\nonumber
	\arg\min\limits_{\mathbf{\tilde z}^{LS},\ \mathbf{\tilde z}^{CS}}\ &\frac{1}{2}\lVert\mathbf{\tilde y} - \mathbf{\Phi}_A \mathbf{\tilde z}^{LS} -\mathbf{\Phi}_B\mathbf{\tilde z}^{CS}\rVert^2\\
	&+ \lambda_1\lVert\mathbf{\tilde z}^{LS}\rVert_{2,1}
	+ \lambda_2\lVert\mathbf{\tilde z}^{CS}\rVert_{2,1},
	\label{FISTA_MMS_LS_CS_original}
	\end{align}
	where $\lambda_1$ and $\lambda_2$ are regularization numbers, determined by the non-zero entries in $\mathbf{\tilde z}^{LS},\mathbf{\tilde z}^{CS}$, that weight the penalty terms.
	The minimum number of the shared channel elements between the consecutive frames is $L_{cm}$ (out of $L$ paths), leading to different scales for $\lVert\mathbf{\tilde z}^{LS}\rVert_{2,1}$ and $\lVert\mathbf{\tilde z}^{CS}\rVert_{2,1}$.
	For a positive constant number $\lambda$, we choose $\lambda_1=\lambda\sqrt{\frac{1}{L_{cm}}}$ and $\lambda_2 = \lambda\sqrt{\frac{1}{L-L_{cm}}}$ since $\lambda_1$ (respectively, $\lambda_2$) is inversely related to the scale of $\lVert\mathbf{\tilde z}^{LS}\rVert_{2,1}$ (respectively, $\lVert\mathbf{\tilde z}^{CS}\rVert_{2,1}$).	
	Letting $\mathbf{\Phi}=[\mathbf{\Phi}_A\ \mathbf{\Phi}_B]$, $\mathbf{x}=[(\mathbf{\tilde z}^{LS})^\top\ (\mathbf{\tilde z}^{CS})^\top]^\top$, we have $f(\mathbf{x})=	\frac{1}{2}\lVert\mathbf{\tilde y} - \mathbf{\Phi} \mathbf{x}\rVert^2$ and $g(\mathbf{x})= \lambda_1\lVert\mathbf{\tilde z}^{LS}\rVert_{2,1}+	\lambda_2\lVert\mathbf{\tilde z}^{CS}\rVert_{2,1}$.
	The problem \eqref{FISTA_MMS_LS_CS_original} can be viewed as a LASSO problem, formulated as
	\begin{equation}
	\arg\min\limits_{\mathbf{x}}\ 
	F(\mathbf{x}) = f(\mathbf{x})
	+ g(\mathbf{x}),
	\end{equation}
	where $f$ is a smooth, convex, and continuously differentiable function.
	For $f(\mathbf{x})$, we derive its gradient as $\nabla{f}(\mathbf{x})\!=\!\mathbf{\Phi}^H(\mathbf{\Phi}\mathbf{x}-\mathbf{\tilde y})$ and the Lipschitz constant equals to $\lVert\mathbf{\Phi}\rVert_2^2$.	
	For the convex function $g:\mathbb{R}^N\rightarrow\mathbb{R}\cup\{\infty\}$ and $\eta\in\mathbb{R}$, the proximal operator of $\mathbf{x}$ associated with $\frac{1}{\eta}g$ is defined by $
	\textrm{prox}_{\frac{1}{\eta}g}(\mathbf{x}) \!=\! \arg\min_{\mathbf{u}\in\mathbb{R}^N}\left\{g(\mathbf{u})\!+\!\frac{\eta}{2}\lVert\mathbf{u}\!-\!\mathbf{x}\rVert_2^2\right\}
	$, which is required to implement the FISTA.
%	is required to implement the FISTA \cite{beck2009fast, tan2014joint}, derived as $	\textrm{prox}_{\frac{1}{\eta}g}(\mathbf{x}) = \arg\min_{\mathbf{u}\in\mathbb{R}^N}\left\{g(\mathbf{u})+\frac{\eta}{2}\lVert\mathbf{u}-\mathbf{x}\rVert_2^2\right\}.	$
%	\nm{shuoldnt this function include the 1 over eta stepsize?}
	Given $\eta\!\geq\! \lVert\mathbf{\Phi}\rVert_2^2$, we derive the proximal operator of $\mathbf{x}$ associated with $\frac{1}{\eta}g$, given by 
	\begin{equation*}
	\text{prox}_{\frac{1}{\eta} g}(\mathbf{x})\!=\!
	\left[\text{prox}_{\frac{1}{\eta} \lambda_1\lVert\cdot\rVert_{2,1}}(\mathbf{\tilde z}^{LS})^\top\ \ \text{prox}_{\frac{1}{\eta} \lambda_2\lVert\cdot\rVert_{2,1}}(\mathbf{\tilde z}^{CS})^\top\right]^\top.
	\end{equation*}
	The operators $\text{prox}_{\frac{1}{\eta} \lambda_1\lVert\cdot\rVert_{2,1}}$ and $\text{prox}_{\frac{1}{\eta} \lambda_2\lVert\cdot\rVert_{2,1}}$ are the group-thresholding operators \cite{tan2014joint}, given by
	\begin{equation*}
	\text{prox}_{\frac{1}{\eta} \lambda_1\lVert\cdot\rVert_{2,1}}(\mathbf{\tilde z}^{LS}_i)=\frac{\mathbf{\tilde z}^{LS}_i}{\lVert\mathbf{\tilde z}^{LS}_i\rVert_2}\max\Big(\lVert\mathbf{\tilde z}^{LS}_i\rVert_2-\frac{1}{\eta} \lambda_1,0\Big),\ 
	\end{equation*}
	for $i=1,\dots,\lvert\Gamma\rvert,$ and
	\begin{equation*}
%	\hspace{0.5em}
	\text{prox}_{\frac{1}{\eta} \lambda_2\lVert\cdot\rVert_{2,1}} (\mathbf{\tilde z}^{CS}_i)=\frac{\mathbf{\tilde z}^{CS}_i}{\lVert\mathbf{\tilde z}^{CS}_i\rVert_2}\max\Big(\lVert\mathbf{\tilde z}^{CS}_i\rVert_2-\frac{1}{\eta} \lambda_2,0\Big),
	\end{equation*}
	for $i=1,\dots,\lvert\Gamma^c\rvert$.
	The vectors $\mathbf{\tilde z}^{LS}_i = [\mathbf{\tilde z}^{LS}(i)\ \mathbf{\tilde z}^{LS}(i+\lvert\Gamma\rvert)\ \cdots\ \mathbf{\tilde z}^{LS}(i+(K_p-1)\lvert\Gamma\rvert)]^\top$ and $\mathbf{\tilde z}^{CS}_i = [\mathbf{\tilde z}^{CS}(i)\ \mathbf{\tilde z}^{CS}(i+\lvert\Gamma^c\rvert)\ \cdots\ \mathbf{\tilde z}^{CS}(i+(K_p-1)\lvert\Gamma^c\rvert)]^\top$ are the subvectors of $\mathbf{\tilde z}^{LS}$ and $\mathbf{\tilde z}^{CS}$, respectively.
	The M-FISTA channel estimator is shown in {\textbf{Algorithm \ref{alg_joint_MMV_LS_CS}}}.
	With the estimated $\mathbf{x}_{u}$,
%	 $\mathbf{x}_{est}=[(\mathbf{\tilde z}_{est}^{LS})^\top\ (\mathbf{\tilde z}_{est}^{CS})^\top]^\top$, 
	the estimates of $\mathbf{\hat z}_k^{LS}$ and $\mathbf{\hat z}_k^{CS}$ can be derived with appropriate rearrangement (line 9).
	The M-FISTA algorithm stops when it reaches $U_{max}$ iterations or the relative error of the objective function $\lVert F(\mathbf{x}_{u})- F(\mathbf{x}_{u-1})\rVert <\epsilon$.		

	\begin{algorithm}[t]\scriptsize{\Large}
		\caption{MMV FISTA-based Channel Estimation (M-FISTA).}
		\label{alg_joint_MMV_LS_CS}
		\begin{algorithmic}[1]
			% \REQUIRE 
			\INPUT measurement $\mathbf{\tilde y}$, dictionary matrix $\mathbf{\Phi}$, initial point $\mathbf{x}_0$, upper bound $\eta\geq\lVert\mathbf{\Phi}\rVert_2^2$
			\OUTPUT estimated channel $\mathbf{\hat{H}}_k$
			\STATE \textbf{Initialization:} $u=1$, $\mathbf{q}_1=\mathbf{x}_0$, $t_1=1$, $\varepsilon=\infty$;
			%			\FOR{$u=1,\dots,U_{max}$}
			\WHILE{$u<U_{max}$ \textbf{and} $\varepsilon >\epsilon$}
			\STATE $\nabla f(\mathbf{q}_u)=\mathbf{\Phi}^H(\mathbf{\Phi}\mathbf{q}_u-\mathbf{\tilde y})$;
			%			\STATE $\nabla h_\mu(\mathbf{q}_u)=\frac{1}{\mu}(\mathbf{q}_u-\text{prox}_{\mu h}(\mathbf{q}_u))$;
			\STATE $\mathbf{x}_u=\text{prox}_{\frac{1}{\eta}g}(\mathbf{q}_u-\frac{1}{\eta}\nabla f(\mathbf{q}_u))$;
			\STATE $t_{u+1}=\frac{1+\sqrt{1+4t_u^2}}{2}$;		
			\STATE $\mathbf{q}_{u+1}=\mathbf{x}_u+\frac{t_u-1}{t_{u+1}}(\mathbf{x}_u-\mathbf{x}_{u-1})$; 
			\STATE $\varepsilon = \lvert F(\mathbf{x}_{u})- F(\mathbf{x}_{u-1})\rvert$; $u\coloneqq u+1$;
			%			\IF{$\lVert F(\mathbf{x}_{u})- F(\mathbf{x}_{u-1})\rVert <\epsilon$} break;
			%			\ENDIF
			\ENDWHILE
			%			\ENDFOR	
			\STATE Derive $\mathbf{\hat{z}}_k^{LS}$ and $\mathbf{\hat{z}}_k^{CS}$ from $\mathbf{x}_{u}$;
			\STATE {The support of $\mathbf{\Theta}_k^{(1)}$:} $\Xi^{(1)}\!=\!\{i: \sum\limits_{k\in\mathcal{P}}\lvert(\mathbf{\hat{z}}_k^{LS} + \mathbf{\hat{z}}_k^{CS})_i\rvert \neq 0\}$;
			\item[]
			\STATE \emph{Enhance the beam resolution by sequential search method}					
			%					\STATE Derive $\Upsilon \!\!=\!\! \{(\hat\psi_{A,\ell},\hat\psi_{B,\ell},\hat\psi_{C,\ell},\hat\psi_{D,\ell}),\ \ell\!=\!1,\dots,L_{t}\}$ from $\Xi^{(1)}$;
			\STATE $\Upsilon=\emptyset$; $\mathcal{R}=\emptyset$; $\mathbf{T}_k=[\ ]$, $\mathbf{r}_k=\mathbf{y}_k,\ k\in\mathcal{P}$;
			\WHILE{$\lvert\Upsilon\rvert< L^{\prime}$}
			\STATE $j_1^*= \arg \max\limits_{j\in\Xi^{(1)}\backslash\Upsilon}
			\sum\limits_{k\in\mathcal{P}}
			\lvert \mathbf{\Theta}_k^{(1)}(:,j)^H\mathbf{r}_{k}\rvert^2$; $\Upsilon:=\Upsilon\cup j_1^*$;	 	
			\STATE Derive $(\hat\psi_{A}^{(1)},\hat\psi_{B}^{(1)},\hat\psi_{C}^{(1)},\hat\psi_{D}^{(1)})$ from $j_1^*$;	
			\STATE Derive $(\hat\psi_{A}^{(M)},\hat\psi_{B}^{(M)},\hat\psi_{C}^{(M)},\hat\psi_{D}^{(M)})$ using \eqref{step_spatial_h_AOD}-\eqref{step_spatial_v_AOA};	
			\STATE Derive $j_2^*$ from $(\hat\psi_{A}^{(M)},\hat\psi_{B}^{(M)},\hat\psi_{C}^{(M)},\hat\psi_{D}^{(M)})$;	$\mathcal{R}:=\mathcal{R}\cup j_2^*$;
			\STATE $\mathbf{T}_k:=\Big[\mathbf{T}_k\ \mathbf{\Theta}_k^{(M)}(:,j_2^*)
			\Big],\ k\in\mathcal{P}$;
			\STATE $\mathbf{r}_k = \mathbf{y}_k-\mathbf{T}_k\mathbf{T}_k^{+}\mathbf{y}_k,\ k\in\mathcal{P}$;
			\ENDWHILE
			\STATE Derive $\tilde{\Xi}^{(M)}$ (i.e. the support of $\mathbf{\Theta}_k^{(M)}$) from $\mathcal{R}$;
			\item[]
			\STATE Reconstruct $\mathbf{\hat{H}}_k$ by \textbf{Algorithm \ref{alg_gain_delay_refine}} with $(\mathbf{Y}_k, \mathbf{\Theta}_k^{(M)},\Tilde{\Xi}^{(M)})$;
		\end{algorithmic}		
	\end{algorithm}

	Since the LASSO approach does not handle highly correlated variables well, we are not able to do the FISTA-based algorithm with dictionary matrices that have very large grid sizes, inducing a higher correlation between the atoms.
	Without additional processing, M-FISTA estimates the channel support with a low-resolution codebook (line 1-10), degrading the estimation accuracy due to the grid mismatch.
%	inducing an inaccurate channel estimation due to the grid-mismatch.
%	Therefore, the M-FISTA algorithm (line 1-9, \textbf{Algorithm \ref{alg_joint_MMV_LS_CS}}) is implemented with the first-level (low-resolution) codebook.
	To address the grid-mismatch issue, we apply a sequential search method with hierarchical codebooks (Section \ref{sec_seq_search}) to enhance the resolution of each estimated path.
	We construct the hierarchical codebook for the index selection on multiple levels, introduced in Section \ref{sec_hierarchical_codebook}.	
	The first-level codebook $\mathbf{\Theta}_k^{(1)}$ is considered the low-resolution codebook, and the $M$-th level ($M\!>\!1$) codebook $\mathbf{\Theta}_k^{(M)}$ refers to the high-resolution codebook.	
%	(The hierarchical codebook is constructed to do the index selection on multiple levels, introduced in Section \ref{sec_hierarchical_codebook}.)
	We have the channel support (of $\mathbf{\Theta}_k^{(1)}$) by combining $\mathbf{\hat z}_k^{LS}$ with $\mathbf{\hat z}_k^{CS}$ as $\Xi^{(1)}\!=\!\{i\!:\! \sum\nolimits_{k\in\mathcal{P}}\lvert(\mathbf{\hat{z}}_k^{LS} + \mathbf{\hat{z}}_k^{CS})_i\rvert \!\neq\! 0\}$ (line 10).
%	\nm{how do you solve this issue? What is the connection between this and the grid mismatch mentioned next?? Feels disconnected and not well motivated}
%	To address the grid-mismatch issue, we apply a sequential search method with hierarchical codebooks (Section \ref{sec_seq_search}) to enhance the resolution of each estimated channel path.
%	Given the estimated $\Xi^{(1)}$, we derive the low-resolution channel support in terms of the indices of AOAs/AODs as $\Upsilon$ (line 12).
	Next, the algorithm is initialized as $\Upsilon=\emptyset$, $\mathcal{R}=\emptyset$, and the residual $\mathbf{r}_{k} = \mathbf{y}_k,\ k\in\mathcal{P}$ (line 12).
	We construct the set $\Upsilon$ by collecting the $L^\prime$ paths that are most correlated with $\mathbf{r}_k$ (line 14).	
%	Then, the set $\Upsilon$ is constructed by collecting the index of the column with the largest correlation with the residual $\mathbf{r}_k$ (line 14), until $L^\prime$ paths are collected (line 13).
	In each iteration, to enhance the resolution of the selected beam index $j_1^*$ (of $\mathbf{\Theta}_k^{(1)}$), the sequential search method derives the beam index $j_2^*$ (of $\mathbf{\Theta}_k^{(M)}$), collected in $\mathcal{R}$ (line 15-17).
	The residual $\mathbf{r}_{k}$ is updated by removing the channel effect of the indices in $\mathcal{R}$ (line 19).
	Lastly, we derive the estimated channel support $\tilde{\Xi}^{(M)}$ from $\mathcal{R}$ in an enhanced beam resolution (line 21).

	\add{To enhance the channel estimation performance, we apply the channel refinement algorithm (\textbf{Algorithm \ref{alg_gain_delay_refine}}, which will be introduced in Section \ref{subsec_gain_delay_refinement}) to reconstruct the estimated channel $\mathbf{\hat H}_k$ by deriving the path gains and time delays of the paths corresponding to $\tilde{\Xi}^{(M)}$ from the received signal jointly on the pilot subcarriers.}

\section{Performance Enhancement and Complexity Analysis}
\label{sec_performance_enhance}
	In this section, we propose additional operations to enhance the channel estimation performance.
%	support tracking-based channel trainings for dual-wideband MIMO-OFDM.
	Section \ref{subsec_gain_delay_refinement} proposes a channel refinement algorithm that leverages the spreading loss structure, and Section \ref{sec_seq_search} proposes a sequential search method using a hierarchical codebook to reduce the complexity.
	Lastly, Section \ref{subsec_complexity} provides the complexity analysis.
	
	\begin{algorithm}[t]\scriptsize{\Large}
		\caption{Channel Refinement Algorithm.}
		\label{alg_gain_delay_refine}
		\begin{algorithmic}[1]
			% \REQUIRE 
			\INPUT measurement $\mathbf{Y}_k$, dictionary matrix $\mathbf{\Theta}_k$, support set ${\Xi}$			
			\OUTPUT ${\mathbf {\hat H}_k}, k=1,\dots, K_o$
			\STATE Derive  $(i_{\theta,\ell},i_{\phi,\ell})$ from $\Tilde{\Xi}$, $\ell=1,\dots, L^\prime$;
			\STATE Derive $\mathbf{\hat q}_k$ by solving \eqref{q_opt}, $k=1,\dots, K_p$;
			\STATE Derive $\hat z_\ell$ by \eqref{z_est_avg}, $\hat \alpha_\ell^{\prime}$ by \eqref{alpha_opt_modified}, $\ell=1,\dots, L^\prime$;			
			\STATE Reconstruct $\mathbf{\hat H}_k$ by \eqref{ch_recon_v2}, $k=1,\dots,K_o$;		
		\end{algorithmic}
	\end{algorithm}

	\subsection{Channel Reconstruction And Path Coefficient/Time Delay Refinement} \label{subsec_gain_delay_refinement}
	Given the channel support ${\Xi}$ estimated by the support tracking-based channel training, the effective path coefficient vector can be directly estimated by the LS estimator as
	\begin{equation} \label{q_opt}
	\arg\min_{\mathbf{q}_k}
	\left\Vert\mathbf{y}_{k}- [\mathbf{\Theta}_k]_{{\Xi}}\ \mathbf{q}_k\right\Vert_F^2,\ k\in\mathcal{P},
	\end{equation}
	where $\mathbf{y}_{k}=\textrm{vec}(\mathbf{Y}_k)$ and $\mathbf{\Theta}_k= (\mathbf{A}_{T,k}^{H}\mathbf{X}_k)^{\top} \otimes(\mathbf{A}_{R,k}^{H}\mathbf{W}_k)^{H}$ as in \eqref{LS_eq}.
	Assuming $\lvert{\Xi}\rvert=L^{\prime}$, the vector $\mathbf{ q}_k= [\gamma_{1,k},\dots,\gamma_{L^{'},k}]\!^\top$ is the effective path coefficient vector. 
	The problem \eqref{q_opt}  is a least squares problem, solved by $\hat{\mathbf{ q}}_k=[\hat\gamma_{1,k},\dots,\hat\gamma_{L^{'},k}]\!^\top=\left([\mathbf{\Theta}_k]_{{\Xi}}\right)^{+} \mathbf{y}_k$. %denoted as \textbf{LS estimator}.
	Note that we have the indices of the estimated AOAs/AODs  as $(i_{\theta,\ell},i_{\phi,\ell}),\ \ell=1,\dots,L^\prime$, corresponding to the indices in ${\Xi}$.
	Thus, the MIMO-OFDM channel is reconstructed as
	$\mathbf{\hat H}_k = \sum_{\ell=1}^{L^\prime} \hat\gamma_{\ell,k}\mathbf{A}_{R,k}(:,i_{\theta,\ell})\mathbf{A}_{T,k}^{H}(:,i_{\phi,\ell})$.

	Note that, however, such LS estimation does not embed any structure (e.g., as in \eqref{freq_MIMO_ch}) in the path coefficients and time delays	across the subcarriers (frequency).
	To leverage such structure, a channel refinement algorithm is proposed in \textbf{Algorithm \ref{alg_gain_delay_refine}} to improve the estimation performance, which proceeds as follows.
	With the generator $\{z_\ell = e^{-j2\pi\frac{B}{K_o}\tau_\ell}\}$, the effective path coefficient is expressed as $\gamma_{\ell,k}\!=\! \sqrt{N_r N_t}\alpha_\ell(\Delta_k)  {z_\ell}^{k-\frac{K_o+1}{2}}$.
	In Section \ref{subsec_freq_sel_channel}, we have proposed and justified the use of the approximation $\alpha_{\ell}(\Delta) \!\approx\! \frac{\alpha_\ell^\prime}{1+\Delta/f_c}$.
	Therefore, the effective path coefficient can be approximated as $\gamma_{\ell,k}\approx \sqrt{N_r N_t} \frac{\alpha_{\ell}^\prime}{1+\Delta_k/f_c} {z_\ell}^{k-\frac{K_o+1}{2}}$.
	Given the estimated $\mathbf{\hat q}_k=[\hat \gamma_{1,k},\dots,\hat \gamma_{L^{'},k}]^\top$ by \eqref{q_opt} and 
	$\frac{\gamma_{\ell,k+1}}{\gamma_{\ell,k}}\approx \frac{1+\Delta_k/f_c}{1+\Delta_{k+1}/f_c}z_\ell$, the estimation of $z_\ell$ can be formulated as
	\begin{equation*} %\label{z_opt}
	\arg \min_{z_{\ell}}\!\sum_{i=1}^{K_p-1}\!\left( z_{\ell} \!-\! \left(\frac{\hat{\gamma}_{\ell,1 + i \delta_p}(1 + \Delta_{1 + i \delta_p}/f_c)} {\hat{\gamma}_{\ell,1+(i-1)\delta_p}(1 + \Delta_{1 + (i-1) \delta_p}/f_c)}\right)^{\frac{1}{\delta_p}}\right)^2
	\end{equation*}	
	with $\delta_p$ as the pilot subcarrier spacing, which is solved by
	\begin{equation} \label{z_est_avg}
	\hat{z}_\ell = \frac{1}{K_p-1}\sum_{i=1}^{K_p-1}\left(\frac{\hat{\gamma}_{\ell,1 + i \delta_p}(1 + \Delta_{1 + i \delta_p}/f_c)} {\hat{\gamma}_{\ell,1+(i-1)\delta_p}(1 + \Delta_{1 + (i-1) \delta_p}/f_c)}\right)^{\frac{1}{\delta_p}}.
	\end{equation}
	The refined time delay is derived by ${\hat{\tau}_\ell = -\frac{K_o}{2\pi B}\measuredangle \hat z_\ell}$, where $\measuredangle \hat z_\ell$ denotes the phase angle of $\hat z_\ell$.
	Next, we estimate the reference path coefficient $\alpha_\ell^{\prime}$ by formulating the optimization problem as
	\begin{equation}\label{alpha_opt_modified}
	\arg\min_{\alpha_\ell^{\prime}}\left\Vert\mathbf{p}_\ell -\mathbf{c}_\ell \alpha_{\ell}^{\prime}\right\Vert_F^2,\ \ell=1,\dots,L,
	\end{equation}
	where $\mathbf{p}_\ell,\mathbf{c}_\ell\in\mathbb{C}^{K_p\times 1}$ with $\mathbf{p}_\ell(k)=\hat \gamma_{\ell,k}$ and $\mathbf{c}_\ell(k)=\frac{\sqrt{N_rN_t}}{1+\Delta_{k}/f_c}{\hat z_\ell}^{k-\frac{K_o+1}{2}}$, $k\in\mathcal{P}$.
	The problem \eqref{alpha_opt_modified} is a least squares problem, solved by $\hat \alpha_\ell^{\prime}=(\mathbf{c}_\ell)^{+}\mathbf{p}_\ell$.
	%\tc{I ran the simulations with this modification. The performance is almost the same as the previous one by averaging all subcarriers. Though, I think it makes sense to formulate it as a least squares problem.}
	%	Next, we estimate the reference path coefficient by averaging over the pilot subcarriers as \nm{why not directly solve the LS problem with fixed z?}
	%	\begin{equation}\label{alpha_opt_modified}
	%	\hat \alpha_{\ell}^\prime = \frac{1}{\lvert\mathcal{P}\rvert}\sum_{k\in\mathcal{P}} 
	%	\hat{\gamma}_{\ell,k}
	%	\frac{ 1+f_k/f_c}{\sqrt{N_rN_t}}  		
	%	{\hat z_\ell}^{-(k-\frac{K_o+1}{2})}.
	%	\end{equation}	
	%	Given the estimated $\hat{z}_\ell$, we then estimate the reference path gains $\alpha_\ell^{\prime}$ by the following optimization problem
	%	\begin{equation}\label{alpha_opt_modified}
	%	\arg\min_{\alpha_{\ell}^{\prime}}\sum_{k\in\mathcal{P}}\Big(\hat{\gamma}_{\ell,k}- \sqrt{N_r N_t}\cdot\frac{\alpha_\ell^\prime}{f_c+f_k}\cdot{\hat z_\ell}^{k-\frac{K_o+1}{2}}\Big)^2,
	%	\end{equation}	
	%	which can be solved by the least square approach, and the refined terms $\hat \alpha_\ell^{\prime}$ are derived.
	%	Note that we could have the indices of the estimated AOAs/AODs  as $(i_{\theta,\ell},i_{\phi,\ell}),\ \ell=1,\dots,L^\prime$, corresponding to the indices in $\Tilde{\Xi}$.
	Given the refined version of reference path coefficients $\hat \alpha_{\ell}^\prime$ and generators $\hat{z}_\ell$ (time delays $\hat{\tau}_\ell$) accompanied with the indices of the estimated AOAs/AODs, we reconstruct the MIMO channel by
	\begin{equation}\label{ch_recon_v2}
	\mathbf{\hat H}_k \!=\! \sqrt{N_r N_t} \sum_{\ell=1}^{L^\prime}\frac{\hat \alpha_\ell^{\prime}}{1+\frac{\Delta_{k}}{f_c}} {\hat z_\ell}^{k-\frac{K_o+1}{2}} \mathbf{A}_{R,k}(:,i_{\theta,\ell})\mathbf{A}_{T,k}^{H}(:,i_{\phi,\ell}).
	\end{equation}

	\begin{figure}[t]
		\centering
		\includegraphics[width=\linewidth]{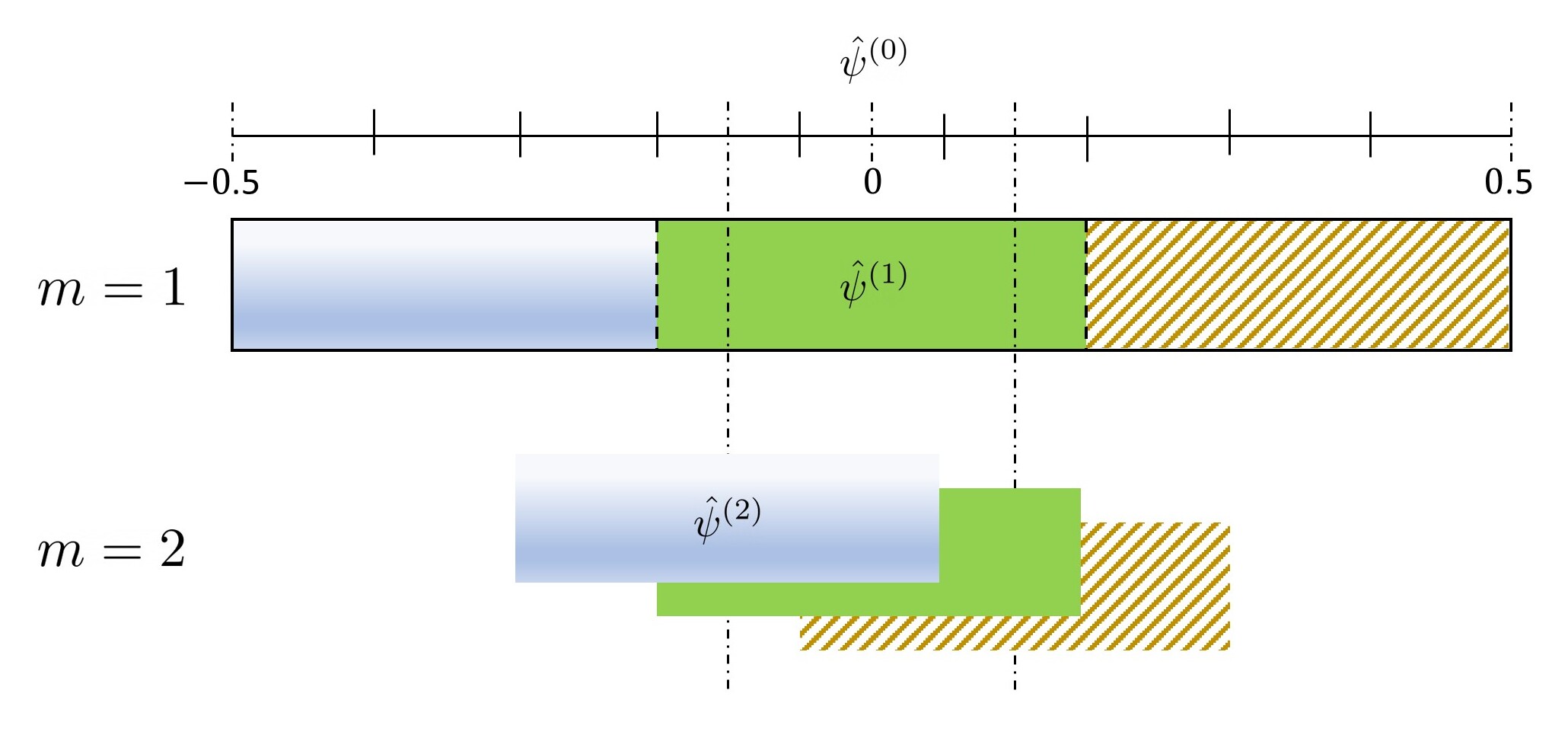}
		\caption{An example of the hierarchical codebook with $(G_{sub},M)=(3,2)$.}
		\label{fig:multi_codebook}
	\end{figure}

	\subsection{Sequential Search Method With Hierarchical Codebook}\label{sec_seq_search}
	We develop a sequential search method using hierarchical codebooks to reduce the computational complexity of the greedy selection in the TS algorithm and enhance the beam resolution in the M-FISTA algorithm.
	Section \ref{sec_hierarchical_codebook} introduces the hierarchical codebooks to divide the index selection into multiple levels.
	Section \ref{sec_seq_search_OMP} proposes the sequential search method.
	
	\subsubsection{Hierarchical Codebook}
	\label{sec_hierarchical_codebook}
	In each spatial dimension, we construct a hierarchical codebook to do the index selection on multiple levels.
	Considering an example of a uniform grid on $[-0.5,0.5)$ of size $G$ along a spatial dimension (same as in \eqref{receive_grid}) as
	%	\begin{equation}
	$\mathcal{G} = \left\{\psi_{i}= \frac{i-\frac{G+1}{2}}{G},\ i = 1,\dots,G\right\},$
	%	\end{equation}
	%	\nm{Is this a subset of the original grid codebook?}
	we divide the index selection of $\mathcal{G}$ into the hierarchical search on $M$ levels, where each level has its sub-codebook of size $G_{sub}$, satisfying $G=G_{sub}^M$.
	We design the $m$-th level sub-codebook as
	\begin{align} 
	%\label{m_level_hierar_cb}
		\nonumber	
	&\mathcal{G}^{(m)}(\hat\psi^{(m-1)}) \\
	&= \Big\{\psi_{(m,i)}=\hat \psi^{(m-1)} + \frac{i-\frac{G_{sub}+1}{2}}{{G_{sub}}^m},\ i = 1,\dots,G_{sub}\Big\},
	\label{eq_hierarchical_search}
	\end{align}
	where $\hat\psi^{(m-1)}$ is the selected codeword from the previous level. 
	The selected codeword is initialized as ${\hat \psi^{(0)}=0}$.
%	The initial value of the selected codeword is assumed as ${\hat \psi^{(0)}=0}$.
	In the first level ($m\!=\!1$), the index search is implemented with the resolution $G_{sub}^{-1}$. %$\frac{1}{G_{sub}}$.
	In the subsequent levels ($m\!>\!1$), the index search is done with a finer resolution $G_{sub}^{-m}$ centered around the previously selected $\hat \psi^{(m-1)}$.	
	Hence, the total number of codewords to be searched is reduced from $G_{sub}^M$ to $MG_{sub}$.	
	In Fig.~\ref{fig:multi_codebook}, an example of the hierarchical codebook with $(G_{sub},M)\!=\!(3,2)$ is illustrated.
	%	Considering an original codebook of size $G=9$, 
	A codebook of size $G\!=\!9$ is divided into $M\!=\!2$ levels, where each level has its sub-codebook with $G_{sub}\!=\!3$.
	%	on which has a subcodebook with $G_{sub}=3$.
	%	, where $G_{sub}=3$ codewords are evaluated in each level.
	Due to the narrow beam of massive MIMO, $G_{sub}$ should be selected to guarantee the beam coverage of $\mathcal{G}^{(1)}$ on all possible spatial angles.
	
	\begin{figure}[t]
		\centering
		\begin{minipage}[t]{0.42\textwidth}
			\centering
			\includegraphics[width=\textwidth]{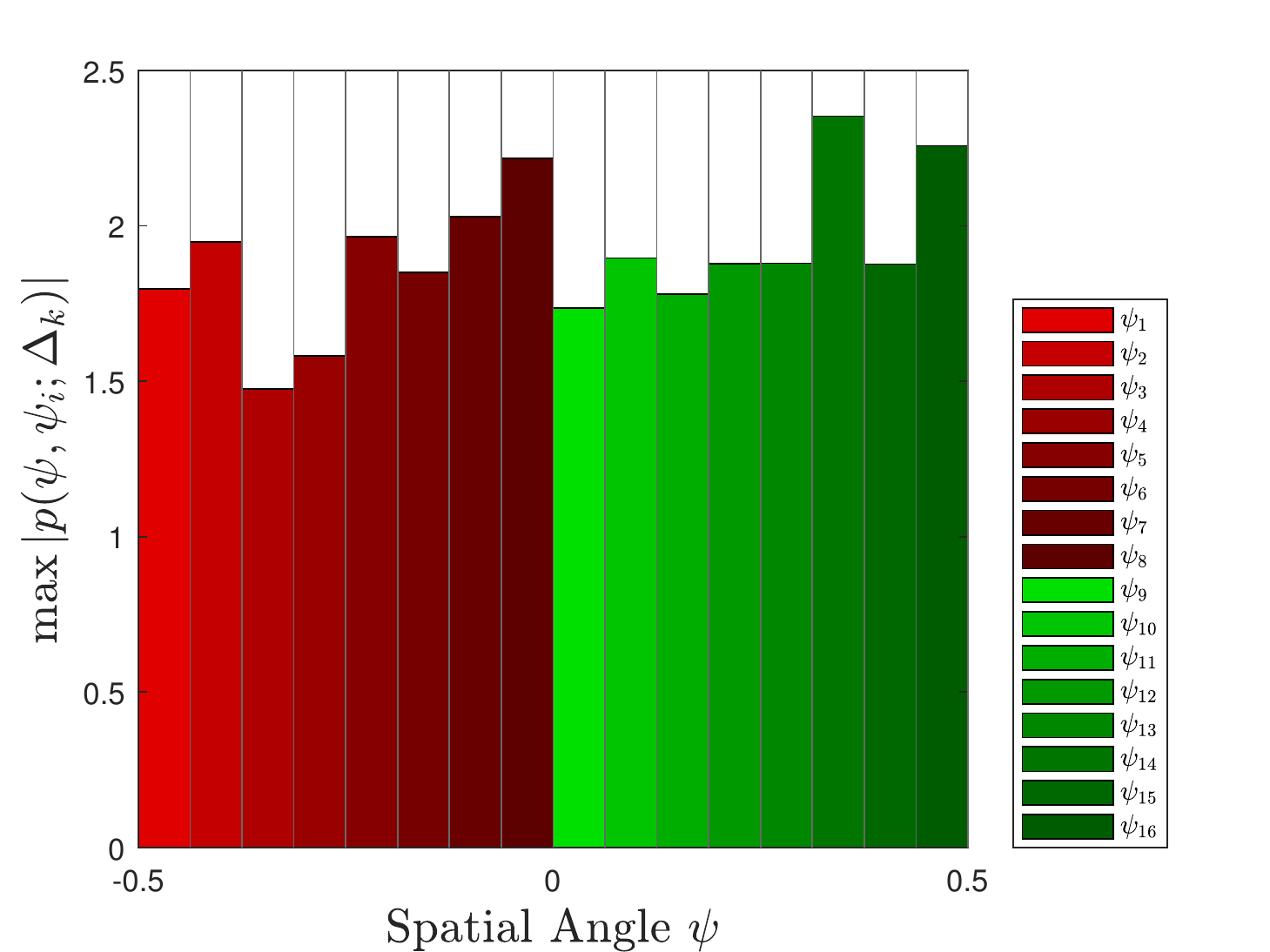}\\
			(a)
			%		\caption{}
			%		\label{fig:beam_forming_pattern} 
		\end{minipage}
		\hfill
		\begin{minipage}[t]{0.42\textwidth}
			\centering
			\includegraphics[width=\textwidth]{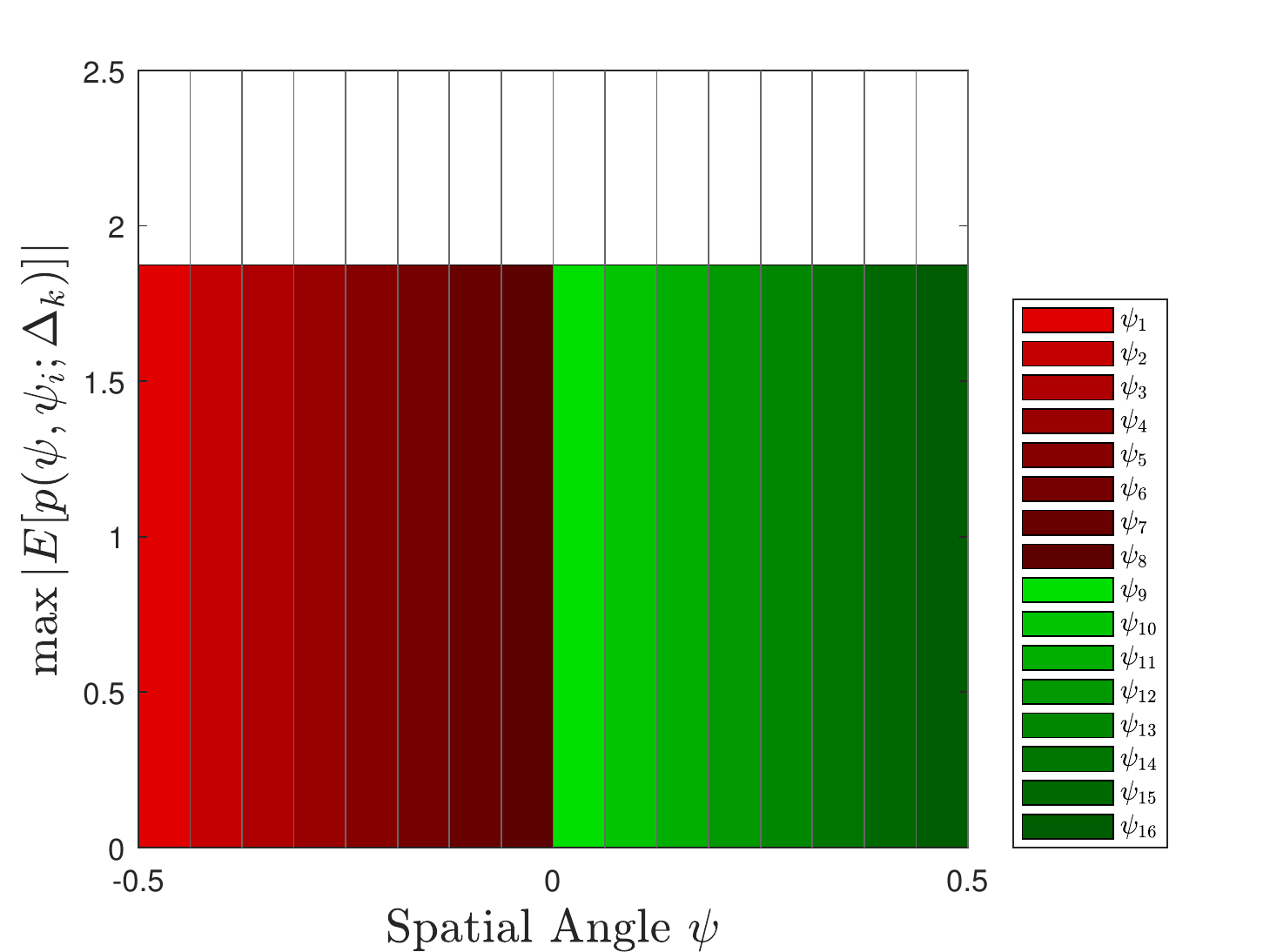}\\
			(b)
			%					\caption{}
			%		\label{fig:expected_beam_forming_pattern}
		\end{minipage}	
		\caption[Two numerical solutions]{(a) Beamforming gain $\max_{\psi} \lvert p(\psi,\psi_i;\Delta_{k})\rvert$ of the effective dictionary vector with $\psi_i \in\mathcal{G}^{(1)}$, given $G_{sub}=N=16$, $Q=30$, $B=8$ GHz, $f_c=142$ GHz, $k=50$ among $K_o=128$ subcarriers.  (b) Expected beamforming gain $\max_\psi \lvert \mathbb{E}[p(\psi,\psi_i;\Delta_{k})]\rvert$.
		}
		\label{fig:beam_forming_pattern}
	\end{figure}
	
	\begin{figure*}[b!]
		\hrulefill
		\begin{equation}
		\tag{33}
		\hat\psi_{A}^{(m)}=\arg\max_{\psi \in \mathcal{G}_{hT}^{(m)}(\hat\psi_{A}^{(m-1)})}\sum\nolimits_{k\in\mathcal{P}}
		\lvert \mathbf{u}_k(\psi,\hat\psi_{B}^{(m-1)},\hat\psi_{C}^{(m-1)},\hat\psi_{D}^{(m-1)})^H\mathbf{r}_{k}\rvert^2,
		\label{step_spatial_h_AOD}
		%	\ m=2,\dots,M;
		\end{equation}	
		%	\vspace{-1.7em}	
		\begin{equation}
		\tag{34}
		\hat\psi_{B}^{(m)}=\arg\max_{\psi \in \mathcal{G}_{vT}^{(m)}(\hat\psi_{B}^{(m-1)})}\sum\nolimits_{k\in\mathcal{P}}
		\lvert \mathbf{u}_k(\hat\psi_{A}^{(m)},\psi,\hat\psi_{C}^{(m-1)},\hat\psi_{D}^{(m-1)})^H\mathbf{r}_{k}\rvert^2,
		\label{step_spatial_v_AOD}
		%	\ m=2,\dots,M;
		\end{equation}
		%	\vspace{-1.7em}			
		\begin{equation}
		\tag{35}
		\hat\psi_{C}^{(m)}=\arg\max_{\psi \in \mathcal{G}_{hR}^{(m)}(\hat\psi_{C}^{(m-1)})}
		\sum\nolimits_{k\in\mathcal{P}}
		\lvert \mathbf{u}_k(\hat\psi_{A}^{(m)},\hat\psi_{B}^{(m)},\psi,\hat\psi_{D}^{(m-1)})^H\mathbf{r}_{k}\rvert^2,
		\label{step_spatial_h_AOA}
		%	\ m=2,\dots,M;
		\end{equation}
		%	\vspace{-1.7em}
		\begin{equation}
		\tag{36}
		\hat\psi_{D}^{(m)}=\arg\max_{\psi \in \mathcal{G}_{vR}^{(m)}(\hat\psi_{D}^{(m-1)})}
		\sum\nolimits_{k\in\mathcal{P}}
		\lvert \mathbf{u}_k(\hat\psi_{A}^{(m)},\hat\psi_{B}^{(m)},\hat\psi_{C}^{(m)},\psi)^H\mathbf{r}_{k}\rvert^2,
		\label{step_spatial_v_AOA}
		%	\ m=2,\dots,M;
		%	\vspace{-0.5em}
		\end{equation}
	\end{figure*}

	To choose $G_{sub}$, we analyze the beamwidth of the effective dictionary vector in \eqref{IO_matrix_beamtraining_virtual}.	
	For ease of exposition, we consider the effective dictionary vector along a spatial dimension as $\mathbf{W}_k^{H}\mathbf{a}_{N}(\psi;\Delta_{k})$, where $\mathbf{W}_k\!\in\!\mathbb{C}^{N\times Q}$ is the measurement matrix and $\mathbf{a}_{N}(\psi;\Delta_{k})\!\in\!\mathbb{C}^{N\times 1}$ is the array response vector (as in \eqref{steering_vector}).	
	The beamforming pattern of $\mathbf{W}_k^{H}\mathbf{a}_{N}(\psi_i;\Delta_{k})$, $\psi_i\!\in\!\mathcal{G}^{(1)}$ is defined as  
	\begin{equation}\setcounter{equation}{30}
	p(\psi,\psi_i;\Delta_{k}) = \left(\mathbf{W}_k^{H}\mathbf{a}_{N}(\psi;\Delta_{k})\right)^H\mathbf{W}_k^{H}\mathbf{a}_{N}(\psi_i;\Delta_{k}).
	\end{equation}
	Note that we adopt the random beamforming method for the channel training, i.e., 	$\mathbf{W}_k{(n,q)}=\frac{e^{j\eta_{n,q}}}{\sqrt{N}}$, $\eta_{n,q}\sim\mathcal{U}(0,2\pi)$.
	The expected beamforming pattern of $\mathbf{W}_k^{H}\mathbf{a}_{N}(\psi;\Delta_{k})$ is $\mathbb{E}[p(\psi,\psi_i;\Delta_{k})] = \frac{Q}{N}\mathbf{a}_{N}(\psi;\Delta_{k})^H\mathbf{a}_{N}(\psi_i;\Delta_{k})$, demonstrating that the expected beamwidth of $\mathbf{W}_k^{H}\mathbf{a}_{N}(\psi;\Delta_{k})$ is the same as the beamwidth of $\mathbf{a}_{N}(\psi_i;\Delta_{k})$, represented by the \text{Half Power Beamwidth} (HPBW) ${=0.5\times \textrm{First Null Beamwidth (FNBW)}}$ \cite{balanis2016antenna}.
	We derive $\textrm{FNBW}=\frac{2}{N(1+\Delta_{k}/f_c)}$ by analyzing the beamforming pattern, so the beamwidth of $\mathbf{a}_{N}(\psi_i;\Delta_{k})$ is $\textrm{HPBW}\!=\!\frac{1}{N(1+\Delta_{k}/f_c)}\!\approx\!\frac{1}{N}$.
	To guarantee the beam coverage of $\mathcal{G}^{(1)}$ on all spatial angles, the resolution ${G_{sub}^{-1}}$ is required to be at least finer than the beamwidth ${N}^{-1}$, so we choose $G_{sub}\!\geq\! N$.	
	In Fig. \ref{fig:beam_forming_pattern}a, we show the beamforming gain of $\mathbf{W}_k^{H}\mathbf{a}_{N}(\psi_i;\Delta_{k})$ with $\psi_i \!\in\!\mathcal{G}^{(1)}$, given $N\!=\!G_{sub}\!=\!16$, $Q\!=\!30$, $B\!=\!8$GHz, $f_c\!=\!142$GHz, and $k\!=\!50$ among $K_o\!=\!128$ subcarriers.
	The expected beamforming gain is shown in Fig. \ref{fig:beam_forming_pattern}b.
	%	 $\left(\mathbf{W}^{H}\mathbf{a}_{N}(\psi_i;f_k)\right)$, for $\psi_{i}= \frac{i-\frac{G_{sub}+1}{2}}{G_{sub}},\ i = 1,\dots,G_{sub}$.

	\subsubsection{Sequential Search Method} \label{sec_seq_search_OMP}
	With the extended virtual representation of the MIMO channel in Section \ref{subsec_extended_virtual_rep}, the grid sizes $(G_{hr},G_{vr}, G_{ht}, G_{vt})$ should be chosen sufficiently large to minimize the grid-mismatch errors between the spatial AOAs/AODs and their quantized values.
	For the TS algorithm (\textbf{Algorithm \ref{alg_tensor_LS_CS}}), however, the greedy selection (line 13) searches over four different grids $(\mathcal{G}_{hT},\mathcal{G}_{vT},\mathcal{G}_{hR},\mathcal{G}_{vR})$ jointly, which can be reformulated into the problem as
	\begin{equation}\label{eq_grid_ori_problem}
	\arg\max_{\substack{(\psi_{A},\psi_{B},\psi_{C},\psi_{D})\\ \in \mathcal{G}_{hT}\times\mathcal{G}_{vT}\times\mathcal{G}_{hR}\times\mathcal{G}_{vR}}}\sum\nolimits_{k\in\mathcal{P}}
	\lvert \mathbf{u}_k(\psi_{A},\psi_{B},\psi_{C},\psi_{D})^H\mathbf{r}_{k}\rvert^2,
	\end{equation}
	where the equivalent dictionary vector is defined as
	\begin{align*}
	\mathbf{u}_k&(\psi_{A},\psi_{B},\psi_{C},\psi_{D})\\
	=&\left((\mathbf{a}_{N_{ht}}(\psi_{A};\Delta_{k})\otimes\mathbf{a}_{N_{vt}}(\psi_{B};\Delta_{k}))^{H}\mathbf{X}_k\right)^{\top}\\ 
	&\otimes
	\left((\mathbf{a}_{N_{hr}}(\psi_{C};\Delta_{k})\otimes\mathbf{a}_{N_{vr}}(\psi_{D};\Delta_{k}))^{H}\mathbf{W}_k\right)^{H}.
	\end{align*} 
	The problem \eqref{eq_grid_ori_problem} is computationally prohibitive for large grid sizes, due to the exhaustive search over $\{i_{hr},i_{vr},i_{ht},i_{vt}\}\in[1,G_{hr}]\times[1,G_{vr}]\times[1,G_{ht}]\times[1,G_{vt}]$. 
	To address the issue, a sequential search method using hierarchical codebooks is developed to reduce the computational overhead of the greedy selection, detailed as below.
	First, the problem \eqref{step_initial_search} is solved by searching over the index set $\mathcal{G}_{hT}^{(1)}\times\mathcal{G}_{vT}^{(1)}\times\mathcal{G}_{hR}^{(1)}\times$ $\mathcal{G}_{vR}^{(1)}$ to maximize the objective function, where $\mathcal{G}_{hT}^{(1)}$, $\mathcal{G}_{vT}^{(1)}$, $\mathcal{G}_{hR}^{(1)}$, $\mathcal{G}_{vR}^{(1)}$ are the first-level hierarchical codebooks of $\mathcal{G}_{hT}$, $\mathcal{G}_{vT}$, $\mathcal{G}_{hR}$, $\mathcal{G}_{vR}$, respectively.
	\begin{align}
		\nonumber
	&(\hat\psi_{A}^{(1)},\hat\psi_{B}^{(1)},\hat\psi_{C}^{(1)},\hat\psi_{D}^{(1)})=\\
	&
	\label{step_initial_search}
	\arg \max_{\substack{(\psi_{A},\psi_{B},\psi_{C},\psi_{D})\\ \in \mathcal{G}_{hT}^{(1)}\times\mathcal{G}_{vT}^{(1)}\times\mathcal{G}_{hR}^{(1)}\times\mathcal{G}_{vR}^{(1)}}}
	\sum\nolimits_{k\in\mathcal{P}}
	\lvert \mathbf{u}_k(\psi_{A},\psi_{B},\psi_{C},\psi_{D})^H\mathbf{r}_{k}\rvert^2;
	\end{align}	
	We assume the hierarchical codebooks are constructed with the same grid size $G_{sub}$. 
	For the distinct spatial dimensions, $G_{sub}$ can be selected as a different value based on the desired resolution.
	Next, we sequentially solve the one-dimensional search problems using the hierarchical codebook, starting with the search on the horizontal spatial AODs \eqref{step_spatial_h_AOD}, the vertical spatial AODs \eqref{step_spatial_v_AOD}, the horizontal spatial AOAs \eqref{step_spatial_h_AOA}, and then the vertical spatial AOAs \eqref{step_spatial_v_AOA}, for the level $m=2,\dots,M$.
	With the sequential search method, we reduce the number of the index sets to be searched from ${G_{ht} G_{vt} G_{hr} G_{vr}=G_{sub}^{4M}}$ to ${G_{sub}^4 + 4(M-1)G_{sub}}$.

	For the M-FISTA algorithm (\textbf{Algorithm \ref{alg_joint_MMV_LS_CS}}), instead of reducing the complexity, the sequential search method enhances the resolution of the estimated channel support.
	M-FISTA first estimates the channel with the low-resolution codebook (line 1-10) inducing the grid-mismatch errors.
	To address such issue, we apply the sequential search method using the codebook as described in \eqref{step_spatial_h_AOD}-\eqref{step_spatial_v_AOA} to estimate the channel support in an enhanced beam resolution (line 11-21).

	\subsection{Computational Complexity} \label{subsec_complexity} 
	We recall that \textbf{TS} is done in two stages, MMV-LS and MMV-CS.
	The complexity of \text{MMV-LS} is dominated by the subset selection and its pseudo-inverse operation, searching over the combinations of $\{i_r,i_t\}\in\hat{\Omega}^{pr}$, with complexity $\mathcal{O}\Big(\lvert\hat{\Omega}^{pr}\rvert K_p(L_{cm})^3\Big)$ \cite{golub2013matrix}.
	The complexity of \text{MMV-CS} is dictated by the greedy selection, searching over all combinations of $\{g_r,g_t\}\in\mathcal{J}$, with complexity $\mathcal{O}\Big(G_r G_t K_pQ_pT_p\hat L\Big)$, where $\hat L$ is the number of estimated paths.
	\add{The TS algorithm requires traversing over all candidate AOA-AOD pairs for the greedy beam selection in MMV-CS, which leads to a prohibitive average execution time.
	It can be reduced by parallel processing because the correlation with the signal of each AOA-AOD pair could be calculated separately, called \textbf{TS with parallel processing}.}
	%	The greedy selection searches over all combinations of $\{g_r,g_t\}\in\mathcal{J}$, with complexity $\mathcal{O}\left(G_r G_t K_pQ_p\hat L\right)$ \cite{golub2013matrix}.
	%	 $\mathcal{O}\left(Q_p(T_p-T_d) K_p G_r G_t P\right)$;
	%	With the sequential search \eqref{eq_seq_search}, the complexity of the greedy selection can be reduced to $\mathcal{O}\left(Q_p(T_p-T_d) G_r G_t + PK_p\right)$.
	%	the least square operation requires the pseudo-inverse operation on each pilot subcarrier, with complexity $\mathcal{O}(K_p{\hat L}^3)$ \cite{golub2013matrix}, where $\hat L$ is the number of estimated paths in \text{MMV-CS}.	
	The proposed sequential search method greatly alleviates the overhead of the greedy selection, leading to the complexity $\mathcal{O}\Big(({G_{sub}^4} + 4(M-1)G_{sub}) K_pQ_pT_p\hat L\Big)$.
%	, where $G_{sub}$ is the sub-codebook size and $M$ is the number of levels.
	For \textbf{M-FISTA}, the complexity is dominated by the gradient calculation $\nabla f$.
	Thanks to the block diagonal structure of $\mathbf{\Phi}_A$ and $\mathbf{\Phi}_B$, the gradient $\nabla f$ can be calculated on each subcarrier separately, which reduces the complexity from $\mathcal{O}\Big(K_p^3G_r^2G_t^2Q_pT_p\Big)$ to
	$\mathcal{O}\Big(K_pQ_pT_p\lvert\Gamma\rvert\lvert\Gamma^c\rvert\Big)$, where $\Gamma$ is the previous channel support.
	\add{Note that M-FISTA is not suitable to operate with parallel processing since it applies the proximal gradient descent method.}
%		applies the proximalgradient descent method, whose implementation is not suitable for parallel processing.	}
	For the \textbf{channel refinement algorithm}, the complexity is dominated by the least squares operations \eqref{q_opt}. %, which has already been calculated in \text{MMV-CS} (line 18, \textbf{Algorithm} \ref{alg_tensor_LS_CS}).
	Assuming $L^\prime$ estimated paths, the update of $\mathbf{q}_k$ requires the pseudo-inverse operations on pilot subcarriers with complexity $\mathcal{O}\left(K_p(L^\prime)^3\right)$.
	\add{With the configuration as in TABLE \ref{table:simulation_setup}, we evaluate the complexity in terms of the average execution time: 
	$30.12$s for TS, 
	$7.12$s for TS with parallel processing, 
	$21.7$s for M-FISTA,
%	$31.2$s for M-FISTA, 
	$16.1$s	for M-FISTA without previous support, as opposed to $34.2$s for GSOMP (state-of-the-art).
	Note that TS with parallel processing starts a pool of $12$ workers in MATLAB for the greedy beam selection, which reduces the average execution time by $76$\% in our configuration.}

	\begin{table}[t]%\small
		\centering	
		\caption{Common Simulation Parameters \cite{rappaport2019wireless,xing2021millimeter}}
		\label{table:simulation_setup}		
		\resizebox{0.45\textwidth}{!}
		{		
			\begin{tabular}{|r    | c|l|} % centered columns (4 columns)
				%			\hline
				\hline %inserts double horizontal lines
				%			&&&&&&\\
				Parameter & Symbol & Value \\ [0.5ex] % inserts table
				%heading
				\hline % inserts single horizontal line
				Carrier frequency & $f_c$  & $142$ GHz \\
				Bandwidth & $B$  & $8$ GHz \\
				Total subcarriers & $K_o$  & {$1024$} \\
				Channel paths & $(L,L_{cm})$  & $(4,3)$ \\	
				%			 & $L_{cm}$  & $3$ \\			
				%			Channel paths & $L$  & $4$ \\	
				%			Common channel paths & $L_{cm}$  & $3$ \\		
				%			Path Delays & $\tau_\ell$& $\mathcal{U}(0,100)\textrm{ns}$\\
				Ref. path coefficient & $\alpha_\ell^\prime$ & $\mathcal{CN}(0,\sigma_\alpha^2)$\\				
				%			\hline %inserts single line			
				Polar AOA/AOD & $\theta_{pr}/\theta_{pt}$  & $\mathcal{U}(-\pi/2,\pi/2)$ \\	
				Azimuth AOA/AOD & $\theta_{ar}/\theta_{at}$  & $\mathcal{U}(-\pi,\pi)$ \\				
				Tx antenna (UPA) & $N_t\ (N_{vt}, N_{ht})$  & $16\ (4, 4)$ \\ 
				Rx antenna (UPA) & $N_r\ (N_{vr}, N_{hr})$  & $256\ (16, 16)$ \\
				%			Tx/Rx RF chain number & $(N^{RF}_t,N^{RF}_r)$  & $(8,8)$ \\ 
				%			Rx RF chain number & $N^{RF}_r$  & $8$ \\
				%			AOD resolution & $G_t$ & $512$ \\
				%			AOA resolution & $(G_{hr},G_{vr})$ & $(512,512)$ \\
				Tx/Rx sub-codebook & $(G_{sub,t},G_{sub,r})$ & $(4,16)$\\							
				Pilot subcarriers & $K_p$  & $10$\\
				Training parameters & $(Q_p,T_p)$  & $(25,25)$\\
				Subframe/Frame duration & $\delta_s/T_{frame}$  & $10\mu\textrm{s}/10\textrm{ms}$\\
				%			[1ex] % [1ex] adds vertical space
				\hline %inserts single line
				
				%			\hline %inserts single line
				%			\hline %inserts single line				
			\end{tabular}						
		}	
	\vspace{-1em}
	\end{table}

	\section{Numerical Results }\label{sec_simulation} 
	We evaluate the performance of the proposed channel estimation in time-varying MIMO-OFDM, and the numerical parameters are listed in Table \ref{table:simulation_setup}.
	We consider the multipath channel having delays $\tau_{\ell}\sim\mathcal{U}(45,55)\textrm{ns}$ with the delay spread $D_s=10\textrm{ns}$ \cite{xing2021millimeter}.
	This yields a coherence bandwidth  $B_c=\frac{1}{2D_s}=50\textrm{MHz}$ \cite{tse2005fundamentals} and the minimum number of subcarriers $\frac{B}{B_c}=160$.
	We consider the total number of subcarriers as $K_o=1024$.
	The receive and transmit array response matrices $\mathbf{A}_{R,k}$, $\mathbf{A}_{T,k}$ are constructed as in \eqref{rx_array_matrix} \eqref{tx_array_matrix}, with the uniform grid $\mathcal{G}_R\!=\!\mathcal{G}_{hR}\times \mathcal{G}_{vR}$ of size $G_r\!=\!G_{hr}G_{vr} $ and the uniform grid $\mathcal{G}_T\!=\!\mathcal{G}_{hT}\times \mathcal{G}_{vT}$ of size $G_t\!=\!G_{ht}G_{vt}$, respectively.
	The sequential search method with hierarchical codebooks as described in Section \ref{sec_seq_search} is applied.
	We consider a transmit hierarchical codebook with $(G_{sub,t},M)$ for the uniform grids $\mathcal{G}_{hT}$, $\mathcal{G}_{vT}$, i.e., $G_{ht}\!=\!G_{vt}\!=\!G_{sub,t}^{M}$.
	Similarly, a receive hierarchical codebook with $(G_{sub,r},M)$ is considered for the uniform grids $\mathcal{G}_{hR}$, $\mathcal{G}_{vR}$, i.e., $G_{hr}\!=\!G_{vr}\!=\!G_{sub,r}^{M}$.
	The receiving signal-to-noise ratio is defined as
	$\textrm{SNR} \!=\!
	\mathbb{E}\left[\frac{\sum_k\lVert \mathbf{W}_k^{H}\mathbf{H}_k \mathbf{X}_k \rVert_F^2}{\sum_k\mathbb{E}\left[\lVert \mathbf{\Tilde V}_k \rVert_F^2 \mid \mathbf{W}_k,\mathbf{X}_k, \mathbf{H}_k, \forall k \right]}\right]$,
	where $\mathbf{W}_k$, $\mathbf{X}_k$,  $\mathbf{H}_k$, and $\mathbf{\Tilde V}_k$ are the measurement, pilot, channel, and combined noise matrices, respectively, as in \eqref{IO_matrix_beamtraining_virtual}.

	We compare the proposed \textbf{TS} and \textbf{M-FISTA} channel estimation (with the \textit{"estimated" previous channel support}) 
	with the estimation schemes listed as follows:
	\begin{itemize}
		\item\textbf{GSOMP} \cite{dovelos2021channel}:	Apply the SOMP to estimate the AOAs/AODs and channel gains (with \textit{frequency-dependent} dictionary matrices) using common support across the pilot subcarriers. GSOMP considers the dual-wideband effect. 
		\item\textbf{DGMP} \cite{gao2016channel}:	Apply the distributed CS to estimate the AOAs/AODs and channel gains (with \textit{frequency-flat} dictionary matrices) using the structured sparsity and the grid mismatch pursuit strategy. We extend this design to the UPA case for comparison. DGMP considers the frequency-wideband effect but neglects the spatial-wideband effect. 
		\item\textbf{Genie-aided LS}: The LS estimator is applied with the \textit{current channel support}. This is considered a performance bound.
	\end{itemize}

	We also evaluate M-FISTA without previous channel support, called \textbf{M-FISTA w/o prev support}, which sets $\Gamma=\emptyset$ and ignores the term $\lambda_1\lVert\mathbf{\tilde z}^{LS}\rVert_{2,1}$ in \eqref{FISTA_MMS_LS_CS_original}.
	Note that we consider the number of training measurements with $(Q_p,T_p)=(25,25)$ (partial training of $15.3\% N_rN_t$ beams), while the work \cite{dovelos2021channel} evaluates the GSOMP estimator under the partial training of $80\%N_rN_t$ beams, which is much larger than our considered setting.
	In addition, the support tracking-based approaches needs to do the initial channel estimation (i.e., \textbf{MMV-CS} for \textbf{TS} or \textbf{M-FISTA w/o prev support} for \textbf{M-FISTA}) if the residual signal power is larger than a predefined threshold. 
	In our simulation, the predefined threshold is chosen such that the number of frames of the initial channel estimation (i.e., $\hat{\Omega}^{pr}=\emptyset$) is less than $10\%$ of all frames. 
	Note that the frames implementing the initial channel estimation are counted in the performance evaluation.
	
	\begin{figure}[t]
		\centering
		\includegraphics[scale=0.45]{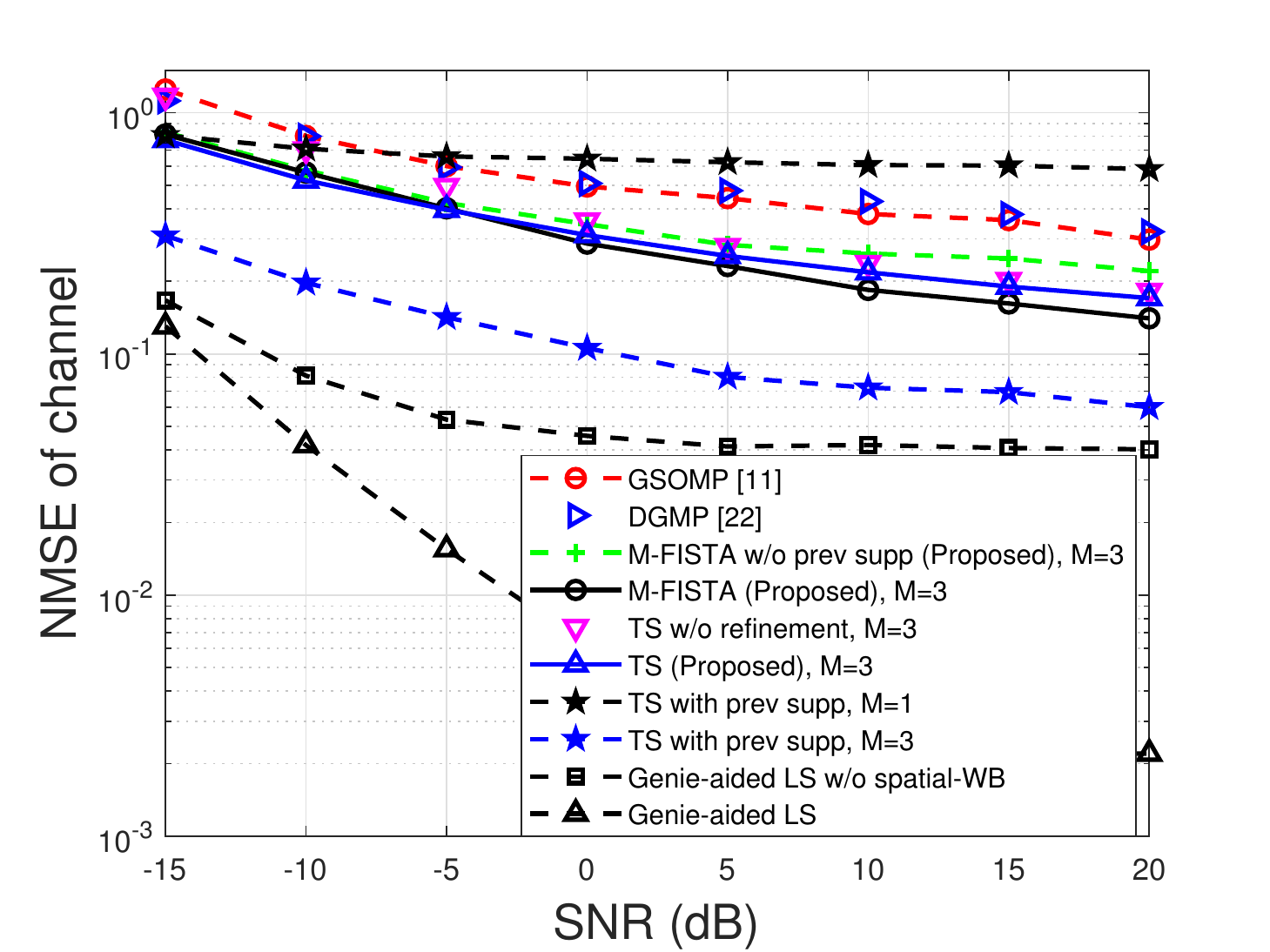}
		\captionof{figure}{The NMSE versus the SNR.}
		\label{fig:Exp1_NMSE_SNR}
	\end{figure}							
	
	%	\nt{NMSE v.s. SNR}	
	To evaluate the channel estimation accuracy, we define the normalized mean squared error (NMSE) of the estimated channel as 
	\begin{equation}\setcounter{equation}{37}
	\textrm{NMSE}=\frac{\sum_{k\in\mathcal{P}}\lVert\mathbf{H}_k - \mathbf{\hat H}_k\rVert_F^2}{\sum_{k\in\mathcal{P}}\lVert\mathbf{H}_k\rVert_F^2}.
	\end{equation}
	In Fig. \ref{fig:Exp1_NMSE_SNR}, we evaluate the NMSE of channel versus the SNR.
	Genie-aided LS attains $\text{NMSE} \!=\! 1.5\times 10^{-2}$ above $\text{SNR} \!=\!-5$dB, which provides a lower bound for the NMSE.
	\textbf{Genie-aided LS w/o spatial-WB} applies the LS estimator with the correct AOAs/AODs, ignoring the spatial-wideband effect in the channel reconstruction.
	Even with the correct AOAs/AODs estimation, it attains only $\text{NMSE}\!=\!4\times 10^{-2}$ when $\text{SNR}\!\geq\! 0$dB due to the spatial-wideband effect.
	To determine a suitable number of levels $M$ of the hierarchical codebook,
	we evaluate TS with the ``\emph{correct}'' previous channel support (called \textbf{TS with prev support}) scheme with different $M$.
	For $M\!=\!1$, TS with prev support attains $\text{NMSE}\!=\!0.6$ with $\textrm{SNR}\!=\!10$dB.
	In the same configuration, the NMSE improves with a larger $M$, attaining $0.07$ for $M\!=\!3$.
	We observe that there is only minor improvement as we increase the number of levels from $M=3$ to $M\!=\!4$.
	%. %since the resolution of the hierarchical codebook is sufficient with $M=3$.
	%	between the cases with $M=3$ and $M=4$.
	Thus, $M\!=\!3$ is chosen for the following experiments.
	Given an $\textrm{SNR}\!=\!20$dB, the proposed TS attains $\textrm{NMSE}=0.17$, as opposed to $0.06$ for \text{TS with prev support}, $0.29$ for \text{GSOMP}, and $0.32$ for \text{DGMP}.
	%	The proposed 
	\text{TS} is inferior to \text{TS with prev support} due to the potential errors in the estimated previous channel support because of the algorithm's reliance on accurate past support estimation, which is difficult at low SNR.
%	dependent on the previous channel estimation performance which tends to be inaccurate at the low SNR region.
%	The	previous estimated channel support depends on the previous channel estimation, which tends to	be inaccurate at the low SNR region.
	With $\textrm{SNR}\!=\!-10$dB, TS attains $\textrm{NMSE}\!=\!0.52$ while the TS w/o refinement only has $\textrm{NMSE}\!=\!0.7$.
	It shows that given	the same estimated AOAs/AODs of the channel paths, the channel refinement algorithm improves the estimation accuracy in the low SNR region.
	In addition, with $\textrm{SNR}\!=\!20$dB, the proposed M-FISTA attains $\textrm{NMSE}\!=\!0.14$, as opposed to $\textrm{NMSE}\!=\!0.22$ for M-FISTA w/o previous support. 
	We observe that M-FISTA has a better NMSE performance than M-FISTA w/o previous support at the high SNR region because the previous channel support can be estimated more accurately, which is beneficial to the LS-CS framework.
	When $\textrm{SNR}\!<\! 0$dB, M-FISTA attains a similar NMSE to M-FISTA w/o prev support.

	\begin{figure}[t]
		\centering
		\includegraphics[scale=0.45]{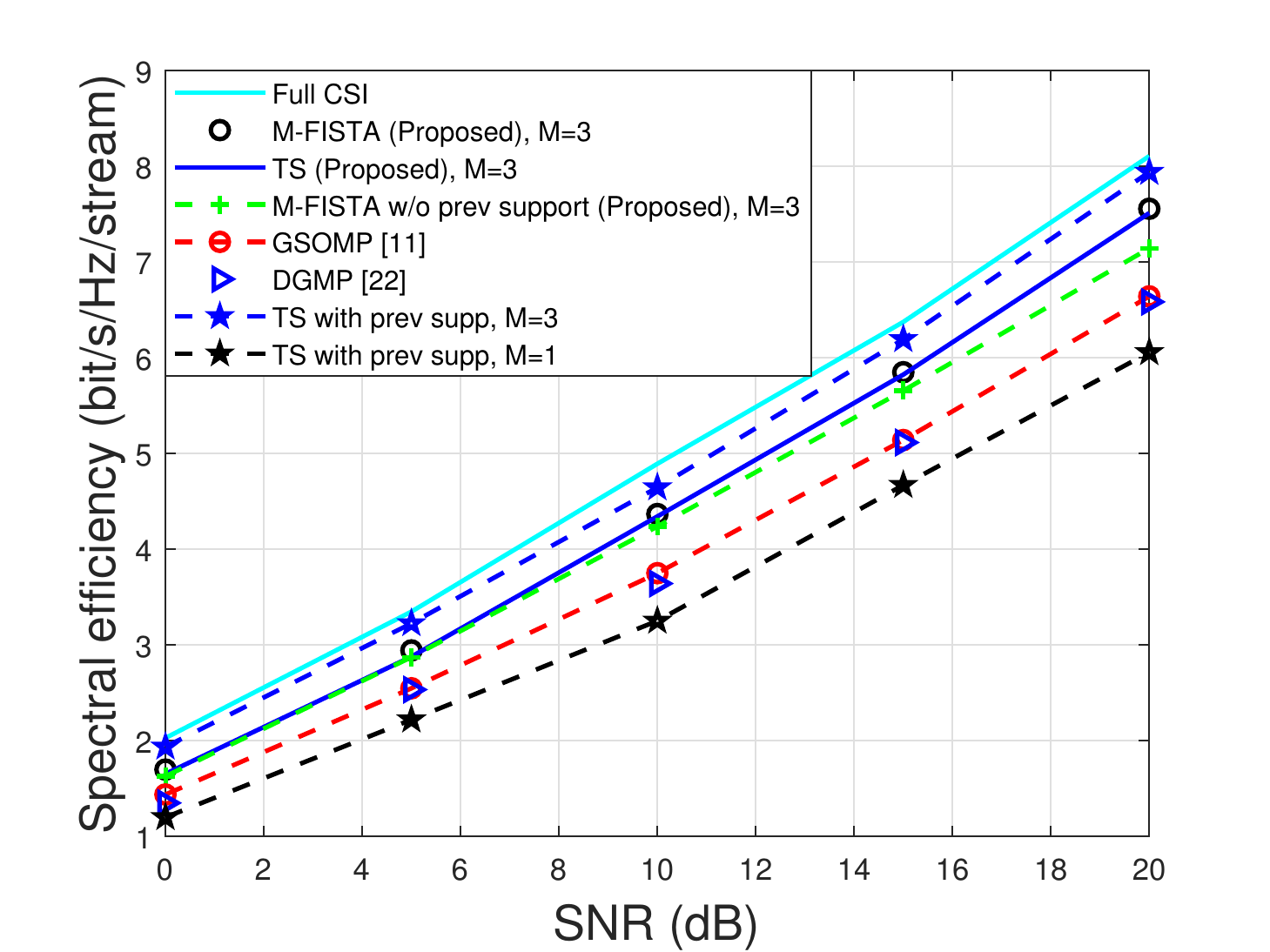}
		\caption{Spectral efficiency versus the SNR, with $N_s=4$. 
		}
		\label{fig:Exp7_SpecEff_SNR}
	\end{figure}

	%	\nt{Spectral Efficiency v.s. SNR}
	Next, we evaluate the spectral efficiency (SE).
	The rate expression maximized by Gaussian signaling \cite{6847111} over the channel is expressed as  
	\begin{equation}\label{eq:R_rate}
	R \!=\! \sum_{k=1}^{K_o}\!\frac{B}{K_o}\log_2 \textrm{det}\left( \mathbf{I}_{N_s}+\frac{P_t}{N_s}\mathbf{R}_{\hat {\mathbf v}_k}^{-1}\mathbf{\hat W}_k^{H}\mathbf{H}_k\mathbf{\hat F}_k\mathbf{\hat F}_k^H\mathbf{H}_k^H\mathbf{\hat W}_k\right),
	\end{equation}
	where $P_t$ is the average transmit power for each transmission.
	The number of data streams is assumed as $N_s=4$.
	The unconstrained combiner $\mathbf{\hat W}_k$ (or the precoder $\mathbf{\hat F}_k$) is derived by the directions of the eigenvectors of $\mathbf{\hat H}_k\mathbf{\hat H}_k^H$ (or $\mathbf{\hat H}_k^H\mathbf{\hat H}_k$).
	%	\nm{this is confusing; in your analysis you assumed F and W to be constant with respect to k!}
	%	\tc{In the current model, I consider the channel training model with frequency-dependent pilot and measurement matrices. Thus, the confusion should be addressed.}
	%	\tc{Moreover, even in the previous model, we allowed the the frequency-dependent hybrid precoder/combiner, but considered the frequency-flat pilot and measurement matrix for the \textit{channel training stage}. Yet, the achievable transmission rate is evaluated for the \textit{data transmission stage}.}
	%	are the eigenbeamforming solution derived from the estimated channel $\mathbf{\hat H}_k$.
	%	The combiner and transmitter $\mathbf{\hat W}_k/\mathbf{\hat F}_k$ are the eigenbeamforming solution derived from the estimated channel $\mathbf{\hat H}_k$.
	The post-processing noise covariance matrix is $\mathbf{R}_{\hat {\mathbf v}_k}=\mathbb{E}[\hat {\mathbf v}_k \hat {\mathbf v}_k^H]$, where $\hat {\mathbf v}_k=\mathbf{\hat W}_k^H \mathbf{v}$ with the additive complex Gaussian noise $\mathbf{v}$.
	The fraction of time for the pilot transmission is $\iota \!=\! \frac{T_{train}}{T_{frame}}$, where the duration of $T_{train}\!=\!T_p\delta_s$ on the $K_p$ pilot subcarriers is the resulting training overhead.
	The data transmission is allowed on the entire bandwidth at the transmission time, and also on the bandwidth other than the pilot subcarriers at the training time.
	We denote the rate on the training time as $R_{train}$, which is constructed as in \eqref{eq:R_rate} without the pilot subcarriers.
	Thus, the SE is defined as $\frac{\iota R_{train} + (1-\iota)R}{B N_s}$  (bit/s/Hz/stream), which includes the loss due to the training overhead.
%	Given a frame duration $T_{frame}$, the training time is assumed as $T_{train}=T_p\delta_s$, and the remaining duration is the transmission time.
%	The data transmission is allowed on whole bandwidth at the transmission time, and also on the bandwidth other that the pilot subcarriers at the training time.		
%	The fraction of time for the data transmission is $\iota=\frac{T_{frame}-T_{train}}{T_{frame}}$, where $T_{train}=T_p\delta_s$ is the resulting overhead.	
	%	The resulting overhead is $T_{train}=T_p\delta_s$, where the subframe duration $\delta_s=10\mu\textrm{s}$.\nm{in the table please!}
	%	The fraction of time for the data transmission is $f_{comm}=\frac{T_{frame}-T_{train}}{T_{frame}}$, where $T_{frame}=10\textrm{ms}$ is the frame duration.\nm{in the table please!}
%	Thus, the SE is defined as $\frac{\iota R}{B N_s}$ (bit/s/Hz/stream), which accounts for the loss due to the training overhead.
	%	In our experiment, we assume the length of training sequence $Tp = 14$, so the fraction of time used for data transmission	is identical for all schemes.
	%	Thus, the spectral efficiency is defined as $\frac{R}{B N_s}$ (bit/s/Hz/stream).
	In Fig.~\ref{fig:Exp7_SpecEff_SNR}, we evaluate the SE versus the SNR.
	\text{Full CSI} attains the largest SE because its $\mathbf{\hat W}_k$ and $\mathbf{\hat F}_k$ are derived
	from the perfect CSI.
	TS with prev support scheme attains a better SE with a larger number of levels, consistent with the NMSE evaluation.
	For $\textrm{SNR}\!=\!20$dB, the SE of M-FISTA and TS both attain $7.6$ bit/s/Hz/stream, while the SE of GSOMP and DGMP are $6.63$bit/s/Hz/stream and $6.58$bit/s/Hz/stream, respectively.
	As opposed to TS with prev support, the SE loss of TS is around $0.42$bit/s/Hz/stream, originating from the potential errors in the estimated previous channel support.
	Even for the case with no previous channel support, our proposed \text{M-FISTA w/o prev support} attains $7.1$bit/s/Hz/stream, which also outperforms the state-of-the-art (GSOMP) by $0.5$bit/s/Hz/stream.

	\begin{figure}[t]
		\centering			
		\includegraphics[scale=0.45]{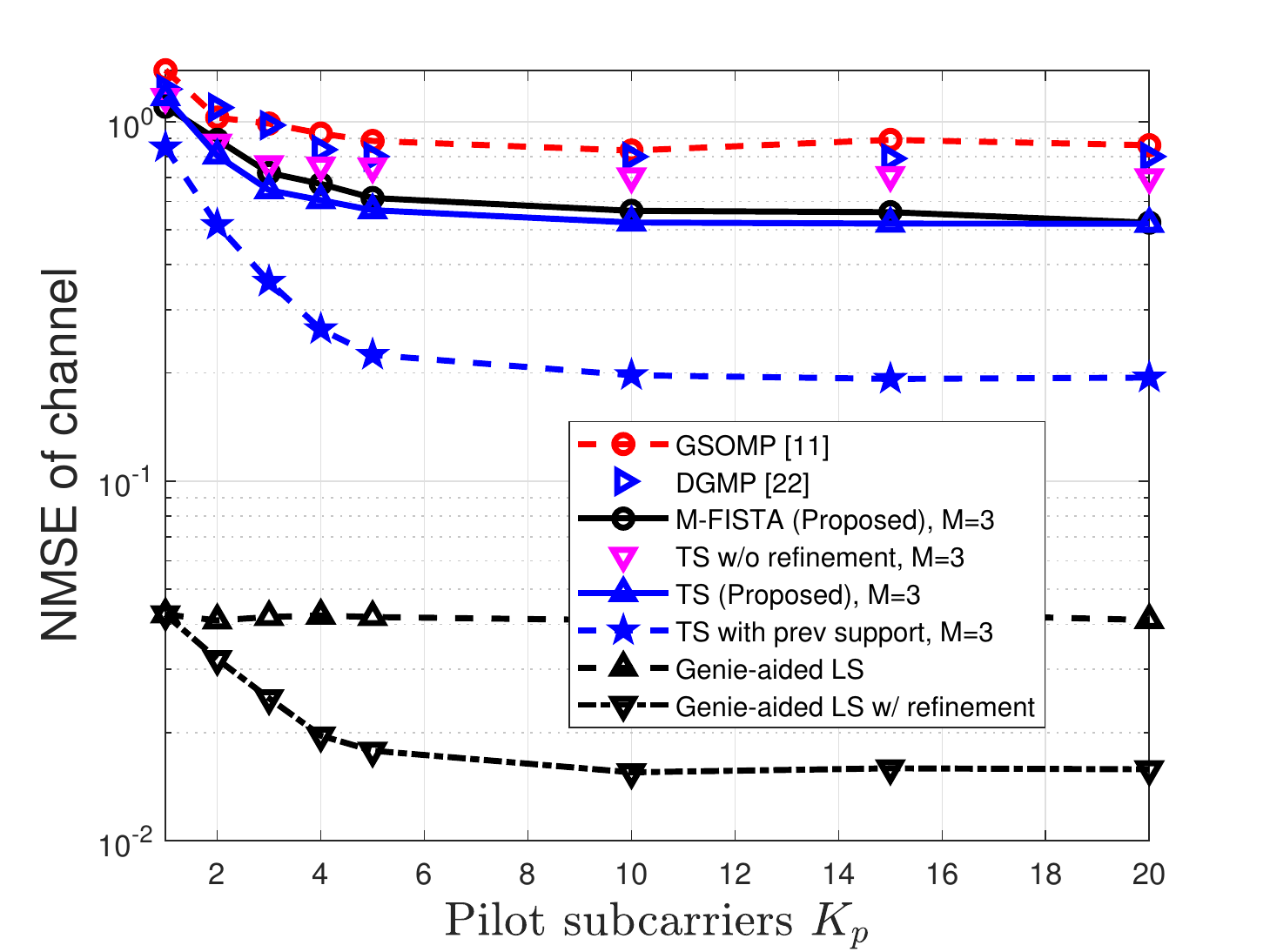}
		\caption{The NMSE versus the number of pilot subcarriers $K_p$, with $\textrm{SNR}= -10$dB. }
		\label{fig:Exp2_NMSE_Kp}
		\vspace{-1em}
	\end{figure}

	%	\nt{NMSE v.s. Kp}
	In Fig. \ref{fig:Exp2_NMSE_Kp}, we evaluate the NMSE versus the number of pilot subcarriers $K_p$, with $\textrm{SNR}=-10$dB.
	For $K_p=1$, we perform LS estimation to reconstruct the channel for \textbf{the schemes with refinement} since there are no multiple subcarriers to do the channel refinement algorithm. %(\textbf{Algorithm \ref{alg_gain_delay_refine}}).
	With $K_p=1$, we have $\textrm{NMSE}=1.1$ for both the TS and M-FISTA,
	as opposed to $\textrm{NMSE}=0.04$ for Genie-aided LS, $\textrm{NMSE}=0.8$ for {TS with prev support}, $\textrm{NMSE}=1.38$ for GSOMP, and $\textrm{NMSE}=1.23$ for DGMP.
	As we increase to $K_p\!=\!5$, we observe that the NMSE of \text{TS} and \text{M-FISTA} significantly improve to $\textrm{NMSE}\!=\!0.56$ and $\textrm{NMSE}\!=\!0.61$, respectively.
	The improvement comes from the MMV using the common channel support across subcarriers.
	%	It is due to the common channel support shared across the pilot subcarriers (MMV), which improves the estimation performance.
	Also, the channel refinement algorithm provides a more accurate estimation, especially in the low SNR region.
	%	Besides, when the pilot subcarrier $K_p$ increases, the channel refinement algorithm provides a more accurate estimation performance by canclnoise effect by averaging o.
	Note that the NMSE of M-FISTA w/o prev support is omitted in Fig. \ref{fig:Exp2_NMSE_Kp} since it is similar to the one of M-FISTA in the low SNR region.
	We observe that {Genie-aided LS with refinement} attains $\textrm{NMSE}\!=\!0.018$ when $K_p\!=\!5$, as opposed to \text{Genie-aided LS} attains $\textrm{NMSE}\!=\!0.04$.
	The improvement arises from fitting the reference path coefficients and time delays across subcarriers.
	For $K_p>5$, all approaches have fairly minor improvement with additional pilot subcarriers.

\begin{figure}[t]
		\centering
		\includegraphics[scale=0.45]{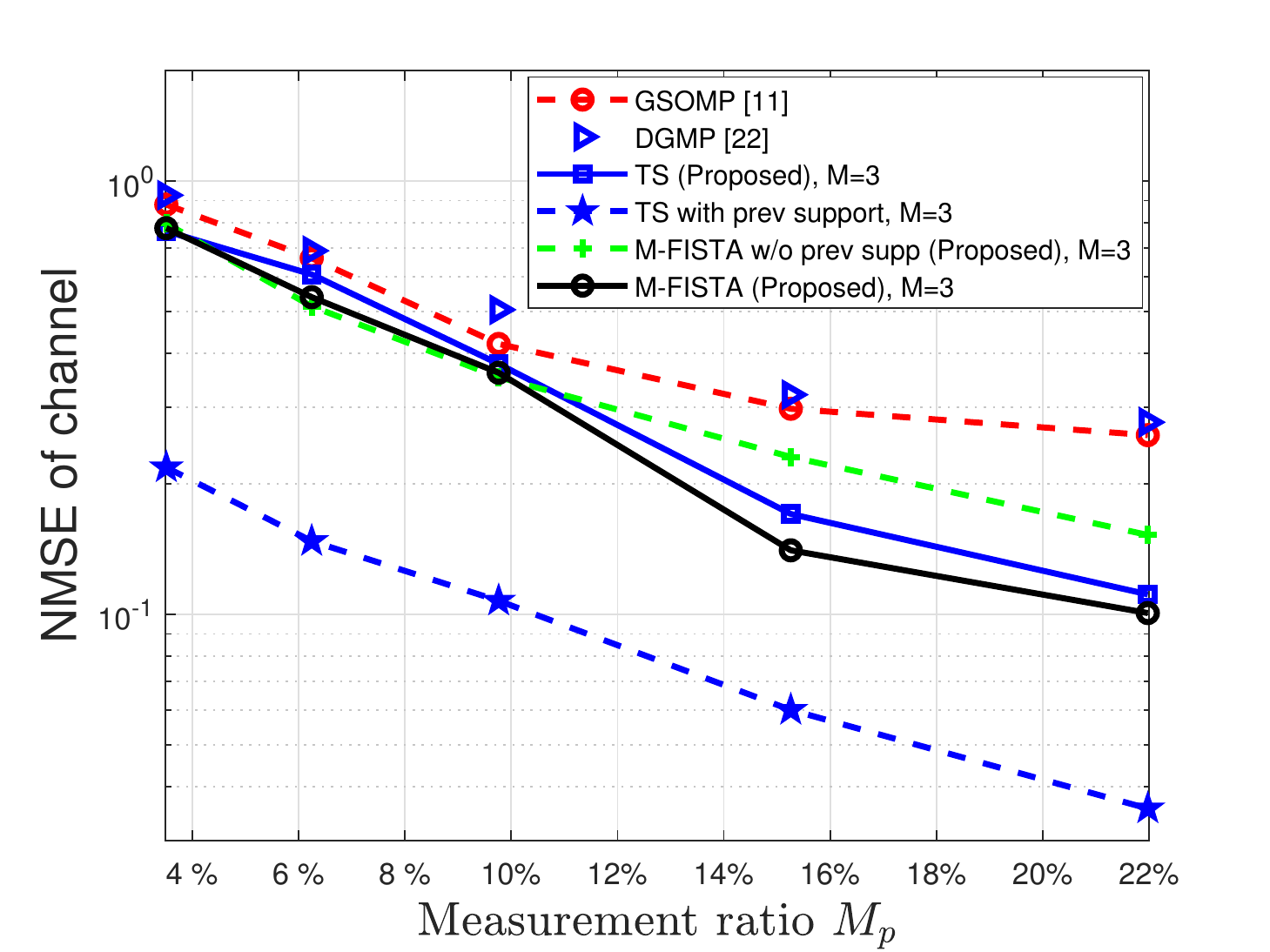}
		\caption{The NMSE versus the measurement ratio $M_p=Q_pT_p/N_rN_t$, with $\textrm{SNR}= 20$dB.}
		\label{fig:Exp8_NMSE_Np}
		\vspace{-1em}
	\end{figure}

	%	\nt{Wordings to be changed.}
	In Fig. \ref{fig:Exp8_NMSE_Np}, with $\textrm{SNR}=20$dB, we evaluate the NMSE of the estimated channel versus the measurement ratio $M_p \!=\! \frac{N_p}{N_rN_t}$, where $N_p$ is the number of training measurements.	
	Given $N_p$ measurements, we assume the training parameters as $Q_p\!=\!T_p\!=\!\sqrt{N_p}$.
	The NMSE of channel decreases with more training measurements.
	\text{TS} and \text{M-FISTA} both attain $\textrm{NMSE}=0.7$ with $M_p=3.5\%$ ($Q_p\!=\!T_p\!=\!12$), as opposed to GSOMP and DGMP achieves $\textrm{NMSE}\!=\!0.90$ and $\textrm{NMSE}\!=\!0.88$ in the same configuration.
	As we increase the training measurements to $M_p \!=\! 15.3\%$ ($Q_p\!=\!T_p\!=\!25$), the NMSE performance improves ($\textrm{NMSE}\!=\!0.17$ for TS, $\textrm{NMSE}\!=\!0.14$ for M-FISTA, $\textrm{NMSE}\!=\!0.23$ for M-FISTA w/o prev support, $\textrm{NMSE}\!=\!0.32$ for DGMP, and $\textrm{NMSE}\!=\!0.29$ for GSOMP), which shows the advantage of our MMV-LS-CS channel training.

\begin{figure}[t]
		\centering
		\includegraphics[scale=0.45]{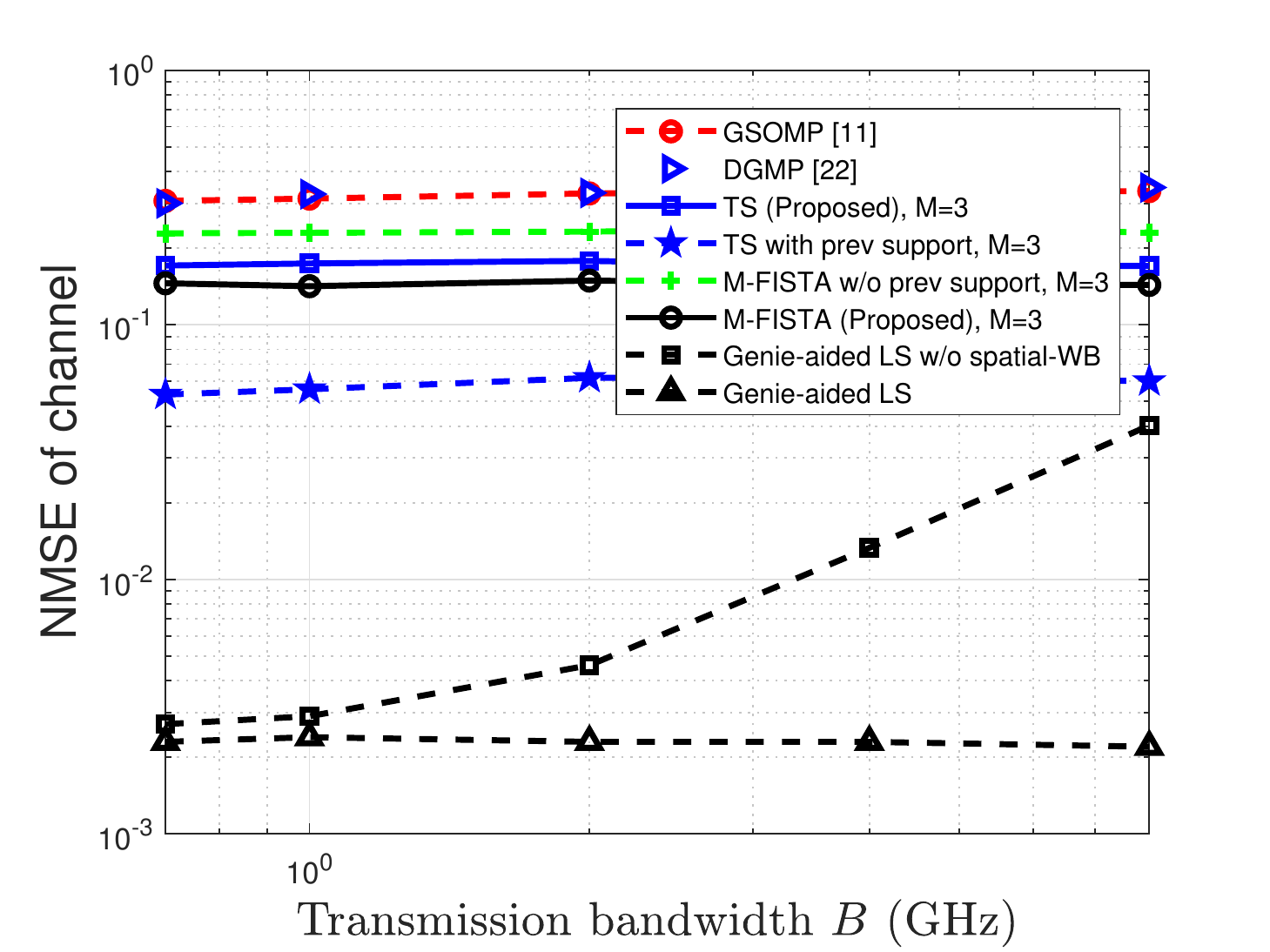}
		\caption{The NMSE versus the transmission bandwidth $B$, with $\textrm{SNR}= 20$dB.}
		\label{fig:Exp5_NMSE_fB}
	\end{figure}

	%	\nt{NMSE v.s. fB}	
	In Fig. \ref{fig:Exp5_NMSE_fB}, we evaluate the NMSE of the estimated channel versus the bandwidth, with $\textrm{SNR}\!=\!2
	0$dB.			
	As the bandwidth increases from $0.7$ to $8$ GHz, the NMSE of channel increases by around $0.005$ for TS and $0.006$ for M-FISTA, as opposed to $0.03$ for GSOMP and $0.04$ for GSOMP.
	The TS, M-FISTA, and GSOMP algorithm are not sensitive to the change of bandwidths owing to the frequency-dependent dictionary matrices.		
%	The pilot subcarrier spacing is ${B}/{K_p}=70$MHz at $B=0.7$GHz, but it requires the scale of ${1}/{D_s} = 100$MHz (delay spread $D_s=10$ns) to identify the time delay.
%	should be at least ${1}/{D_s} = 100$MHz (delay spread $D_s=10$ns) to identify the time delay while it is only ${B}/{K_p}=70$MHz at $B=0.7$GHz.
	In this configuration, the Genie-aided LS w/o spatial-WB scheme deteriorates from $\textrm{NMSE}\!=\!2.7\times 10^{-3}$ to $\textrm{NMSE}\!=\!4\times 10^{-2}$ because the dual-wideband effect becomes more severe with a larger bandwidth.
	We observe that the NMSE of channel increases even with the accurate AOAs/AODs estimation if the spatial-wideband effect is neglected, which shows the necessity of using the frequency-dependent dictionary matrices.

	\section{Conclusion}\label{sec_conclusion} 
	We proposed a CS channel training for time-varying sub-THz dual-wideband MIMO-OFDM systems.
	We constructed the frequency-dependent array response matrices to preserve the common channel support among subcarriers and enabled the MMV channel recovery.
	For slowly-varying channels, we employed the previous channel support to develop an MMV-LS-CS framework, and proposed two channel estimation algorithms (TS and M-FISTA). 
	The TS algorithm adopts a two-stage procedure to do the MMV-CS on the MMV-LS residual signal. The M-FISTA algorithm solved a joint MMV-LS-CS using the framework of FISTA.
	We proposed a channel refinement algorithm to reconstruct the channel by estimating the time delays and path coefficients jointly on the subcarriers, leveraging the spreading loss structure. 
	To reduce the computational complexity and enhance the beam resolution, we proposed a sequential search method using hierarchical codebooks.
	Numerical results showed that the TS and M-FISTA algorithms provide improved estimation accuracy and SE over the state-of-the-art techniques.
	
	%	\ifCLASSOPTIONcaptionsoff	\newpage	\fi
	
%	{
		\bibliographystyle{IEEEtran}
    	\bibliography{IEEEabrv,reference}
%}

	\begin{IEEEbiography}[{\includegraphics[width=1.25in,height=1.25in,clip,keepaspectratio]{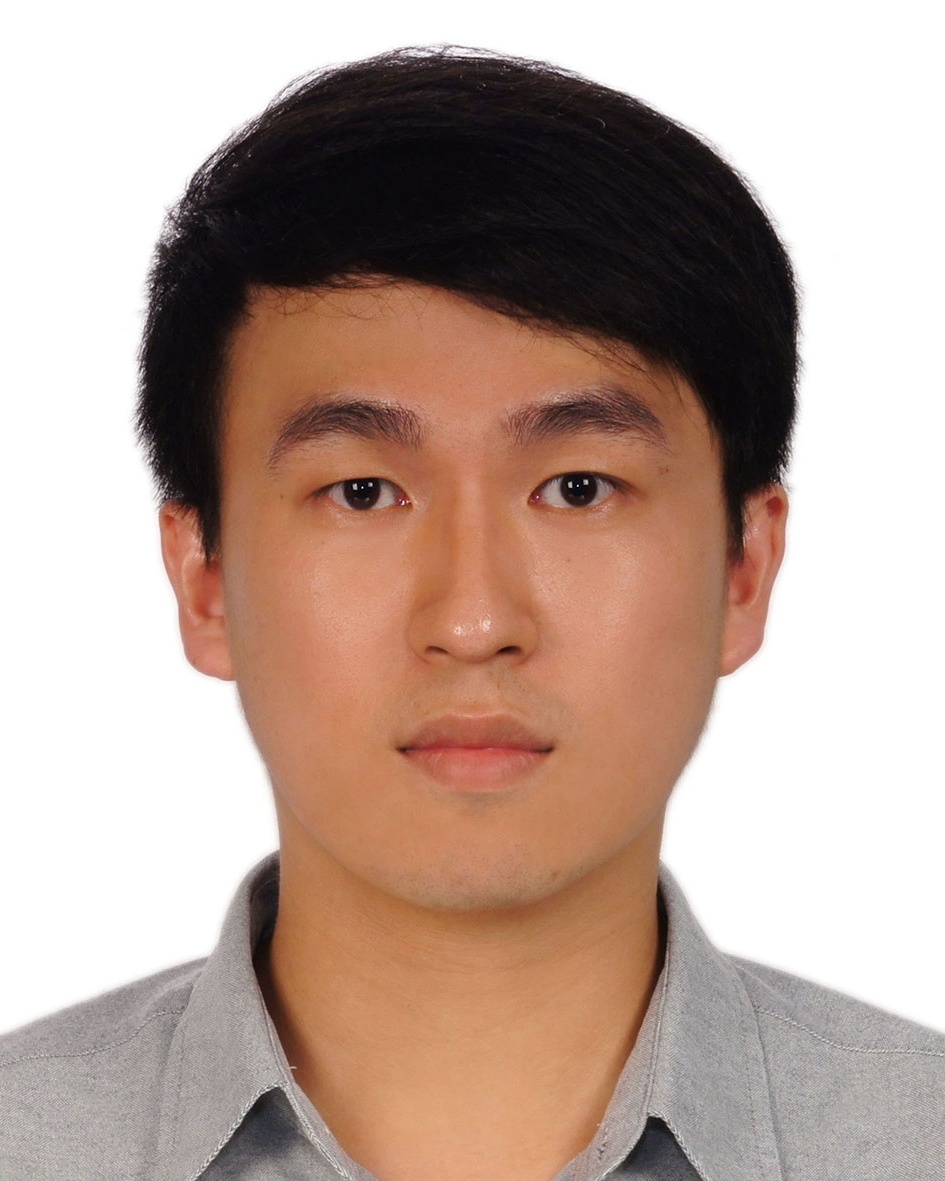}}]{Tzu-Hsuan Chou}(Member, IEEE)
	received the B.S. degree in electrical engineering and the M.S. degree in electrical and control engineering from National Chiao Tung University, Hsinchu, Taiwan, in 2011 and 2013, respectively, and the Ph.D. degree in electrical and computer engineering from Purdue University, West Lafayette, IN, USA, in 2022.
%	received the B.S. degree in electrical engineering and the M.S. degree in electrical and control engineering from National Chiao Tung University, Hsinchu, Taiwan, in 2011 and 2013, respectively. 
%	He is currently pursuing the Ph.D. degree with the School of Electrical and Computer Engineering, Purdue University, West Lafayette, IN, USA.
	He is currently a Senior Engineer with Qualcomm Inc., San Diego, CA, USA.
	From 2014 to 2017, he was a Software Engineer with MediaTek, Hsinchu, Taiwan.	
	His research interests include massive MIMO, wireless communications, compressed sensing, and tensor decomposition for signal processing.
\end{IEEEbiography}	
\vskip -0.8\baselineskip plus -1fil
\begin{IEEEbiography}[{\includegraphics[width=1in,height=1.25in,clip,keepaspectratio]{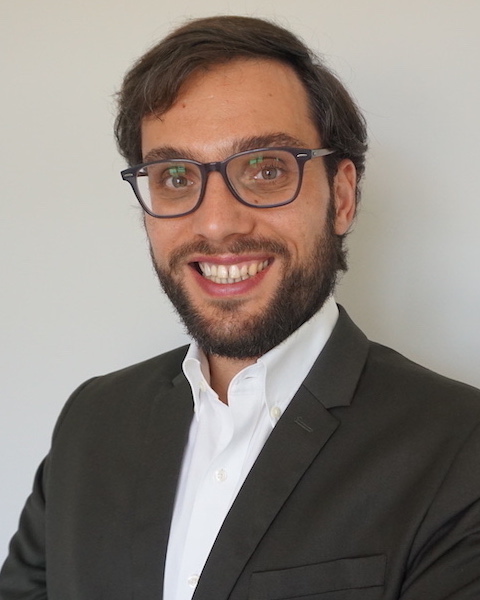}}]
	{Nicol\`{o} Michelusi} (Senior Member, IEEE) received the B.Sc. (with honors), M.Sc. (with honors), and Ph.D. degrees from the University of Padova, Italy, in 2006, 2009, and 2013, respectively, and the M.Sc. degree in telecommunications engineering from the Technical University of Denmark, Denmark, in 2009, as part of the T.I.M.E. double degree program. From 2013 to 2015, he was a Postdoctoral Research Fellow with the Ming-Hsieh Department of Electrical Engineering, University of Southern California, Los Angeles, CA, USA, and from 2016 to 2020, he was an Assistant Professor with the School of Electrical and Computer Engineering, Purdue University, West Lafayette, IN, USA. He is currently an Assistant Professor with the School of Electrical, Computer and Energy Engineering, Arizona State University, Tempe, AZ, USA. His research interests include 5G wireless networks, millimeter-wave communications, stochastic optimization, distributed optimization, and federated learning over wireless systems. He served as Associate Editor for the IEEE TRANSACTIONS ON WIRELESS COMMUNICATIONS from 2016 to 2021, and currently serves as Editor for the IEEE TRANSACTIONS ON COMMUNICATIONS. He was the Co-Chair for the Distributed Machine Learning and Fog Network workshop at the IEEE INFOCOM 2021, the Wireless Communications Symposium at the IEEE Globecom 2020, the IoT, M2M, Sensor Networks, and Ad-Hoc Networking track at the IEEE VTC 2020, and the Cognitive Computing and Networking symposium at the ICNC 2018. He is the Technical Area Chair for the Communication Systems track at Asilomar 2023. He received the NSF CAREER award in 2021 and the 2022 Early Achievement Award of the IEEE Communications Society - Communication Theory Technical Committee (CTTC).
\end{IEEEbiography}
%\vskip -0.5\baselineskip plus -1fil
\begin{IEEEbiography}
	[{\includegraphics[width=1in,height=1.25in,clip,keepaspectratio]{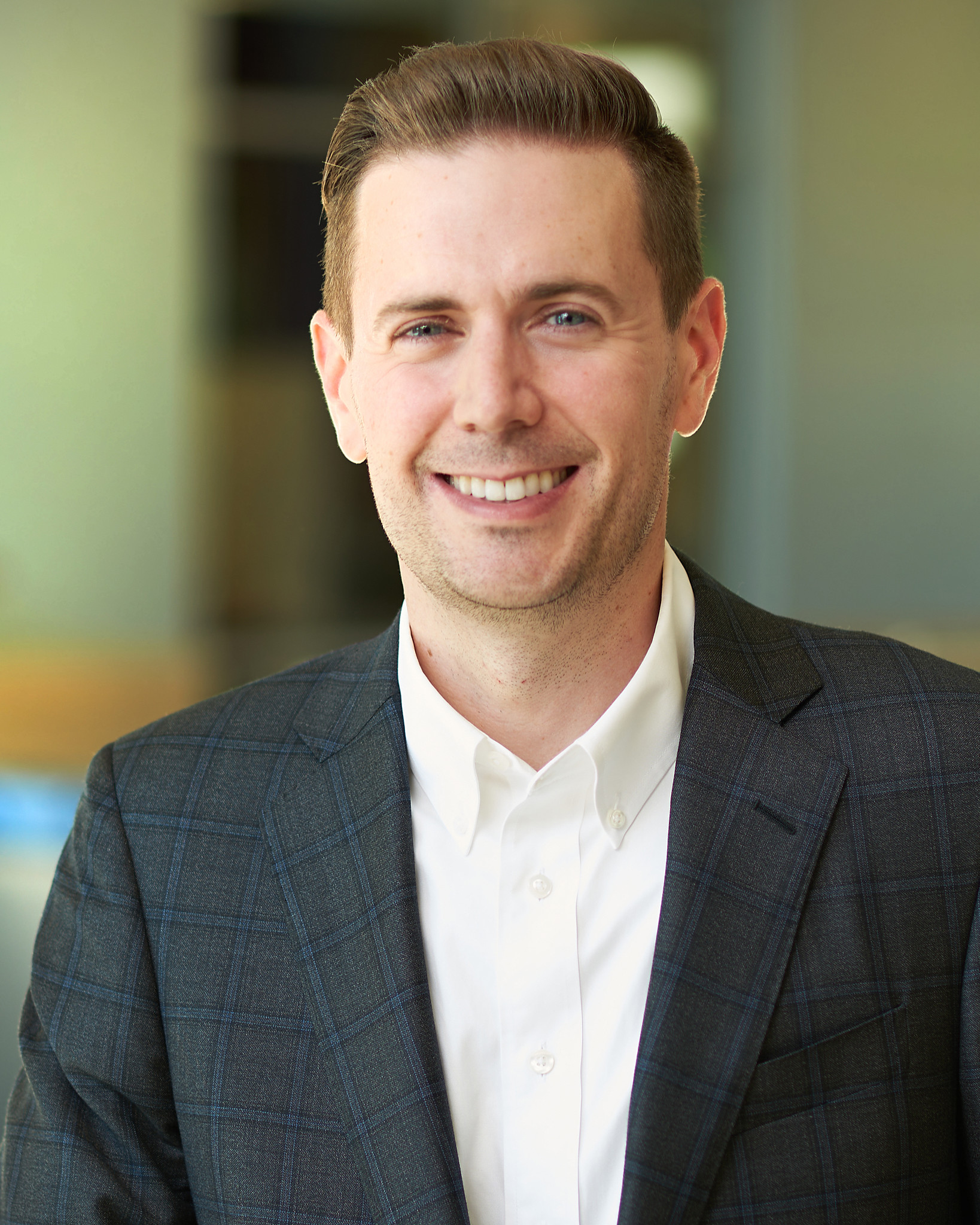}}]		
	{David J. Love}	(Fellow, IEEE) 
%	(S'98 - M'05 - SM'09 - F'15) 
	received the B.S. (with highest honors), M.S.E., and Ph.D. degrees in electrical engineering from the University of Texas at Austin in 2000, 2002, and 2004, respectively. Since 2004, he has been with the Elmore Family School of Electrical and Computer Engineering at Purdue University, where he is now the Nick Trbovich Professor of Electrical and Computer Engineering. He served as a Senior Editor for IEEE Signal Processing Magazine, Editor for the IEEE Transactions on Communications, Associate Editor for the IEEE Transactions on Signal Processing, and guest editor for special issues of the IEEE Journal on Selected Areas in Communications and the EURASIP Journal on Wireless Communications and Networking. He was a member of the Executive Committee for the National Spectrum Consortium. He holds 32 issued U.S. patents. His research interests are in the design and analysis of broadband wireless communication systems, beyond-5G wireless systems, multiple-input multiple-output (MIMO) communications, millimeter wave wireless, software defined radios and wireless networks, coding theory, and MIMO array processing.
	
	Dr. Love is a Fellow of the American Association for the Advancement of Science (AAAS) and was named a Thomson Reuters Highly Cited Researcher (2014 and 2015). Along with his co-authors, he won best paper awards from the IEEE Communications Society (2016 Stephen O. Rice Prize and 2020 Fred W. Ellersick Prize), the IEEE Signal Processing Society (2015 IEEE Signal Processing Society Best Paper Award), and the IEEE Vehicular Technology Society (2010 Jack Neubauer Memorial Award).				
\end{IEEEbiography}
\begin{IEEEbiography}[{\includegraphics[width=1.0in,height=1.3in,clip,keepaspectratio]{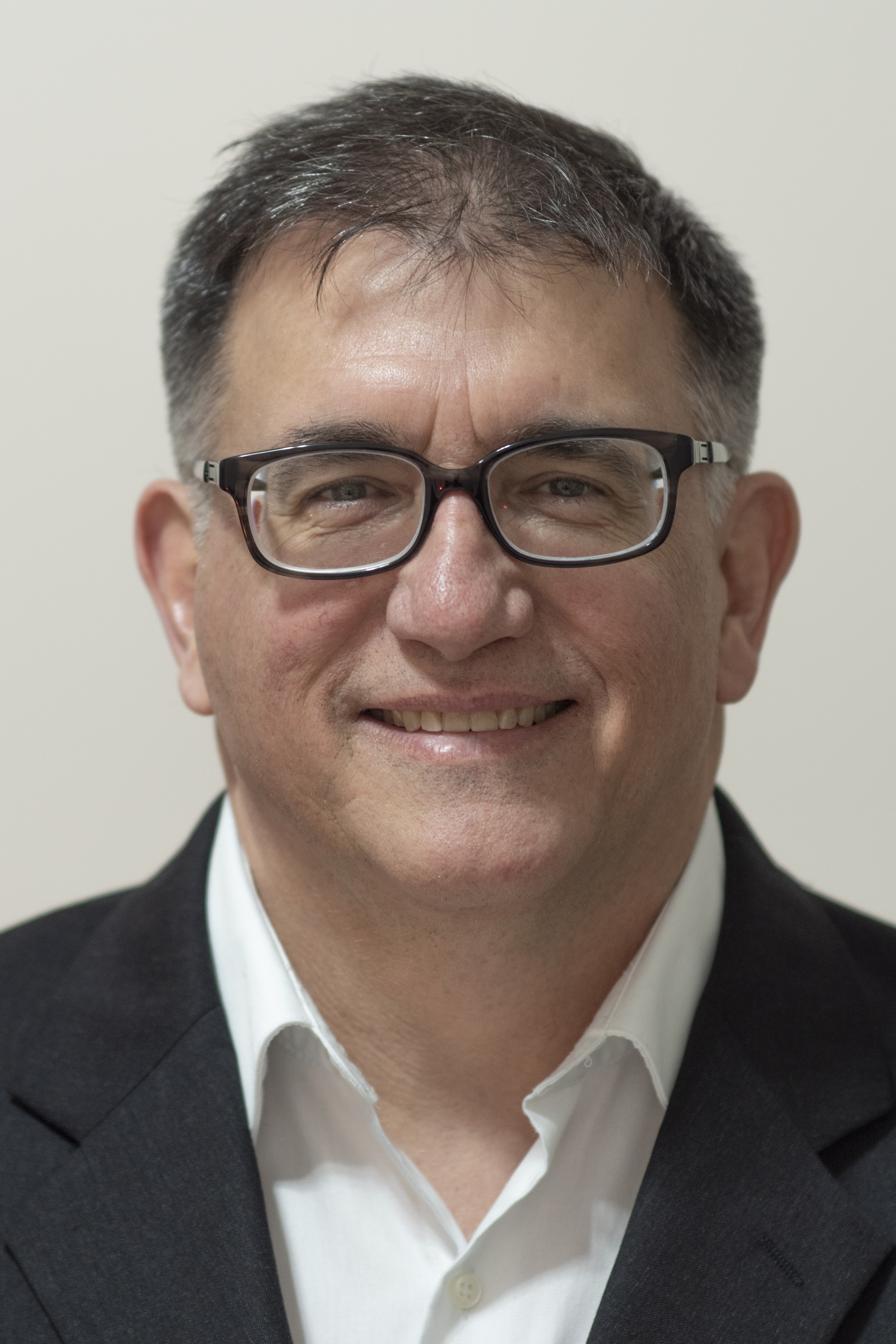}}]
	{James V. Krogmeier} (Senior Member, IEEE) received the B.S.E.E. degree from the University of Colorado Boulder, Boulder, CO, USA, and the M.S. and Ph.D. degrees from the University of Illinois at Urbana-Champaign, Champaign, IL, USA. He has industry experience in telecommunications and is a Founding Member of two software startup companies. He is currently a Professor of electrical and computer engineering with Purdue University, West Lafayette, IN, USA. He has authored or coauthored many technical papers in refereed journals and conference proceedings of the IEEE, the ASABE, and the Transportation Research Board, and is a Co-Inventor of five U.S. patents. His research interests include the applications of statistical signal and image processing in agriculture, intelligent transportation systems, sensor networking, and wireless communications. His research has been funded by the USDA-NIFA, the NSF, the DARPA, the Indiana Department of Transportation, the Federal Highway Administration, and industry. He was on a number of IEEE technical program committees and an Associate Editor for several IEEE journals.	
\end{IEEEbiography}
	
\end{document}